\documentclass[twocolumn,nofootinbib,superscriptaddress,preprintnumbers,floatfix]{revtex4}

\usepackage{graphicx}
\usepackage{amsmath}
\usepackage[bookmarks=false]{hyperref}

\newcommand{\eq}[1]{Eq.~\eqref{eq:#1}}
\newcommand{\eqs}[2]{Eqs.~\eqref{eq:#1} and \eqref{eq:#2}}
\renewcommand{\sec}[1]{Sec.~\ref{sec:#1}}
\newcommand{\subsec}[1]{Sec.~\ref{subsec:#1}}
\newcommand{\app}[1]{App.~\ref{app:#1}}

\newcommand{\apps}[2]{Apps.~\ref{app:#1} and \ref{app:#2}}

\newcommand{\abs}[1]{\lvert#1\rvert}
\newcommand{\mae}[3]{\langle#1\lvert#2\rvert#3\rangle}
\newcommand{\vevB}[1]{\langle #1 \rangle_{\! B}}
\newcommand{\ket}[1]{\lvert#1\rangle}
\newcommand{\ord}[1]{\mathcal{O}(#1)}

\newcommand{\inte}[1]{\int\! \df #1 \,}
\newcommand{\intlim}[3]{\int_{#1}^{#2}\! \df #3 \,}

\newcommand{\e}{\epsilon}
\newcommand{\w}{\omega}
\newcommand{\V}{V}

\newcommand{\df}{\mathrm{d}}

\newcommand{\vp}{\vec{p}}
\newcommand{\cL}{\mathcal{L}}

\newcommand{\hC}{\widehat{C}}
\newcommand{\hP}{\widehat{P}}
\newcommand{\hF}{\widehat{F}}
\newcommand{\hFmod}{\widehat{F}^{\rm mod}}

\newcommand{\beq}{\begin{equation}}
\newcommand{\eeq}{\end{equation}}
\newcommand{\beqa}{\begin{eqnarray}}
\newcommand{\eeqa}{\end{eqnarray}}
\newcommand{\nn}{\nonumber}

\newcommand{\lqcd}{\Lambda_\mathrm{QCD}}
\newcommand{\GeV}{\,\mathrm{GeV}}

\newcommand{\pxp}{\ensuremath{p_X^+}}
\newcommand{\pxm}{\ensuremath{p_X^-}}
\newcommand{\pxpm}{\ensuremath{p_X^\pm}}
\newcommand{\inv}{\mathrm{i}}
\newcommand{\Fname}{$F$}

\allowdisplaybreaks[4]

\begin{document}


\preprint{\vbox{\hbox{arXiv:0807.1926}\hbox{MIT--CTP 3955}}}

\title{\boldmath Treating the $b$ quark distribution function with reliable uncertainties}

\author{Zoltan Ligeti}
\affiliation{Ernest Orlando Lawrence Berkeley National Laboratory,
University of California, Berkeley, CA 94720}

\author{Iain W.\ Stewart}
\affiliation{Center for Theoretical Physics, Massachusetts Institute of
Technology, Cambridge, MA 02139\vspace{4pt}}

\author{Frank J.\ Tackmann}
\affiliation{Ernest Orlando Lawrence Berkeley National Laboratory,
University of California, Berkeley, CA 94720}

\begin{abstract}

  The parton distribution function for a $b$ quark in the $B$ meson (called the
  shape function) plays an important role in the analysis of the $B\to
  X_s\gamma$ and $B\to X_u\ell\bar\nu$ data, and gives one of the dominant
  uncertainties in the determination of $|V_{ub}|$.  We introduce a new
  framework to treat the shape function, which consistently incorporates its
  renormalization group evolution and all constraints on its shape and moments
  in any short distance mass scheme.  At the same time it allows a reliable
  treatment of the uncertainties. We develop an expansion in a suitable complete
  set of orthonormal basis functions, which provides a procedure for
  systematically controlling the uncertainties due to the unknown functional
  form of the shape function. This is a significant improvement over fits to
  model functions.  Given any model for the shape function, our
  construction gives an orthonormal basis in which the model occurs as the first
  term, and corrections to it can be studied.  We introduce a new short distance
  scheme, the ``invisible scheme", for the kinetic energy matrix element,
  $\lambda_1$.  We obtain closed form results for the differential rates that
  incorporate perturbative corrections and a summation of logarithms at any
  order in perturbation theory, and present results using known
  next-to-next-to-leading order expressions.  The experimental implementation of
  our framework is straightforward.

\end{abstract}

\maketitle

\section{Introduction}

The determination of the Cabibbo-Kobayashi-Maskawa (CKM) matrix element
$|V_{ub}|$ from inclusive semileptonic $B \to X_u \ell\bar\nu$ decays suffers
from large $B \to X_c \ell\bar\nu$ backgrounds.  In most regions of phase space
where this background is kinematically forbidden, the hadronic physics enters
via unknown nonperturbative functions, so-called shape functions.  At leading
order in $\lqcd/m_b$, there is only one such function, which can be extracted
from the photon energy spectrum in $B\to X_s\gamma$~\cite{Neubert:1993ch,
Bigi:1993ex} and used to predict various $B\to X_u \ell\bar\nu$ spectra.  In
$B\to X_s\gamma$ it is the main uncertainty in the effect of the cut on the
photon energy, which is near the border of the region where a local operator
product expansion (OPE) is applicable. Due to experimental cuts, the shape
function is also important for $B\to X_s\ell^+\ell^-$~\cite{Lee:2005pw,
Lee:2005pk}.

The determination of $|V_{ub}|$ received renewed attention recently, since the
measurement of $\sin2\beta$ favors a somewhat smaller value of $|V_{ub}|$ than
its determination from inclusive decays.  One of the most sensitive tests of the
standard model flavor sector comes from comparing the sides and angles of the
unitarity triangle, so it is important to determine $|V_{ub}|$ with minimal
model dependence.  Refined calculations of the $B\to X_s\gamma$
rate~\cite{Misiak:2006zs} also provide stringent constraints on new physics.

To obtain $|V_{ub}|$ as precisely as possible, one should combine all existing
information on the shape function.  The shape function is constrained by the
measurements of the shape of the $B\to X_s\gamma$ photon energy
spectrum~\cite{Chen:2001fja, Abe:2008sx, Aubert:2007my} and the $m_X$ spectrum
in $B\to X_u\ell\bar\nu$~\cite{Aubert:2004bq}, and its moments are related to
the $b$ quark mass, $m_b$, and nonperturbative matrix elements of local
operators in the OPE, which are constrained by fits to $B\to X_c \ell\bar\nu$
decay distributions~\cite{Bauer:2004ve, Buchmuller:2005zv, Barberio:2007cr}. In
addition, the tail of the shape function as well as its renormalization group
evolution (RGE) can be calculated perturbatively.  A problem is that an arbitrarily
small renormalization group running of the shape function develops a
perturbative tail whose moments diverge~\cite{Balzereit:1998yf}.

Currently, there are several approaches to determine $|V_{ub}|$.  Often a model
for the shape function is chosen, which has a fixed functional form roughly
consistent with the $B\to X_s\gamma$ spectrum, and a few adjustable parameters
that are fixed by imposing constraints on the first few moments of the shape
function. A proposal~\cite{Bosch:2004th} used by many experimental analyses
involves defining moments of the shape function with a cutoff, and a particular
procedure to attach a perturbative tail to the model.  Unfortunately, there is
no clear way to disentangle the shape function and $m_b$ dependencies in this
approach, and experimental uncertainties in the shape of the measured $B\to
X_s\gamma$ spectrum are not easily incorporated. The issues related to modeling
the shape function can be avoided using model independent relations between
$B\to X_u \ell\bar\nu$ partial rates and weighted integrals of the $B\to
X_s\gamma$ spectrum~\cite{Leibovich:1999xf, Leibovich:2000ey, Hoang:2005pj,
  Lange:2005xz}, which use the measured $B\to X_s\gamma$ spectrum directly as
input to predict the $B\to X_u\ell\bar\nu$ rates.  Although this weighting
method allows one to take into account the experimental uncertainties from $B\to
X_s\gamma$ straightforwardly, it is hard to combine several measurements, and
there is no way to include the additional constraints on $m_b$ and the heavy
quark effective theory (HQET) matrix elements. Phase space cuts for which the rate has
only subleading dependence on the shape function are also
possible~\cite{Bauer:2000xf}, at the expense of increasing the size of the
expansion parameter.

A fully consistent method to combine all experimental constraints on both the
shape and moments of the shape function, while also incorporating its known
perturbative and nonperturbative behavior, has not yet been given.  The
framework proposed in this paper provides such a method.  It also allows one to
obtain reliable error estimates by (i) taking into account all experimental and
theoretical uncertainties and correlations; and (ii) estimating the uncertainty
related to the unknown functional form of the shape function in a systematic
fashion.

The shape function, $S(\w, \mu)$, contains nonperturbative physics and obeys the
renormalization group equation
\begin{equation} \label{eq:S_RGE}
S(\w,\mu_i) = \int\! \df\w'\, U_S(\w-\w', \mu_i, \mu_\Lambda)\,
  S(\w', \mu_\Lambda)
\,,
\end{equation}
where the evolution kernel $U_S(\w, \mu_i, \mu_\Lambda)$ sums logarithms between
the two scales $\mu_i > \mu_\Lambda$.
The question is how to determine the function $S(\w,\mu)$ reliably, which can
then be used to extract $\abs{V_{ub}}$ from $B\to X_u\ell\bar\nu$, and to
analyze $B\to X_s\gamma$ or $B\to X_s\ell^+\ell^-$ in the low-$q^2$ region.  In
Ref.~\cite{Lee:2005pw} we used the construction
\begin{equation} \label{eq:S_construction}
S(\w, \mu_\Lambda) = \int \df k\, C_0(\w-k, \mu_\Lambda)\, F(k)\,,
\end{equation}
where $C_0(\w, \mu_\Lambda)$ is the $b$ quark matrix element of the
shape function operator calculated in perturbation theory, and $F(k)$ is a
nonperturbative function that can be extracted from data.
Equation~\eqref{eq:S_construction} has many advantages
compared to earlier treatments of the shape function.  It ensures that:
\begin{enumerate}\vspace*{-2pt}\itemsep -2pt
\item $S(\w, \mu_\Lambda)$ has the correct $\mu_\Lambda$ dependence and RGE.
\item $S(\w, \mu_\Lambda)$ has the correct perturbative tail at large $\w$,
  while for small $\w$ it is determined by $F(k)$.
\item The moments of $F(k)$ exist without a
cutoff and $F(k)$ falls off exponentially at large $k$.
\item Information about matrix elements of local operators in any short distance
  scheme can be incorporated via constraints on moments of $F(k)$.
\end{enumerate}\vspace*{-4pt}
A construction similar to \eq{S_construction} was also used to treat the soft
function that describes nonperturbative radiation in jet production in
Ref.~\cite{Hoang:2007vb}.

The outline of this paper is as follows.  In \sec{bsg} we set up our notation
and discuss how the shape function enters the decay rates.  Our new treatment of
the shape function based on \eq{S_construction} is discussed in \sec{SF},
including the procedure for incorporating moment constraints in any short
distance scheme and an analysis of perturbative corrections in the shape
function and decay rate up to two-loop order with a resummation of large
logarithms at next-to-next-to-leading-logarithmic (NNLL) order. In \sec{basis} we
introduce a systematic expansion of $F(k)$ in terms of a suitably chosen set of
orthonormal basis functions, which allows one to control the uncertainties
arising from its unknown functional form.  In \sec{fit}, we summarize our
proposal of how to use all the available data to extract the function $F(k)$ and
determine the shape function $S(\w,\mu)$ with reliable uncertainties, which can
then be used for the extraction of $|V_{ub}|$ from $B\to X_u\ell\bar\nu$.
Section~\ref{sec:conc} contains our conclusions. Details on perturbative
corrections and the invisible scheme for the kinetic energy matrix element are
summarized in three appendices.

\section{\boldmath The $B\to X_s\gamma$ and $B\to X_u\ell\bar\nu$ Rates in
the Shape Function Regions}
\label{sec:bsg}

We use the kinematic variables $p_X^\pm = E_X \mp \abs{\vp_X}$.  We also define
the partonic variable $p^-= p_X^- + m_b - m_B$.  In $B\to X_s\gamma$, $\pxm =
m_B$ ($p^-=m_b$) and $\pxp = m_B-2E_\gamma$, while in $B\to X_u\ell\bar\nu$ they
are independent variables with $p_X^+ \leq p_X^- \leq m_B$. There are three
cases where it is known how to carry out a systematic expansion of the decay
rate
\begin{align} \label{Expansions}
1)\,\ & \text{Nonpert. shape function :}   & \lqcd & \sim p_X^+ \ll p_X^- \,,
  \nn \\
2)\,\ & \text{Shape function OPE :} & \lqcd &\ll  p_X^+ \ll p_X^- \,,
  \nn \\
3)\,\ & \text{Local OPE :}   & \lqcd &\ll p_X^+ \sim p_X^- \,.
\end{align}
The region 2) was first studied in
$B\to X_u\ell\bar\nu$ in Refs.~\citetext{\citealp{Bauer:2003pi}, \citealp{Bosch:2004th}}, and
for the $B\to X_s\gamma$ rate in Refs.~\cite{Neubert:2004dd, Becher:2006pu}.
In the SCET regions 1) and 2), where $\pxp \ll \pxm$, the decay rates
$\Gamma_s\equiv\Gamma(B\to X_s\gamma)$ and $\Gamma_u\equiv\Gamma(B\to
X_u\ell\bar\nu)$ are given by the factorization theorems~\cite{Korchemsky:1994jb, Bauer:2001yt}
\begin{align} \label{eq:factor}
\frac{\df\Gamma_s}{\df E_\gamma} &= 2\Gamma_{0s}\, H_s(\pxp,\mu_i)
\\ & \quad
  \times \inte{\w} m_b J(m_b\, \w, \mu_i)\, S(\pxp-\w, \mu_i)
\,,\nn\\
\frac{\df\Gamma_u}{\df E_\ell\, \df\pxp \df\pxm} &= \Gamma_{0u}\,
 H_u(E_\ell,\pxm,\pxp,\mu_i)
\nn\\ & \quad
  \times \int \!\df\w\, p^- J(p^-\w, \mu_i)\, S(\pxp-\w,\mu_i) \nn
\,,\end{align}
where
\begin{equation}\label{Gamma0su}
\Gamma_{0s} = \frac{G_F^2\, m_b^5}{8\pi^3}\,
  \frac{\alpha_\mathrm{em}}{4\pi}\, \abs{V_{tb} V_{ts}^*}^2
\,,\quad
  \Gamma_{0u} = \frac{G_F^2\, m_b^5}{192\pi^3}\, \abs{V_{ub}}^2
\,.\end{equation}
Corrections are suppressed by ${\Lambda}_{\rm QCD}/m_b$ and it is known how to
include them in \eq{factor}. Here we focus on the leading term since
the procedure to incorporate the subleading terms follows the same method.  The
integration limits are implicit in the support of the $S$ and $J$ functions in
\eq{factor}, which are nonzero when their first argument is positive.
Both, the hard functions, $H_u$ and $H_s$, and the jet function,
$J$, in \eq{factor} are calculable in a perturbation series in $\alpha_s$. We
summarize results for them in \app{pert} up to two-loop order.  Only $H_u$ and
$H_s$ are process dependent, and they can sum logarithms between the hard scale
and $\mu_i^2 \sim p_X^+ p_X^-$.

The shape function $S(\omega,\mu_i)$ in \eq{factor} sums logarithms
between $\mu_i$ and $\mu_\Lambda$ through \eq{S_RGE}, where in case 1)
in Eq.~(\ref{Expansions}) $\mu_\Lambda\sim 1\GeV$, while in case 2)
$\mu_\Lambda\sim p_X^+$.  In case 1) the shape function is nonperturbative,
while in case 2) it can be computed with an OPE to separate the scales
$\Lambda_{\rm QCD}\ll p_X^+$.  A key feature of \eq{S_construction} is
that it makes the expressions for the decay rates in \eq{factor}
simultaneously valid both for cases 1) and 2). For example, it allows
an analysis of the photon energy cut in $B\to X_s\gamma$ without having to
rely on the expansion in region 2) in $\lqcd/\pxp$, as was done in
Refs.~\cite{Neubert:2004dd, Becher:2006pu}.
This is important, since in practice, if the momenta in region 2) are not well
separated numerically, the utility of expanding in $\lqcd/\pxp$ is unclear.

It is possible to make \eq{factor} valid for region 3) of
Eq.~(\ref{Expansions}), by including appropriate power suppressed and
perturbative corrections.  At tree level this was carried out in
Refs.~\cite{Mannel:2004as, Tackmann:2005ub}, and all results presented below are
valid in region 3) at this order.  At the level of perturbative corrections
these issues were studied in Ref.~\cite{Lange:2005yw}. We do not include the
additional perturbative corrections needed in region 3), since our primary
interest is to study the SCET regions 1) and 2).  However, it is important that for the
perturbative corrections to be correct in region 3), it is required to scale up
the different $\mu$'s so that $\mu_\Lambda = \mu_i = \mu_b \sim m_b$, because
there is only one scale $\mu$ in the local OPE.  A procedure to carry out this
scaling of the $\mu$'s is discussed around Eq.~(\ref{rscale}).
A dedicated study of the transition to region 3) is left for future work.

Combining Eqs.~(\ref{eq:S_RGE}), (\ref{eq:S_construction}), and
\eqref{eq:factor}, and switching the order of convolutions we arrive at
\begin{align} \label{eq:dG}
\frac{\df\Gamma_s}{\df E_\gamma}
&= 2 \Gamma_{0s}\, H_s(\pxp,\mu_i)
\nn\\ & \quad
\times \inte{k} P(m_b, \pxp - k, \mu_i)\, F(k)
\,,\nn\\
\frac{\df\Gamma_u}{\df E_\ell\, \df\pxp \df\pxm}
&= \Gamma_{0u}\, H_u(E_\ell,\pxm,\pxp,\mu_i)
\nn\\ & \quad
\times \inte{k} P(p^-, \pxp -k, \mu_i)\, F(k)
\,.\end{align}
Here, the perturbatively calculable function $P$ is process independent,
\begin{align}
\label{eq:P_def}
P(p^-, k, \mu_i) &= \inte{\w}\! \inte{\w'} p^-\, J[p^-(k - \w), \mu_i]
\nn\\
& \quad\times U_S(\w - \w', \mu_i, \mu_\Lambda)\, C_0(\w', \mu_\Lambda)
\,.\end{align}
At lowest order in perturbation theory
\begin{equation}
P(p^-, k, \mu_i) = \delta(k) + \ord{\alpha_s}
\,,\end{equation}
and the result for $P$ up to ${\cal O}(\alpha_s^2)$ with NNLL  resummation is given
in \app{pert}.

Equation~\eqref{eq:dG} can be used to determine the $F$ function by fitting to
experimental $B\to X_s\gamma$ and $B\to X_u\ell\nu$ data. We return to this in
\sec{fit}.  In the next two sections we explore \eq{S_construction} in detail
and construct a complete orthonormal basis for $F(k)$ that is suitable for
carrying out these fits.

\section{General Treatment of the Shape Function}
\label{sec:SF}

\subsection{Master formula and OPE constraints}

The shape function $S(\w,\mu)$ is the $B$ meson matrix element
\begin{equation}
\label{eq:S_def}
S(\w,\mu) = \mae{B}{O_0(\w,\mu)}{B} \equiv \vevB{O_0(\w,\mu)}
\,,\end{equation}
of the operator
\begin{equation}
\label{O0def}
O_0(\w, \mu) = \bar b_v\, \delta(i D_+ - \delta  + \w)\, b_v
\,,\end{equation}
where we defined
\begin{equation}
\delta = m_B-m_b \,.
\end{equation}
Here, $b_v$ is the HQET $b$ quark field and $\ket{B}$ is the full QCD $B$ meson
state, respectively.  (If we used the HQET $\ket{B_v}$ state, this would
correspond to absorbing time-ordered products of $O_0(\w)$ with all power
corrections in the HQET Lagrangian into the definition of $S(\w, \mu)$.) Also,
$D_+ = n \cdot D$, $v$ is a timelike vector, and $n$ is a lightlike vector
with $n\cdot v = 1$.  For our application, $v = p_B/m_B$ is the four-velocity of
the $B$ meson, and $\vec n=\vec p_X/\lvert\vec p_X\rvert$ is the direction of
the light-quark jet.  The $\mu$ dependencies of $S(\w,\mu)$ and $O_0(\w,\mu)$
are the same.  In Eq.~\eqref{O0def}, our use of $\delta = m_B-m_b$ and the
$\ket{B}$ state ensures that $S(\w,\mu)$ has support for $\w \geq 0$ with any
mass scheme for $m_b$~\cite{Tackmann:2005ub}.  Note that $\delta$ explicitly
depends on the mass scheme. We will use the pole mass scheme first, and discuss
converting to short distance schemes in the next subsection.

The information about the shape function that can be obtained from perturbation
theory arises from the fact that when integrated over a large enough region, $0
\leq \w \leq \Lambda$, such that perturbation theory is reliable at the scale
$\Lambda$, the operator $O_0(\w)$ can be expanded in a sum of local operators,
\begin{align}
\label{eq:O_OPE}
O_0(\w,\mu)
 &= \sum_{n=0}^2 C_n(\w,\mu)\, Q_n + \ldots
\nn\\*
 &=  \sum_{n=0}^2 C_n(\w-\delta,\mu)\, \widetilde Q_n + \ldots
\,,\end{align}
where we use either of the operator bases
\begin{equation}
Q_n = \bar b_v\, (i{D}_+-\delta)^n\, b_v\,, \qquad
\widetilde Q_n= \bar b_v\, (i{D}_+)^n\, b_v\,.
\end{equation}
The Wilson coefficients $C_n(\w)$ for these two bases are equivalent, because
$O_0(\w)$ only depends on the combination $i D_+ -\delta + \w$.  The ellipses in
\eq{O_OPE} represent operators of dimension six and higher (where four-quark
operators first appear).  Taking the $B$ meson matrix element of \eq{O_OPE},
\begin{equation} \label{eq:S_OPE}
S(\w,\mu) = \sum_{n=0}^2 C_n(\w,\mu)\, \vevB{Q_n} + \ldots
\,,\end{equation}
the moments of $S(\w,\mu)$ with an upper cutoff can be computed
\begin{equation}
\intlim{0}{\Lambda}{\w}\, \w^k\, S(\w,\mu)
= \sum_n \vevB{Q_n} \!\intlim{0}{\Lambda}{\w}\, \w^k\, C_n(\w,\mu) + \ldots
\,,\end{equation}
and are determined by the local matrix elements $\vevB{Q_n}$ plus the
perturbative information in the $C_n$~\cite{Bosch:2004th}.  For the first few
matrix elements, we have
\begin{align} \label{MeltQi}
\vevB{Q_0} = 1 \,,\quad
\vevB{Q_1} = - \delta \,,\quad
\vevB{Q_2} = - \frac{\lambda_1}{3} + \delta^2
\,,\end{align}
where $\lambda_1 \equiv \mae{B}{\bar b_v\, (i D)^2\, b_v}{B}$, with the matrix
element defined in dimensional regularization. With this definition and the
pole mass, the matrix elements in Eq.~\eqref{MeltQi} are $\mu$ independent.

The matching coefficients $C_n(\w,\mu)$ in the OPE in \eq{O_OPE} can be
determined at fixed order in perturbation theory by taking a partonic matrix
element of both sides of \eq{O_OPE}. Consider~\cite{Bauer:2003pi}
\begin{equation}
\mae{b_v}{O_0(\w+\delta,\mu)}{b_v}
= \sum_n C_n(\w,\mu)\, \mae{b_v}{\widetilde Q_n}{b_v} = C_0(\w,\mu)
,\end{equation}
where the $b_v$ states have zero residual momentum, and we used
$\mae{b_v}{\widetilde Q_n}{b_v} = \delta_{0n}$. The $n \geq 1$ matrix elements
vanish in $\overline {\rm MS}$ because there is no dimensionful quantity they
can be proportional to.  To determine $C_n(\w,\mu)$ for $n \geq 1$, consider the
matrix element between $b_v$ states with residual momentum $k^\mu$ where $v\cdot
k=0$ but $k_+ = n\cdot k\ne 0$. The right-hand side of \eq{O_OPE}
gives~\cite{Bauer:2003pi}
\begin{align}
&\mae{b_v(k_+)}{O_0(\w+\delta,\mu)}{b_v(k_+)}
\nn\\ &\quad
= \mae{b_v(0)}{O_0(\w + k_++\delta,\mu)}{b_v(0)}
\nn\\&\quad
= C_0(\w + k_+, \mu)
= \sum_n \frac{k_+^n}{n!}\, \frac{\df^n C_0(\w,\mu)}{\df\w^n}
\,.\end{align}
Comparing this with the left-hand side of \eq{O_OPE},
\begin{align} \label{OPEmeltk}
& \mae{b_v(k_+)}{O_0(\w+\delta,\mu)}{b_v(k_+)}
\nn\\* &\quad
= \sum_{n = 0}^2 C_n(\w,\mu)\, \mae{b_v(k^+)}{ \widetilde Q_n }{b_v(k^+)} + \ldots
\nn\\* &\quad
= \sum_{n = 0}^2 C_n(\w,\mu)\, (k^+)^n + \ldots
\,,\end{align}
gives for $n = 0,1,2$
\begin{equation} \label{eq:Cn_dimreg}
C_n(\w,\mu) = \frac{1}{n!}\, \frac{\df^n C_0(\w,\mu)}{\df\w^n}
\,.\end{equation}
In Eq.~\eqref{OPEmeltk}, the matrix elements of $\widetilde Q_n$ for $n \leq 2$
in $\overline {\rm MS}$ are given by their tree-level values, $k_+^n$, because
loop graphs have dimension $\geq 1$, but are scaleless and vanish.

The coefficients of $Q_1$ and $Q_2$ are related by \eq{Cn_dimreg} to the same
perturbative coefficient function as $Q_0$ to all orders in perturbation theory.
This is no longer the case at dimension six and higher, where the operator basis
includes four-quark operators, and more than one matrix element must be
computed.

Our key point is to write the renormalized shape function at the scale $\mu$ as
in \eq{S_construction},
\begin{equation} \label{eq:F_def}
S(\w,\mu) = \inte{k} C_0(\w - k, \mu)\, F(k)
\,,\end{equation}
where the function $C_0(\w, \mu) = \mae{b_v}{O_0(\w+\delta,\mu)}{b_v}$ has
an expansion in $\alpha_s$ and contains perturbatively accessible
information about $S(\w,\mu)$. Equation~\eqref{eq:F_def} defines the function
$F(k)$, which is a nonperturbative object that can be extracted from data.
To see that \eq{F_def} uniquely
specifies $F(k)$, note that in Fourier space $\widetilde S(y,\mu) = \widetilde
C_0(y,\mu)\widetilde F(y)$, so $\widetilde F(y)= \widetilde S(y,\mu)/\widetilde
C_0(y,\mu)$.

An important feature of \eq{F_def} is that it is consistent with
the OPE result in \eq{S_OPE}.  Expanding its right-hand side in $k$ gives
\begin{equation}
S(\w,\mu) = \sum_n \frac{1}{n!}\, \frac{\df^n C_0(\w,\mu)}{\df\w^n}
\inte{k} (-k)^n F(k)
\,,\end{equation}
and comparing with \eqs{S_OPE}{Cn_dimreg} one finds for $n=0,1,2$
\begin{equation} \label{eq:Fmoments}
\inte{k}\, k^n\, F(k) = (-1)^n\, \vevB{Q_n}
\,.
\end{equation}
Thus, the first few moments of $F(k)$ are determined by the matrix elements of
the local operators, $\vevB{Q_n}$, reproducing the OPE for these terms.  (The
decomposition in \eq{F_def} can be extended such that it works for the
OPE terms with $n \geq 3$ as well, although we will not do so explicitly here.)
Unlike for $S(\w,\mu)$, for $F(k)$ all moments without a cutoff exist, so $F(k)$
falls faster than any power of $k$ at large $k$. Furthermore, the characteristic
width over which $F(k)$ has substantial support is of order $\lqcd$.

Equation~\eqref{eq:F_def} also ensures that $S(\w,\mu)$ has the correct
dependence on $\mu$ in the $\overline{\rm MS}$ scheme, since it is determined by
$C_0(\w,\mu)$, which satisfies the shape function RGE in \eq{S_RGE}. As shown in
\eq{C0_evolve} of \app{pert}, a simple formula valid to all orders in $\alpha_s$
can be derived, which combines $S(\w,\mu_\Lambda)$ from \eq{F_def} with the
evolution from $\mu_\Lambda$ up to $\mu_i$ in \eq{S_RGE},
\begin{align} \label{eq:Swrun}
S(\w,\mu_i) &=
E_S(\w, \mu_i, \mu_\Lambda) \sum_{j=-1}^\infty\, \sum_{\ell = -1}^{j+1} \V^j_\ell(\eta)
\\* &\quad\times
S_j\Bigl[\alpha_s(\mu_\Lambda), \frac{\w}{\mu_\Lambda}\Bigr]\,
\intlim{0}{1}{z} \cL_\ell^\eta(z)\, F[\w (1-z)]
\,.\nn
\end{align}
Here, $\cL_\ell^\eta(z) = [\ln^\ell(z)/z^{1-\eta}]_+$ is defined in
\eq{cLna_def}, and $S_j$, $E_S$, $\eta = \eta(\mu_i,\mu_\Lambda)$, and
$V_\ell^j(\eta)$ are given in Eqs.~\eqref{eq:JS2var}, \eqref{eq:US},
\eqref{eq:etaK_def}, and \eqref{eq:Vkna_def}, respectively.   The $S_j$
coefficients are determined by partonic fixed-order calculations of
$C_0(\w,\mu)$, and at ${\cal O}(\alpha_s^2)$ the sum is bounded by $j\le 3$. The
RGE factors $\eta$ and $E_S$ are determined using the anomalous dimensions at
various orders. The RGE here has Sudakov double logarithms, implying that the
``cusp'' anomalous dimension, $\Gamma_{\rm cusp}$, must be included at one
higher order than the standard anomalous dimensions, $\gamma_x$.  There is no
large hierarchy between the scales $\mu_b>\mu_i>\mu_\Lambda$, so the optimal
procedure for combining fixed order and resummation is debatable. We adopt the
following conventions for our analysis:
\begin{align} \label{eq:RGEconventions}
& {\rm LL:}  & &\text{$1$--loop $\Gamma_{\rm cusp}$, tree-level matching}\,;
  \\*
& {\rm NLL:}  & &\text{$2$--loop $\Gamma_{\rm cusp}$, $1$--loop $\gamma_x$,
   $1$--loop matching}\,; \nn\\*
& {\rm NNLL:}  & &\text{$3$--loop $\Gamma_{\rm cusp}$, $2$--loop $\gamma_x$,
   $2$--loop matching} \nn \,.
\end{align}

Our construction ensures that $S(\w, \mu_i)$ has the correct perturbative tail
for large $\w$, specified by both the constant and logarithmic terms in
$C_0(\w,\mu_i)$, while for small $\w$ it is controlled by the nonperturbative
function $F(k)$. Thus \eqs{F_def}{Swrun} build in all
the information that can be obtained from considering region~2) in
Eq.~(\ref{Expansions}) {\em without} having to carry out an expansion of the
decay rate in this region in $\lqcd/\pxp$.

\begin{figure}[t]
\includegraphics[width=\columnwidth]{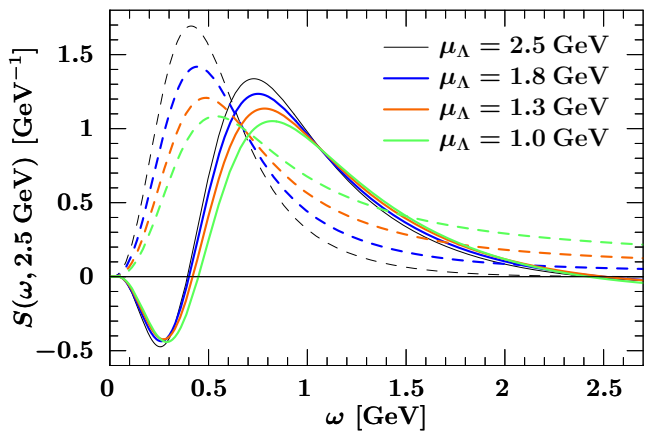}
\caption{Scale independence of our shape function construction.  The thin black
  dashed curve with the highest peak shows the model function $S(\w,\mu_i) =
  \hFmod(\w)$ given in \eq{Fmodel}. The other dashed curves show the result of
  taking $S(\w, \mu_\Lambda) = \hFmod(\w)$ at $\mu_\Lambda = 1.0,\, 1.3,\,
  1.8\GeV$ (from bottom to top near the peaks) and running up to $\mu_i=2.5\GeV$
  with NNLL accuracy.  The solid curves use \eq{Swrun} with $\mu_\Lambda =
  1.0,\, 1.3,\, 1.8,\, 2.5\GeV$ and running up to $\mu_i=2.5\GeV$ at NNLL. Note
  the stable tails of the solid curves. The dip at small $\w$ is discussed in
  the next section.}
\label{fig:running}
\end{figure}

A common approach for modeling a parton distribution function is to specify a
model for $S(\w, \mu_\Lambda)$ at a fixed scale $\mu_\Lambda$ and then run it up
to a higher $\mu_i$. In this case, $\mu_\Lambda$ must be treated as a model
parameter, and changing it can cause significant changes in the model.
Furthermore, for $\w\gg \lqcd$ the perturbative tail of $S(\w, \mu_i)$ obtained
in this approach will not be consistent with carrying out the OPE at $\mu_i$.
The perturbative logarithms are reproduced, but the constant terms are not (and
the latter are sizeable contributions to the OPE for the scales considered
here). In Fig.~\ref{fig:running} we compare this approach (dashed curves), with
the superior approach of describing the shape function via Eq.~\eqref{eq:Swrun}
(solid curves) at NNLL.  In each case we use a model for $F(k)$, $\hFmod(k)$
given below in \eq{Fmodel}, whose first three moments correspond to $m_b =
4.7\GeV$ and $\lambda_1 = -0.31\GeV^2$. The thin black dashed curve shows
$S(\w,\mu_i) = \hFmod(\w)$. The other three dashed curves show the result
of fixing $S(\w,\mu_\Lambda)=\hFmod(\w)$ at $\mu_\Lambda = 1.0,\, 1.3,\,
1.8\GeV$ (from bottom to top near the peaks) and running up to $\mu_i=2.5\GeV$
using \eq{S_RGE}.  The resulting tails at large $\w$ are clearly inconsistent
with each other.  The solid curves show the result of our approach, using
\eq{Swrun} to obtain $S(\w,\mu_\Lambda)$ at $\mu_\Lambda = 1.0,\,1.3,\,
1.8,\,2.5\GeV$ and running up to $\mu_i=2.5\GeV$.  In our approach,
$S(\w,\mu_i)$ is independent of the initial scale $\mu_\Lambda$, up to
subleading corrections in $\alpha_s(\mu_\Lambda)$, and the tails at large $\w$
are consistent with one another. The negative dip at small $\w$ is an artifact
of using the pole mass scheme, and will be removed by switching to short
distance schemes in the next section.  Another feature of the solid curves in
Fig.~\ref{fig:running} is that their tails become negative for $\w\gtrsim
2.5\GeV$.   It was noted in Ref.~\cite{Bosch:2004th} that most of this negative
tail is canceled by the perturbative corrections from the jet function.  We
discuss in \sec{num} that this negative tail also disappears if $\mu_\Lambda$ is
increased as $\w$ increases.

To obtain the correct perturbative tail, the procedure used by
BLNP~\cite{Lange:2005yw} for $|V_{ub}|$ analyses is to take the perturbative
computation of $C_0(\omega,\mu)$ for $\omega\ge \omega_0$ and a model for
$S(\omega,\mu)$ for $\omega\le \omega_0$, and these two pieces are glued
together, choosing $\omega_0$ so that the result is continuous.  The advantage
of our construction in \eq{F_def} is that the tail automatically turns on in a
smooth manner when it dominates over the nonperturbative function $F(k)$ and
provides the proper $\mu$ dependence for $S(\omega,\mu)$ at any $\w$.

Imposing the moment constraints on $F(k)$ in \eq{Fmoments} provides a clean way
to incorporate the information on the local OPE matrix elements, $m_b$,
$\lambda_1$, etc., from $B\to X_c\ell\bar\nu$. It is possible to use a shape
function scheme in which moments of $S(\w,\mu)$ with a cutoff define the
nonperturbative parameters~\cite{Bosch:2004th}.  Our approach has the advantage
of allowing one to use any desired short distance scheme, as we discuss next.

\subsection{Short distance schemes}

The most precise information on the matrix elements in Eq.~(\ref{MeltQi}) is
provided by fitting OPE results to $B\to X_c\ell\bar\nu$ decay distributions.
This directly constrains $F(k)$ through \eq{Fmoments}. Ideally, we would like to
incorporate these constraints on $F$ in a manner that is independent of the
order in perturbation theory used to calculate $P$ in Eq.~(\ref{eq:P_def}).
However, if we define the moment parameters $\delta$ and $\lambda_1$ from
Eq.~\eqref{MeltQi} in an infrared sensitive manner such as the pole mass scheme,
then the dependence on the order in $\alpha_s$ will not be small --- infrared
renormalon ambiguities in the perturbation series will cancel against
ambiguities in the parameters $\delta$ and $\lambda_1$. In practice, this means
that the values of $\delta$ and $\lambda_1$ may change substantially when the
fit in $B\to X_c\ell\bar\nu$ is done at different orders in perturbation
theory.  In $B$ physics, cancellations between perturbative corrections are
significant already at low orders in perturbation theory. Thus, it is preferable
to define $F(k)$ and $C_0(\w,\mu)$ so that they are individually free of
renormalon ambiguities, which should make the values of $\delta$ and $\lambda_1$
more stable to the inclusion of perturbative corrections.

Consider shifting to a new perturbative kernel
$\hC_0(\w,\mu)$ and nonperturbative function $\hF(k)$ that are free from
renormalons. To implement this we let
\begin{align} \label{shiftCF}
C_0(\w) &= \hC_0(\w) + \delta C_0(\w)
\,, \nn\\
F(k) &= \hF(k) + \delta F(k)
\,,\end{align}
such that
\begin{equation}  \label{eq:Swhat}
S(\w) = \int\! \df k\, C_0(\w-k) F(k) = \int\! \df k\, \hC_0(\w -k)\, \hF(k)
\,.\end{equation}
These shifts move a series of perturbative corrections between $F(k)$ and
$C_0(\w,\mu)$. The shift $\delta F(k)$ will be chosen such that the moments of
$\hF(k)$ are given by renormalon-free parameters, and \eq{Swhat} then
determines the corresponding shift $\delta C_0(\w)$.

We switch from the pole mass $m_b$ and $\lambda_1$ to short distance parameters,
$\widehat m_b$ and $\widehat \lambda_1$,
\begin{equation}\label{delmbl1}
m_b = \widehat m_b + \delta m_b
\,,\qquad
\lambda_1 = \widehat \lambda_1 + \delta \lambda_1
\,,\end{equation}
where $\delta m_b$ and $\delta \lambda_1$ consist of series in $\alpha_s(\mu)$
with the same renormalon as $m_b$ and $\lambda_1$, respectively. The freedom to
choose these series corresponds to the freedom to choose different schemes for
$\widehat m_b$ and $\widehat\lambda_1$. To obtain $\hF(k)$ with moments only
depending on $\widehat m_b$ and $\widehat \lambda_1$, we pick $\delta F(k)$ such
that
\begin{align} \label{eq:delF}
\delta F(k) &= F(k) - F(k-\delta m_b) - \Delta F(k)
\,, \nn\\
\hF(k) &= F(k-\delta m_b) + \Delta F(k)
\,.\end{align}
Shifting the argument allows us to switch to $\widehat m_b$.  To implement
$\widehat\lambda_1$, $\Delta F(k)$ has to satisfy
\begin{align} \label{eq:DelFconst}
\int\! \df k\, \Delta F(k) &= \int\! \df k\, k\, \Delta F(k) = 0
\,, \nn\\
\int\! \df k\, k^2\, \Delta F(k) &= \frac{\delta\lambda_1}{3}
\,.\end{align}
The most general solution to \eq{DelFconst} is $\Delta F(k) = (\delta\lambda_1/
6)\, F_1''(k)$, where $F_1(k)$ is an arbitrary function, normalized as $\int\!
\df k\, F_1(k) =1$.

Using \eqs{delF}{DelFconst} with \eq{Fmoments}, one can easily check that
$\hF(k)$ has renormalon-free moments, as desired, 
\begin{align} \label{eq:Fhatmoments}
 &  \int\! \df k\, \hF(k) =  1
  \,,\nn\\
 &  \int\! \df k\, k\, \hF(k) = \delta + \delta m_b = \widehat\delta
  \,,\nn\\
 &  \int\! \df k\, k^2\, \hF(k) = -\frac{\widehat\lambda_1}{3}
  + \widehat\delta^{\,2}
\,,
\end{align}
where $\widehat\delta = m_B -\widehat m_b$.  Thus the experimental values of the
$b$ quark mass and $\widehat\lambda_1$ extracted in any short distance scheme
can be used as inputs in our framework via the moment constraints in
\eq{Fhatmoments}.

To determine the corresponding shift in $C_0(\w)$, we note that \eq{Swhat}
implies $\int\! \df k\, [ C_0(\w - k)\, \delta F(k) + \delta C_0(\w -
k)\, \hF(k) ]=0$. To solve for $\delta C_0(\w)$, we take
the Fourier transform, $\delta \widetilde C_0(y) = -\widetilde C_0(y)\, \delta
\widetilde F(y) /\,\widetilde{\!\hF}(y)$, and thus
\begin{equation}
\delta \widetilde C_0(y) =
  \biggl[1 - e^{i y\, \delta m_b} - \frac{\delta\lambda_1}{6}\, y^2\,
  \frac{ \widetilde F_1(y)}{\,\widetilde{\!\hF}(y)}
  \biggr] \widetilde C_0(y)
\,.\end{equation}
Here, any $\delta m_b\, \delta \lambda_1$ cross terms only have higher-order renormalon ambiguities and are dropped. Any choice of $F_1(k)$ is equally good for incorporating the
$\delta\lambda_1$ shift. We adopt the simplest choice $F_1(k) =
\hF(k)$, which is unique in that it keeps $\hC_0(\w)$ independent of the precise form
of $\hF(k)$.\footnote{There are other possible choices. For example, taking
$F_1(k)\propto k^2 \hF(k)$ would ensure that the $\delta\lambda_1$ shift does
not change the small-$k$ behavior of $\hF(k)$. However, if $F_1(k) \neq \hF(k)$,
the shift $\delta C_0(\w)$ depends on $\hF(k)$, and must be recomputed each time
$\hF(k)$ changes. Hence other choices are more difficult to implement when
performing a fit to data to extract $\hF(k)$.}
In this case, in momentum space, we have
\begin{align} \label{hatC0}
\hC_0(\w)
&= C_0(\w+\delta m_b) - \frac{\delta\lambda_1}{6}\, \frac{\df^2}{\df\w^2}\, C_0(\w)
\nn\\*
&= \biggl[1 + \delta m_b\,\frac{\df}{\df \w} + \bigg(\frac{(\delta m_b)^2}{2}
  - \frac{\delta\lambda_1}{6} \bigg) \frac{\df^2}{\df\w^2}\biggr] C_0(\w)
\nn\\*
&\quad + \ldots \,,
\end{align}
where $C_0(\w)$ is determined by the perturbative calculation of
$\mae{b_v}{O_0(\w+\delta,\mu)}{b_v}$, and the ellipsis denotes terms that are
either $\ord{\alpha_s^3}$ or beyond the order we are working at for the moments
of $\hF(k)$.  The same strategy to determine $\hF(k)$ and derive the
corresponding $\hC_0(\w)$ can be applied to higher moments if in the future
terms with $n\geq 3$ in \eq{Fmoments} are included in the analysis.

In the remainder of this paper we use the convolution formula for $S(\w,\mu)$ in
terms of $\hC_0(\w,\mu)$ and $\hF(k)$ in \eq{Swhat}.  We shall regard $\hF(k)$
as the fundamental nonperturbative object to be extracted from data. In
particular, in \sec{basis} we will build a complete basis of functions for
$\hF(k)$.

\begin{table}[t]
\tabcolsep 6pt
\begin{tabular}{c|c}
\hline\hline
Parameter & Value \\
\hline
$\alpha_s(m_Z)$ \cite{pdg} & $0.1176$  \\
$\alpha_s(4.7\GeV)$ & $0.2155$  \\
$m_B$ & $5.279\GeV$ \\
\hline
$m_b^{1S}$ \cite{Barberio:2007cr} & $4.70\GeV$ \\
$\lambda_1$ \cite{Barberio:2007cr} & $-0.31\GeV^2$ \\
\hline
$\lambda_1^\inv$ & $-0.32\GeV^2 $ \\
\hline
$m_b^\mathrm{kin}(1\GeV)$ & $4.57\GeV$ \\
$\lambda_1^\mathrm{kin}(1\GeV)$ & $-0.47\GeV^2$ \\
\hline\hline
\end{tabular}
\caption{Central values of the input parameters.  The last three entries are
computed from $m_b^{1S}$ and $\lambda_1$ at order $\alpha_s^2$.}
\label{tab:num}
\end{table}

\subsection{\boldmath Numerical results for $S(\w,\mu)$}
\label{sec:num}

To illustrate the effect of using our method to include the perturbative
corrections to the shape function through $\widehat C_0$ we choose a model
for $\hF(k)$,
\begin{equation} \label{eq:Fmodel}
\hFmod(k) = \frac{1}{\lambda}\,
  \biggl[ \sum_{n=0}^2 c_n\, f_n\Bigl(\frac{k}{\lambda}\Bigr) \biggr]^2
\,.\end{equation}
Here $\lambda=0.8\GeV$, $f_n(x)$ is given below in \eq{ourbasis}, and the three
parameters, $c_0$, $c_1$, and $c_2$, are fixed to satisfy the constraints in
\eq{Fhatmoments} for the appropriate short distance scheme. For our numerical
analysis we use the input values collected in Table~\ref{tab:num}.  With
$\{m_b^{1S},\, \lambda_1\}$ we have $\{c_0,\, c_1,\, c_2\} = \{0.949,\,
-0.309,\, 0.064\}$, while for $\{m_b^{1S},\, \lambda_1^\inv \}$ we have
$\{c_0,\, c_1,\, c_2\} = \{0.949,\, -0.307,\, 0.075\}$, and for $\{m_b^{\rm
kin},\, \lambda_1 \}$ we have $\{c_0,\, c_1,\, c_2\} = \{0.988,\, -0.120,\,
-0.095\}$.

\begin{figure}[t]
\includegraphics[width=\columnwidth]{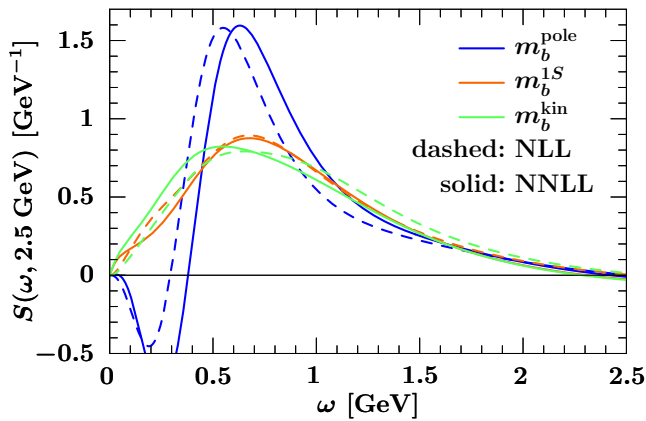}
\caption{$S(\w,\mu_i)$ obtained from $\hC_0(\w,\mu_\Lambda)$ with $\mu_\Lambda =
  1.3\GeV$ and run up to $\mu_i = 2.5\GeV$ at NLL (dashed) and NNLL (solid) order.
  Shown are results using the pole mass scheme and the $1S$ and kinetic short distance
  mass schemes. Switching to the short distance schemes, the result becomes
  more stable going from NLL to NNLL, the negative dip at small $\w$ in the pole
  scheme is removed, while the perturbative tail at large $\w$ remains unchanged.}
\label{fig:schemes}
\end{figure}

First, we switch from the pole mass to a short distance mass.  We use the $1S$
mass~\cite{Hoang:1998ng} and the kinetic mass~\cite{Czarnecki:1997sz} schemes,
\begin{align}
  \widehat\delta &\equiv \delta^{1S} = m_B - m_b^{1S} \,,\nn\\*
  \widehat\delta &\equiv \delta^\mathrm{kin} = m_B - m_b^\mathrm{kin}
\,,\end{align}
to fix the first moment in \eq{Fhatmoments}, and $\lambda_1$ to determine the
second moment.  The choice of mass scheme enters $\widehat C_0(\omega)$ through
the $\delta m_b$ and $\delta\lambda_1$ in Eq.~(\ref{hatC0}), which must be
expanded in $\alpha_s$ to avoid the renormalons.  Details of the implementation
of short distance schemes and the  expressions for $\delta m_b^{1S}$ and $\delta
m_b^\mathrm{kin}$ are discussed in Appendix~\ref{subsec:scheme}.  In
Fig.~\ref{fig:schemes} we show the result for $S(\w,\mu_i)$ obtained from
$C_0(\w,\mu_\Lambda)$ with $\mu_\Lambda = 1.3\GeV$, run up to $\mu_i = 2.5\GeV$
at NLL order (dashed) and NNLL order (solid).  The blue (dark), orange
(medium), and green (light) curves show the results in the pole, $1S$, and
kinetic mass schemes, respectively.  In both short distance mass schemes the
negative dip present in the pole scheme is removed, while the perturbative tail
at large $\omega$ remains unchanged. The removal of the negative dip is similar
to what was observed for the soft function for jets in Ref.~\cite{Hoang:2007vb}.

\begin{figure}[t!]
\includegraphics[width=\columnwidth]{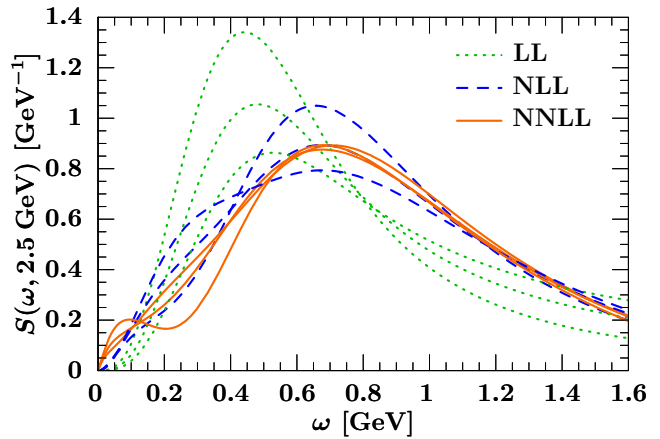}
\caption{$\mu_\Lambda$ dependence of $S(\w,2.5\GeV)$ in the peak region obtained
  from $\hC_0(\w,\mu_\Lambda)$ at LL (dotted green), NLL (dashed blue), and NNLL (solid orange)
  order, using $m_b^{1S}$ and
  $\lambda_1$. The three curves in each case are for $\mu_\Lambda = 1.0,\,
  1.3,\, 1.8\GeV$. The $\mu_\Lambda$ dependence is significantly reduced at each higher
  order.}
\label{fig:1Sschemepeak}
\end{figure}

In Fig.~\ref{fig:1Sschemepeak} we illustrate the perturbative convergence and residual $\mu_\Lambda$ scale
dependence of the result of \eq{Swrun} order by order, using the $1S$
mass scheme and $\lambda_1$. We show
$S(\w,\mu_i=2.5\GeV)$ run up from $\mu_\Lambda$ at LL (dotted green), NLL
(dashed blue) and NNLL (solid orange).  For each order, the three curves
correspond to $\mu_\Lambda = 1.0,\, 1.3,\, 1.8\GeV$. As expected, the
$\mu_\Lambda$ dependence is significantly reduced by going from LL to NLL to
NNLL. For the lowest scale, $\mu_\Lambda =1.0\,{\rm GeV}$, an oscillation begins
to build up at small $\omega$, which is clear at NNLL where the curves for the
two larger scales are quite stable.  Although we continue to explore
$\mu_\Lambda$ as low as $1.0\GeV$ in this section, we take this as evidence that
slightly larger values of $\mu_\Lambda$ should be used to ensure a convergent
expansion for $\hC_0$. Therefore, we will use $\mu_\Lambda = 1.2,\, 1.5,\,
1.9\,\GeV$ in \subsec{pertrate} for the decay rate.

\begin{figure}[t]
\includegraphics[width=\columnwidth]{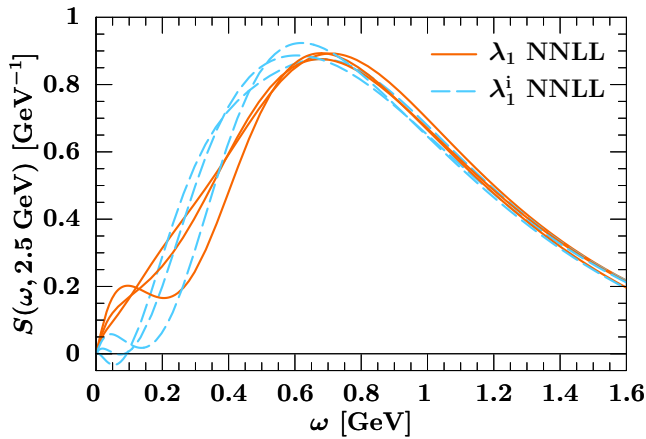}
\caption{The effect of using the invisible scheme for $\lambda_1$ on
  $S(\w,2.5\GeV)$ at NNLL order. The orange solid lines are the same as those in
  Fig.~\ref{fig:1Sschemepeak}, using $m_b^{1S}$ and $\lambda_1$, while the light
  blue dashed lines use $m_b^{1S}$ together with $\lambda_1^\inv$. The three
  curves in each case are for the same values of $\mu_\Lambda$ as in
  Fig.~\ref{fig:1Sschemepeak}.}
\label{fig:la1inv}
\end{figure}

Next, we switch to short distance schemes for $\lambda_1$. Note that
the NLL results in Fig.~\ref{fig:schemes} with $\lambda_1$ defined in
dimensional regularization are already quite stable.  Unlike for the pole mass,
there is not much numerical evidence for the importance of switching to a short
distance scheme for $\lambda_1$, and adding a sizable $\delta\lambda_1 \sim
{\cal O}(\alpha_s)$ correction may oversubtract.  The $u=1$ renormalon in
$\lambda_1$ is related to the large-order behavior of perturbation theory, and
it is unclear how much numerical impact it has on the perturbative coefficients
computed at ${\cal O}(\alpha_s)$ and ${\cal O}(\alpha_s^2)$.  In \app{invla1} we
show that schemes with $\delta\lambda_1 = {\cal O}(\alpha_s)$ appear to
oversubtract, causing large oscillations in the shape function at small
$\omega$. This is shown explicitly for the kinetic scheme $\lambda_1^{\rm kin}
(= -\mu_\pi^2)$. Therefore, in \app{invla1} we introduce a new short distance
scheme with $\delta\lambda_1 = \ord{\alpha_s^2}$, which we call the
``invisible'' scheme and denote by $\lambda_1^\inv$. The expression for
$\delta\lambda_1^\mathrm{i}$ is given in Eq.~\eqref{lam1inv}. In this short
distance scheme, the NLL results are unchanged. In Fig.~\ref{fig:la1inv} we
compare results at NNLL for $m_b^{1S}$ and $\lambda_1$ versus using $m_b^{1S}$
and $\lambda_1^\inv$. One sees that the invisible scheme has only a small effect
on the NNLL shape function, which is entirely at small $\w$. It damps the
oscillation that occurs when $\mu_\Lambda =1.0\,{\rm GeV}$ and modifies the
slope.

\begin{figure}[t!]
\includegraphics[width=\columnwidth]{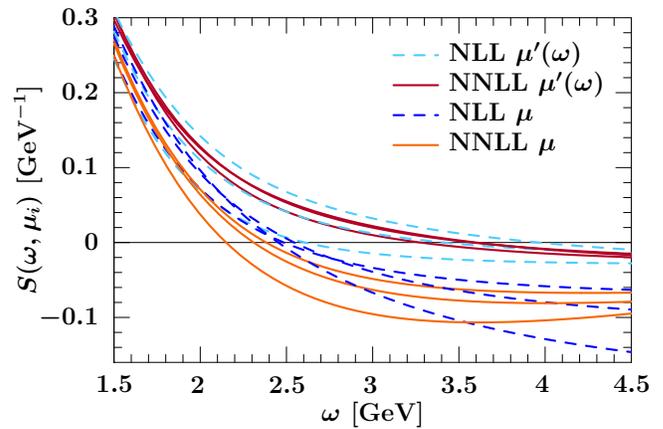}
\caption{$\mu_\Lambda$ dependence of $S(\w,2.5\GeV)$ in the tail region (lower
  six curves) and of $S(\w,\mu_i')$ (upper six curves) at NLL (dashed)
  and  NNLL (solid) order, using $m_b^{1S}$ and $\lambda_1^\inv$. For the lower
  six curves we vary $\mu_\Lambda = 1.0,\, 1.3,\, 1.8\GeV$, while for the upper
  ones we use the running scale parameters $\mu'_\Lambda(\w)$ in
  Eq.~(\ref{rscale}) and $\mu'_i(\w)$ described in the text.}
\label{fig:1Sschemetail}
\end{figure}

In Fig.~\ref{fig:1Sschemetail} we show the $\mu_\Lambda$ scale dependence at NLL (dashed)
and NNLL (solid) order in the tail region. The lower six curves use the same scale variation as
in Fig.~\ref{fig:1Sschemepeak}. Here, as one uses the SCET expansion, but enters
the local OPE region, the scale dependence increases with $\w$, and the tail becomes negative.
This is due to increasing $\ln(\w/\mu_\Lambda)$ terms,
for which the above choices of $\mu_\Lambda$ are inappropriate. To avoid
potentially large logarithms, we can increase $\mu_\Lambda$ as we increase $\w$ by
taking, for example,
\begin{equation} \label{rscale}
\mu'_\Lambda(\w) = a + b\, \arctan\bigl(\w -2.5\,{\rm GeV} \bigr)
\,.\end{equation}
To vary the scales, we take three functions of this form, with $a$ and $b$
chosen such that for $\w=0$ they give $\mu'_\Lambda=1.0,\, 1.3,\, 1.8\GeV$, and
for $\w=4.7\GeV$ they give $\mu'_\Lambda = 2.35,\, 4.7,\, 9.4\GeV$. In
Fig.~\ref{fig:1Sschemetail}, the six upper curves show $S(\omega,\mu_i)$
obtained with these $\mu'_\Lambda(\w)$ choices, taking $\mu_i = \mu_i'(\w)
=[\mu'_\Lambda(\w)\, m_b^{1S}]^{1/2}$ with the central $\mu'_\Lambda(\w)$.
Comparing the upper and lower curves shows that increasing $\mu_\Lambda$ with
$\w$ significantly reduces the $\mu_\Lambda$ scale dependence to a similar
level as in the peak region shown in Fig.~\ref{fig:1Sschemepeak}. (We have checked that
using the running $\mu_\Lambda'(\w)$ does not have an effect on the size of the
scale uncertainty in the peak region.) The upper curves in Fig.~\ref{fig:1Sschemetail}
also have a much less negative tail, which is caused predominantly by
increasing $\mu_\Lambda$ with $\w$, and only partially by increasing the reference
scale $\mu_i$. This means that the dominant part of the
negative tail in the lower curves is caused by large logarithms of $\w/\mu_\Lambda$.
Hence, increasing $\mu_\Lambda$ with $\w$ will be important to obtain
a positive rate in the tail region.
The importance of increasing $\mu$ with a kinematic variable in the tail region was also pointed
out for jet production in Ref.~\cite{Fleming:2007xt}, where it was required to
avoid a negative tail for the cross section.

In the decay rate the $\mu_i$ dependence of $S(\omega,\mu_i)$ cancels against
corrections from the jet function, so the intermediate scale, $\mu_i$, should
also run with increasing $\omega$. This is accomplished by taking $\mu_i =
\mu_i'(\w) = [\mu'_\Lambda(\w)\, \mu_b]^{1/2}$ for each of the three running
$\mu_\Lambda'(\w)$'s from Eq.~(\ref{rscale}). This treatment of the tail has the
advantage that we obtain $\mu_\Lambda = \mu_i = \mu_b$ for $\omega \sim m_b$,
and by construction the standard factor of two scale variation, $m_b/2<\mu_b<
2m_b$. Hence it is consistent with the local OPE treatment for region 3) in
Eq.~(\ref{Expansions}). The rate at which equality is approached can be
controlled by multiplying the argument of the $\arctan$ in Eq.~(\ref{rscale}) by
a scaling factor, and the center of the transition region can be adjusted by
modifying the $2.5\,{\rm GeV}$ value shown there.

\subsection{\boldmath Perturbative results for the $B\to X_s\gamma$ spectrum}
\label{subsec:pertrate}

In this section we explore the scale dependence of the perturbative corrections
to the $B\to X_s\gamma$ spectrum in the SCET regions 1) and 2) in
Eq.~(\ref{Expansions}). The rate depends on three scale parameters $\mu_\Lambda
\gtrsim 1\GeV$, $\mu_i \sim \sqrt{m_b\, \Lambda_{\rm QCD}}$, and
$\mu_b \sim m_b$. In a short distance scheme the decay rate in \eq{dG} becomes
\begin{align}
 \frac{\df\Gamma_s}{\df \pxp} & = \Gamma_{0s}\, H_s(p_X^+,\mu_b)\,
 U_H(m_b,\mu_b,\mu_i)
 \nn\\ & \quad\times
 \int\! \df k \, \widehat P(m_b,k,\mu_i) \, \widehat F(p_X^+ - k) \,,
\end{align}
where $H_s$ and $U_H$, are given in \eqs{Hs}{UH},
and the notation $\widehat P$ and $\widehat F$ indicate that these are
given in a short distance scheme. From the analysis in
\app{pert}, all perturbative corrections can be organized into a simple series
of plus distributions $\cL_j^\eta(z)$, defined in
Eqs.~(\ref{eq:cLna_def}) and (\ref{Lm1}).  The integral of the perturbative
function $P$ with the $F$ function is then
\begin{align} \label{PFrate}
  &\inte{k} P(p^-, k, \mu_i)\, F(\pxp - k)
\\*
 & \quad
= \sum_{j = - 1}^\infty P_j(p^-, \pxp, \mu_i,\mu_\Lambda)
  \int_0^1\! \df z\, \cL_j^\eta(z)\, F[\pxp(1-z)]
 \,,\nn
\end{align}
where explicit results for the coefficients $P_j$ are given in
Eq.~(\ref{eq:Pj}).  The additional terms generated by transforming
Eq.~(\ref{PFrate}) into the integral over the short distance $\widehat P$ and
$\widehat F$ are described in detail in App.~\ref{subsec:scheme}.

\begin{figure}[t!]
\includegraphics[width=\columnwidth]{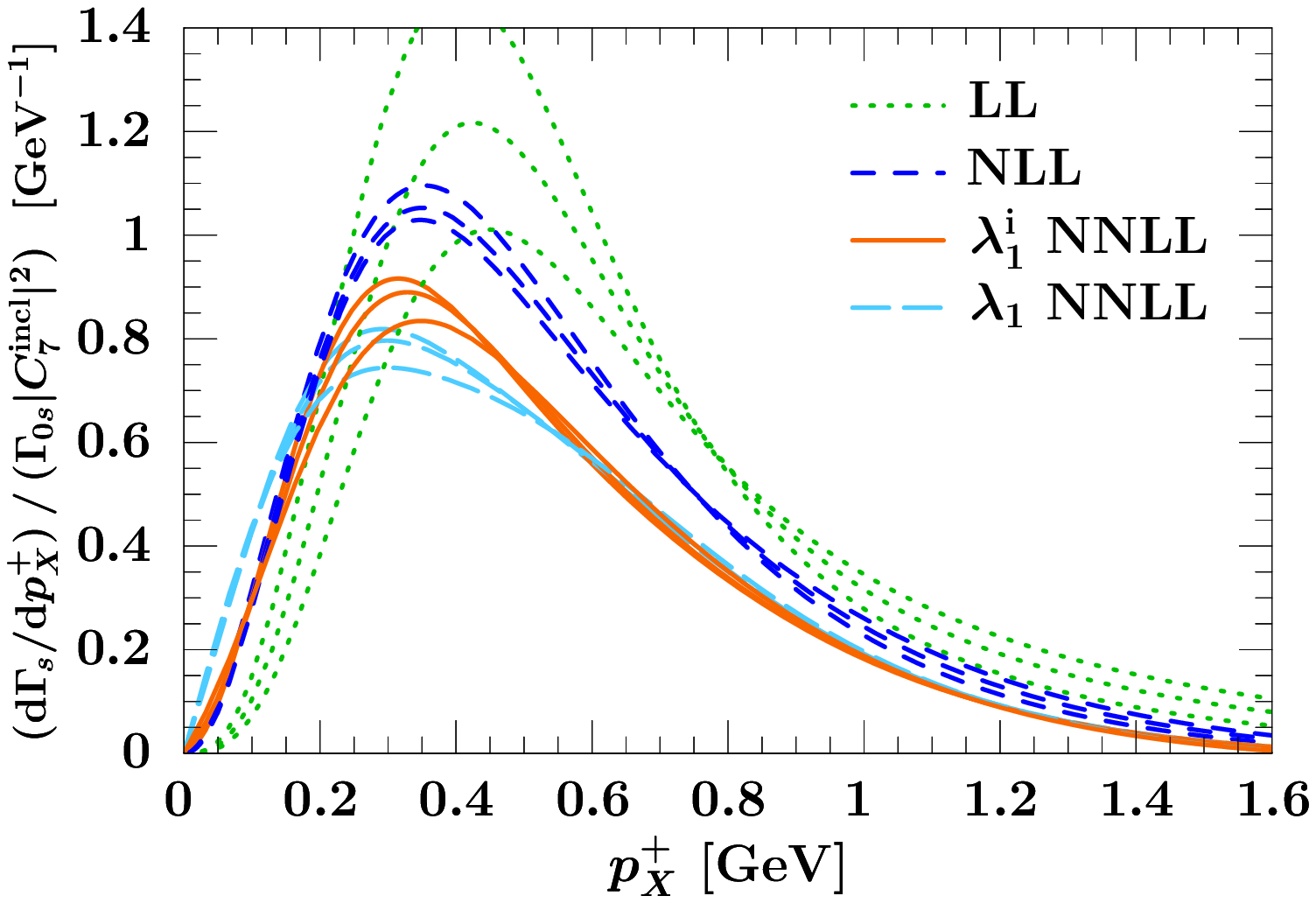}%
\vspace*{-1.5ex}
\caption{$\mu_\Lambda$ dependence of the $B\to X_s\gamma$ spectrum in terms of
  $m_b^{1S}$ and $\lambda_1^\inv$, using $\hFmod(k)$.  Shown are the LL
  (dotted), NLL (dashed), and NNLL (solid) spectra for $\mu_\Lambda =
  1.2,\, 1.5,\, 1.9\GeV$.  The long dashed curves are the NNLL spectra with
  $m_b^{1S}$ and $\lambda_1$, showing less convergence at small $\w$.}
\label{fig:rate1Sinv1}
\vspace*{2ex}

\includegraphics[width=\columnwidth]{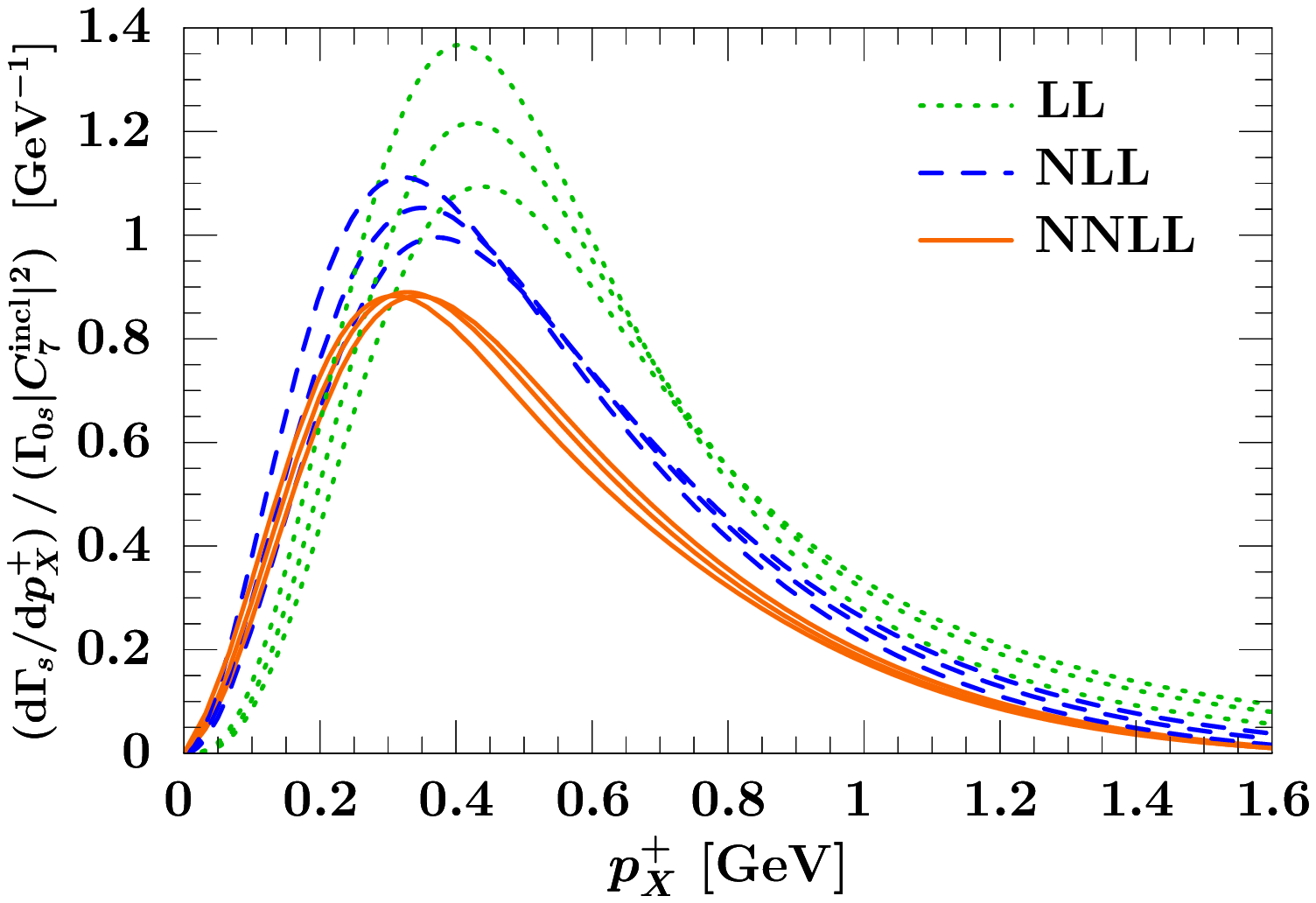}%
\vspace*{-1.5ex}
\caption{$\mu_i$ dependence of the $B\to X_s\gamma$ spectrum in terms of
  $m_b^{1S}$ and $\lambda_1^\inv$, using $\hFmod(k)$, at LL (dotted),
  NLL (dashed), and NNLL (solid) order for $\mu_i = 2.0,\, 2.5,\, 3.0\GeV$.}
\label{fig:rate1Sinv2}
\vspace*{2ex}

\includegraphics[width=\columnwidth]{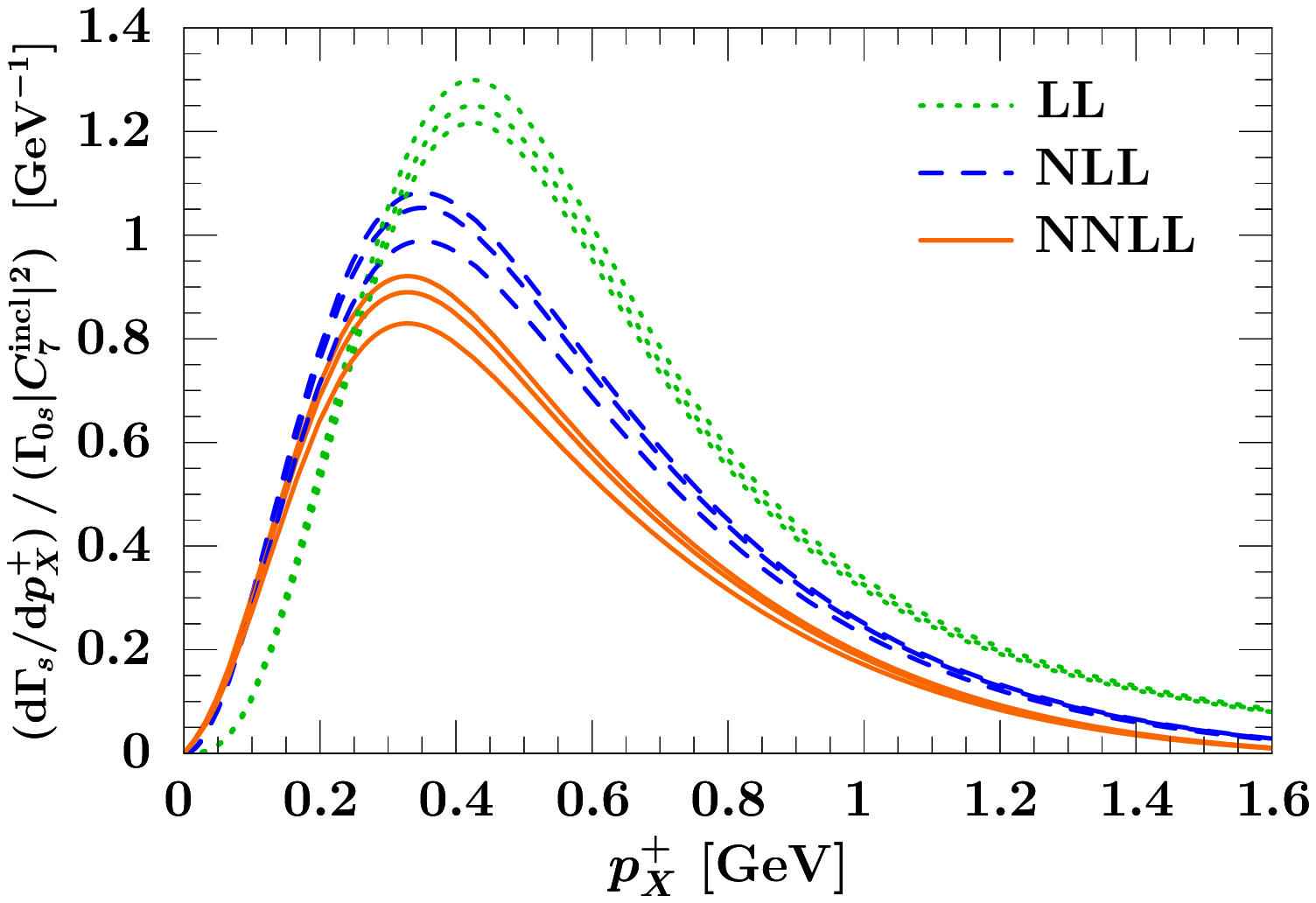}%
\vspace*{-1.5ex}
\caption{$\mu_b$ dependence of the $B\to X_s\gamma$ spectrum in terms of
  $m_b^{1S}$ and $\lambda_1^\inv$, using $\hFmod(k)$, at LL (dotted), NLL
  (dashed), and NNLL (solid) order for $\mu_b = 2.35,\, 4.7,\, 9.4\GeV$.}
\label{fig:rate1Sinv3}
\vspace*{-6ex}
\end{figure}

In Figs.~\ref{fig:rate1Sinv1},~\ref{fig:rate1Sinv2}, and~\ref{fig:rate1Sinv3} we
study the $\mu_\Lambda$, $\mu_i$, and $\mu_b$ dependencies of the $p_X^+$
spectrum for $B\to X_s\gamma$, namely $(\df\Gamma_s/\df\pxp) / \bigl[\Gamma_{0s}\,
|C_7^{\rm incl}(0)|^2\bigr]$. Recall that this has a simple relation to the
photon energy spectrum, $p_X^+ = m_B-2E_\gamma$.  We show the $p_X^+$ spectrum
to facilitate easier comparison with the results for the shape function in the
previous section. Since we are interested in studying the perturbative
corrections, we keep $\hFmod(k)$ fixed to be our default model in
Eq.~(\ref{eq:Fmodel}).  In each of Figs.~\ref{fig:rate1Sinv1},
\ref{fig:rate1Sinv2}, and \ref{fig:rate1Sinv3} we show results at LL
(dotted green curves), NLL (dashed blue curves), and NNLL (solid
orange curves) order, for three different values of the scales, and using the $1S$ mass
and $\lambda_1^\inv$ scheme.  The central values are $\mu_\Lambda=1.5\GeV$,
$\mu_i=2.5\GeV$, and $\mu_b=4.7\GeV$, two of which are held fixed in each plot.
In Fig.~\ref{fig:rate1Sinv1} the three curves at each order show
$\mu_\Lambda=1.2,\, 1.5,\, 1.9\GeV$, in Fig.~\ref{fig:rate1Sinv2} they show
$\mu_i=2.0,\, 2.5,\, 3.0\GeV$, and in Fig.~\ref{fig:rate1Sinv3} they show
$\mu_b=2.35,\, 4.7,\, 9.4\GeV$. Since the scales should obey the hierarchy
$\mu_\Lambda < \mu_i < \mu_b$, it is not possible to vary each by a factor of
two. For illustration, we do vary $\mu_b$ by a factor of two, but keep $\mu_\Lambda$ and $\mu_i$
in ranges suitable to the physical shape function and jet function regions,
respectively. The largest effect going from LL to NLL to NNLL is the change in
the normalization.  Although it is not captured by our range of scale variation,
the normalization still exhibits reasonable convergence.  Rescaled to
a common normalization, the shape of the spectrum shows very nice
convergence, and the range of scales used is clearly suitable here. We find a
reduction in the scale dependence order by order for most values of $p_X^+$.

In Fig.~\ref{fig:rate1Sinv1} we show three additional NNLL curves (long
dashed light blue) that use $m_b^{1S}$ and $\lambda_1$ instead of
$\lambda_1^\inv$. Comparing these curves to the orange curves ($m_b^{1S}$ and
$\lambda_1^\inv$), we see that the overall effect of using $\lambda_1^\inv$
instead of $\lambda_1$ is small. However, one can clearly observe that the
invisible scheme improves the perturbative convergence of the spectrum at small
$p_X^+$.

Although we do not show results here for the spectrum in the local OPE region 3) in Eq.~\eqref{Expansions},
we checked that the same effects as in the tail
of $S(\w,\mu_i)$ in Fig.~\ref{fig:1Sschemetail} appear in the tail of the spectrum.
To avoid a large $\mu_\Lambda$ dependence in the tail, the scales must be increased
with $\w$ as discussed below Eq.~\eqref{rscale}, and this also helps
to obtain a positive tail for the spectrum in region 3).

It is important to emphasize that in the common approach to model distribution functions
corresponding to the dashed curves in Fig.~\ref{fig:running}, where a fixed model for $S(\w, \mu_\Lambda)$ is specified at the scale $\mu_\Lambda$ and then run up to $\mu_i$, the spectrum will have a large dependence on $\mu_\Lambda$, which must be considered a model parameter.
Thus, there is no analog to the $\mu_\Lambda$ independence of the rate obtained in our construction, and illustrated in Fig.~\ref{fig:rate1Sinv1}. Note that the dependence of the $B\to X_s\gamma$ rate for $\pxp \leq (\pxp)^\mathrm{cut}$ on the soft and intermediate scales, $\mu_\Lambda$ and $\mu_i$, has so far been studied only by performing an expansion in $\lqcd/(\pxp)^\mathrm{cut}$ in region 2) of Eq.~\eqref{Expansions}~\cite{Neubert:2004dd, Becher:2006pu}, as in \eqs{O_OPE}{S_OPE}. In this case, one can combine the same perturbative ingredients
as in our approach, depending on the three scales $\mu_\Lambda$, $\mu_i$, and $\mu_b$.
As mentioned before, the advantage of our framework is that it does not rely on
performing an expansion in region 2).

\section{\boldmath Expansion of the \Fname\ Function}
\label{sec:basis}

\subsection{Complete orthonormal basis}
\label{subsec:orthonorm}

So far in the literature the uncertainty related to the unknown functional
form of the shape function has been either neglected or estimated by using 
a few model functions for $S(\w, \mu)$, and varying their parameters or 
distorting them~\cite{Lee:2005pw, Lange:2005yw, DeFazio:1999sv, Gambino:2007rp},
subject to constraints on their first few moments.
To obtain a systematic estimate of the
uncertainty related to the unknown functional form of the shape function, we
construct a suitable set of complete orthonormal basis functions for the
function $\hF(k)$ defined by \eq{Swhat}.  The uncertainty in the functional form
is then determined by the uncertainty in the coefficients of this basis.  This
expansion will also be convenient to extract $\hF(k)$ from experimental data,
including experimental and theoretical uncertainties and correlations.

Since $\hF(k)$ has mass dimension $-1$, it is convenient to introduce a
dimension-one parameter, $\lambda$, and use the dimensionless variable
$x=k/\lambda$.  By power counting, $\lambda\sim \lqcd$.  We expect on physical
grounds that $\hF(k)$ is positive, so we can expand its square root,
\begin{equation} \label{eq:expdef}
\hF(\lambda\, x)
  = \frac{1}{\lambda}\,\biggl[ \sum_{n = 0}^\infty c_n\, f_n(x) \biggr]^2
\,,\end{equation}
where $f_n(x)$ are a complete set of orthonormal functions,
\begin{equation}
\int_0^\infty \df x\, f_m(x)\, f_n(x) = \delta_{mn}
\,.\end{equation}
Since $\hF(k)$ is normalized to unity, the coefficients $c_n$ satisfy
\begin{equation}
\label{eq:cn_constraint}
1 = \inte{k} \hF(k) = \inte{x} \biggl[\sum_n c_n\, f_n(x) \biggr]^2
= \sum_n c_n^2
\,.\end{equation}
Although $\hF(k)$ is independent of the choice of basis functions $f_n(x)$ when
summing over all $n$, in practice only a finite number of terms can be kept.
Therefore, we want to choose basis functions, $f_n(x)$, such that the first few
terms in \eq{expdef} provide a good approximation to $\hF(k)$. Unfortunately,
most of the well-known orthonormal functions on $[0,\infty)$ become broader with
increasing $n$, and hence have moments whose values increase with $n$~\cite{book}.  By
dimensional analysis, the $n$-th moment of $\hF(k)$ scales as $\lqcd^n$, and we
would like the basis functions to satisfy this constraint, at least for the
low-$n$ moments.

To construct a suitable orthonormal basis $f_n(x)$ on $[0,\infty)$, we consider
orthonormal functions $\phi(y)$ on $[-1,1]$ and a variable transformation
$y(x)$, which maps $x\in [0,\infty)$ to $y\in [-1,1]$.  We choose $y(x)$ to be
increasing, $y'(x)>0$.  Then, for any orthonormal basis $\phi_n(y)$ on $[-1,1]$,
\begin{equation}
\label{fphi}
f_n(x) = \sqrt{y'(x)}\; \phi_n[y(x)]
\,.\end{equation}
provides an orthonormal basis on $x\in[0,\infty)$. Choosing different $\phi_n$'s
and different $y(x)$'s allows us to change the basis functions.  It is natural to
choose $\phi_n(y)$ to be polynomials of degree $n$, and we find it convenient to
use the normalized Legendre polynomials
\begin{equation}\label{legendre}
\phi_n(y) = \sqrt{\frac{2n+1}{2}}\, P_n(y)
\,,\quad
P_n(y) = \frac{1}{2^n n!}\, \frac{\df^n}{\df y^n} (y^2 - 1)^n
.\end{equation}

To determine $y(x)$, note that for a positive definite function $Y(x)$, such that
$\int_0^\infty\!\df x\, Y(x) = 1$, the function
\begin{equation}\label{eq:Ydef}
y(x) = -1 + 2 \intlim{0}{x}{x'} Y(x')
\,,\end{equation}
satisfies $y(0)=-1$ and $y(\infty)=+1$.  Since $y'(x) = 2Y(x)$, the first basis
function is simply
\begin{equation} \label{eq:firstfn}
[f_0(x)]^2 = y'(x)\, \phi_0^2[y(x)]= Y(x)
\,.\end{equation}
To obtain a good approximation to $\hF(k)$ with the first few terms in
\eq{expdef}, one should choose $Y(x)$ to be ``similar'' to $\lambda\,
\hF(\lambda x)$.  This gives an intuition about suitable choices, and once
$Y(x)$ and $\lambda$ are fixed, the full basis is specified.  A convenient choice for $Y(x)$ is
\begin{equation}\label{eq:ourY}
Y(x,p) = \frac{(p+1)^{p+1}}{\Gamma(p+1)}\, x^p\, e^{-(p+1) x}
\,\end{equation}
for any $p> 0$ real parameter. Constructing a basis from $Y(x,p)$ yields basis
functions for which the first and second moments of $f_m(x) f_n(x)$ are order
one or smaller. Therefore, the terms in the expansion in \eq{expdef} have $n$-th
moments of order $\lambda^n$ or smaller. Choosing the scaling parameter
$\lambda\sim\lqcd$, the first few terms in the expansion in the resulting basis
gives a good approximation to the shape function.

The function in \eq{ourY} is similar to those used to model $S(\w, \mu)$ in the
literature. In fact, by choosing $Y(x)$ to be a specific model shape function,
our construction allows one to expand about it, and systematically study
corrections to an assumed functional form.  We emphasize, however, that in our
approach the only role of $\phi_n(y)$ and $Y(x)$ is to specify the basis
functions, $f_n(x)$.  The functional form for $Y(x)$ affects how quickly the
expansion in \eq{expdef} converges, but not the fact that it is a convergent
expansion.  The completeness of the basis $\phi_n(y)$ on $[-1,1]$ implies that
any square integrable function on $[0,\infty)$ can be expanded in terms of the
bases $f_n(x)$ resulting from the above construction.

\begin{figure}[t]
\includegraphics[width=\columnwidth]{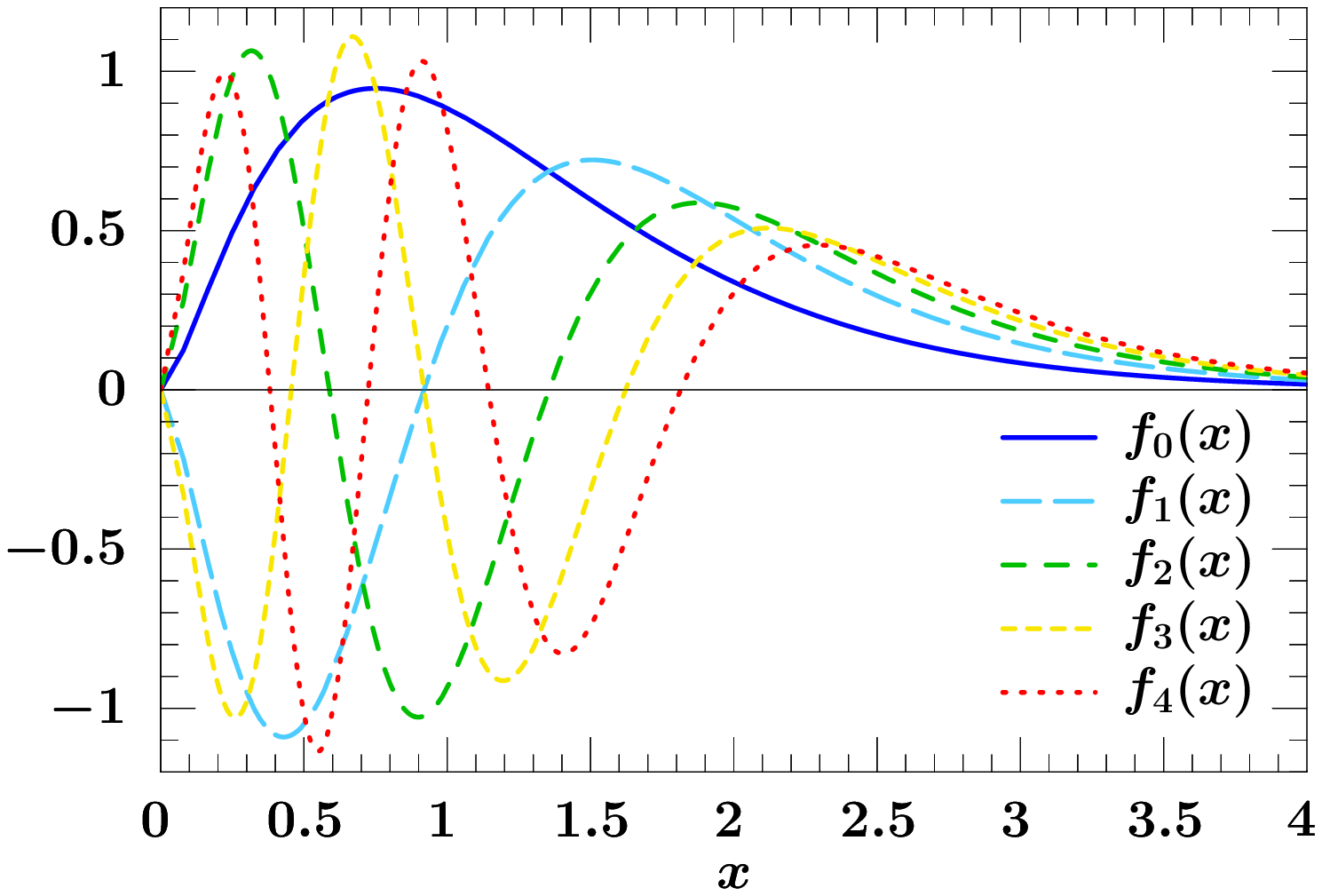}
\caption{The first 5 orthonormal basis functions in \eq{ourbasis}.}
\label{fig:basis}
\end{figure}

\begin{figure}[t]
\includegraphics[width=1\columnwidth]{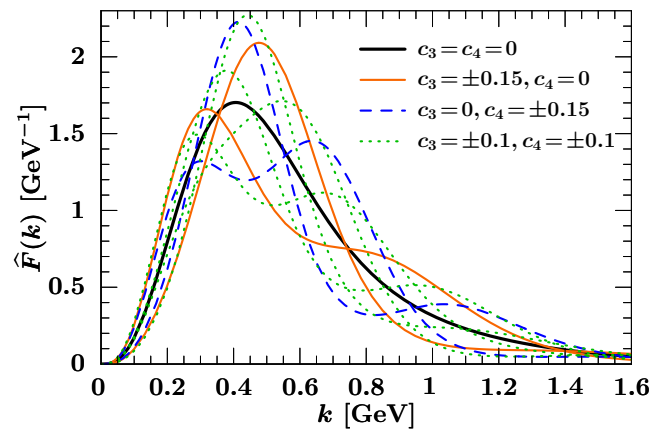}
\caption{Nine functions with identical first three moments. For each curve, we
fix $c_3$ and $c_4$ and then choose $c_0$, $c_1$, $c_2$ to satisfy the moment
constraints in \eq{Fhatmoments} with $m_b^{1S}$ and $\lambda_1^\inv$. The thick
solid curve ($c_3 = c_4 = 0$) corresponds to the default model used in
\sec{SF}.}
\label{fig:expansionmom}
\end{figure}

Using \eq{ourY} gives basis functions that behave as $f_n^2(x)\sim x^p$ as $x\to
0$. Due to the short distance subtractions in Eq.~(\ref{hatC0}), to ensure that
$S(\w,\mu)$ goes to zero at $\w=0$, we need $\hF(k)$ to go to zero at least
as $k^3$ for $k\to 0$.  Thus, we find it convenient to use $Y(x,3)$ as our
default choice, with
\begin{equation}
y(x,3) = 1 - 2 \bigg(1+4x+8 x^2+\frac{32}{3}x^3\bigg)\, e^{-4x}
\,,\end{equation}
which gives the orthonormal functions
\begin{equation}
\label{eq:ourbasis}
f_n(x) = 8\, \sqrt{\frac{2x^3\, (2n+1)}{3}}\, e^{-2x}\, P_n[y(x,3)]
\,.\end{equation}
For numerical calculations we use $\lambda = 0.8\GeV$ as default. These
functions are also convenient because they allow analytic calculations of the
decay spectra.  The first five basis functions in \eq{ourbasis} are shown in
Fig.~\ref{fig:basis}. In Fig.~\ref{fig:expansionmom} we use these to illustrate
the uncertainty remaining in $\hF(k)$ if its first three moments are fixed.  We
fix $c_3$ and $c_4$ to 9 different combinations, and choose the first three coefficients $c_{0,1,2}$ to
satisfy the moment constraints in \eq{Fhatmoments}.  All the 9 functions shown
have $0.92 < c_0 < 0.95$, $-0.35 < c_1 < -0.28$, and $0.01 < c_2 < 0.14$. Even
this plot makes the uncertainties look smaller than they are, since at small $k$
the  $k^3$ behavior of these models appears to imply a small uncertainty.
Including the short distance subtractions from Eq.~(\ref{hatC0}), these models
yield a significantly wider variation in $S(\w, \mu)$ for small $\w$. Figure~\ref{fig:expansionmom} shows that even with small errors of the $B\to
X_c\ell\bar\nu$ moments, more information on the shape function can be extracted
from the $B\to X_s\gamma$ or $B\to X_u\ell\bar\nu$ data.

\subsection{Truncation uncertainties}
\label{subsec:truncation}

\begin{figure*}[tb]
\includegraphics[width=\columnwidth]{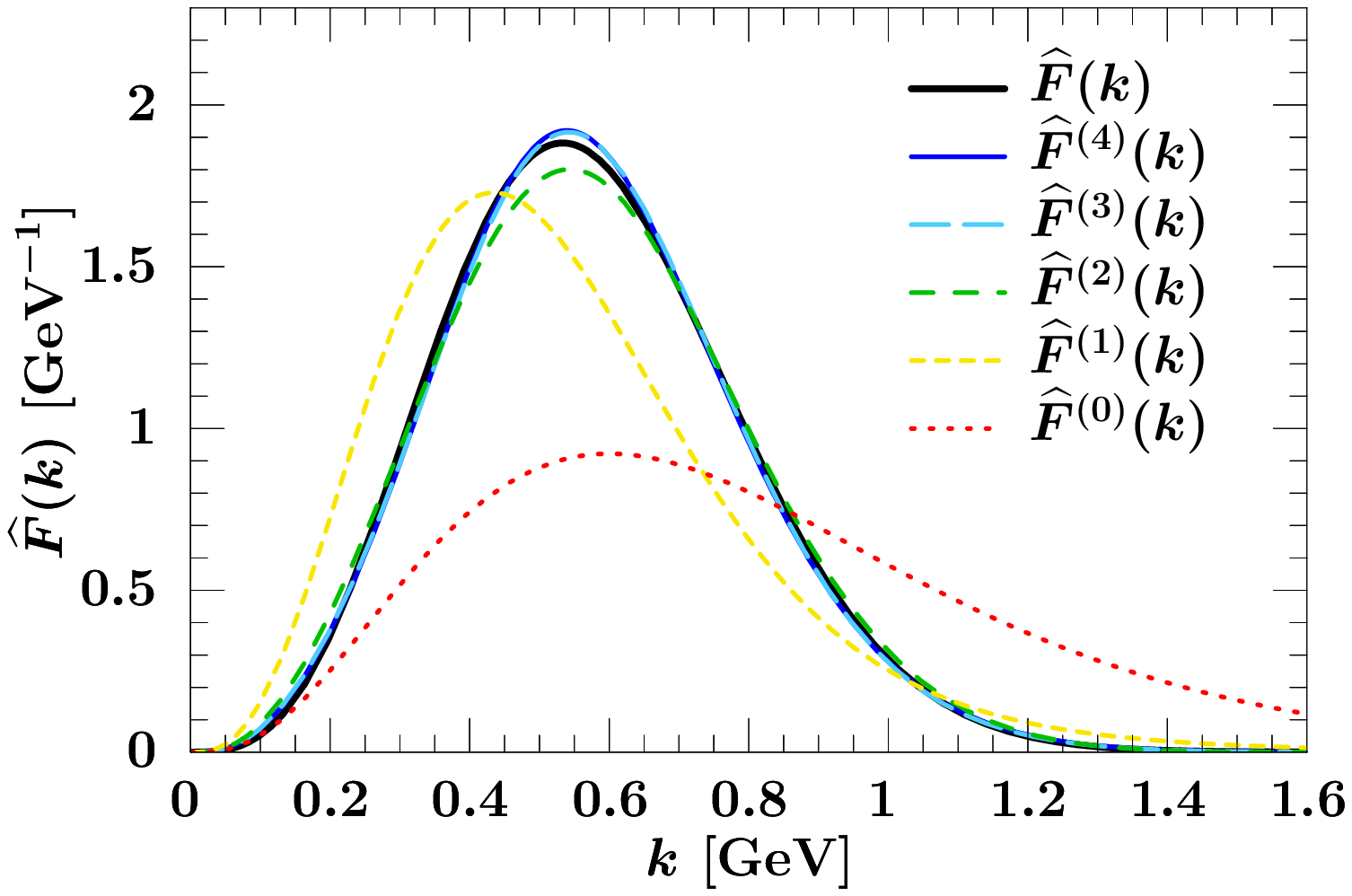}%
\hspace{\columnsep}%
\includegraphics[width=\columnwidth]{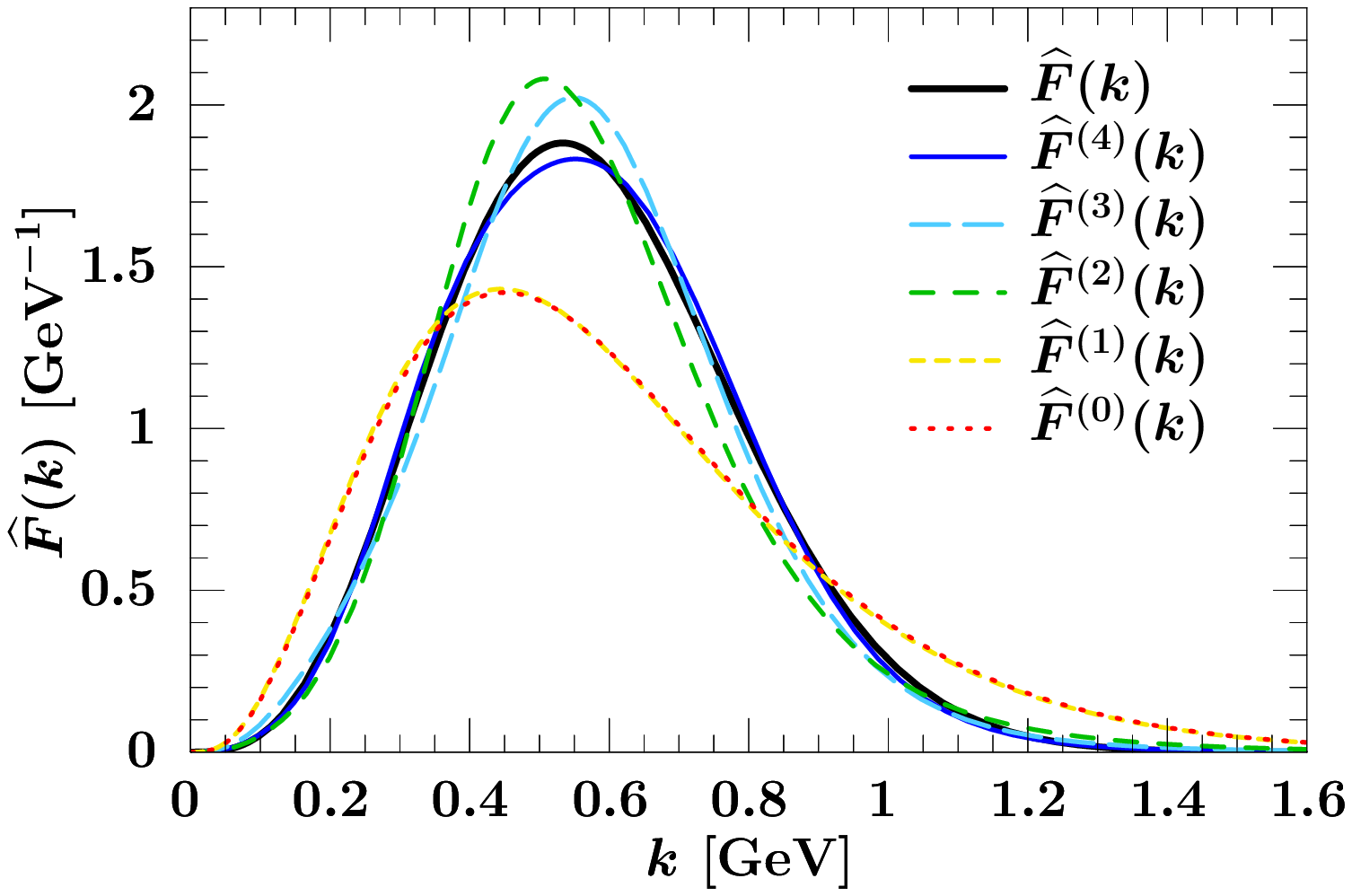}
\caption{Left: Expansion of a Gaussian model function $\hF^\mathrm{Gauss}(k) =
(2/a)(k/a)^3 \exp[-(k/a)^2]$ with $a = 0.639\GeV$, corresponding to $m_b =
4.7\GeV$, using the first 5 basis functions, $f_n(x)$, in \eq{ourbasis}. Right:
Same as on the left, except that in the definition of the basis $\lambda$ is
changed from $0.8\GeV$ to $0.6\GeV$.}
\label{fig:modeling}
\end{figure*}

Equation~(\ref{eq:expdef}) provides a model independent description of $\hF(k)$
for any choice of the basis.  Since the basis is complete for any value of
$\lambda$, we will regard $\lambda$ along with the function $Y(x)$ as part of
the convention that defines the basis. In practical applications one has to
truncate the series in \eq{expdef} after the first $N+1$ terms,
\begin{align} \label{eq:truncate}
\hF^{(N)}(k)
& \equiv \frac{1}{\lambda}\, \bigg[ f^{(N)}\Big(\frac k\lambda\Big) \bigg]^2
= \frac{1}{\lambda}\, \biggl[\sum_{n = 0}^N c_n\,
  f_n\Bigl(\frac{k}{\lambda}\Bigr) \biggr]^2 \nn\\*
&= \frac{1}{\lambda}\, \sum_{m,n = 0}^N c_m c_n\, f_m
  \Bigl(\frac{k}{\lambda}\Bigr)\, f_n\Bigl(\frac{k}{\lambda}\Bigr)
\,,\end{align}
and use $\hF^{(N)}(k)$ in the actual calculations.

Figure~\ref{fig:modeling} illustrates how the expansion converges.  Both plots
show a toy Gaussian model function (black solid curve), which we expand
in terms of the first $n$ basis functions in \eq{ourbasis} for $n \leq 4$. In
the left panel, we use the default value $\lambda = 0.8\GeV$ to define the basis.
The truncated series in \eq{truncate}
quickly approaches the model function, even for small values of $N$. In the right
panel, we show the same expansion using $\lambda = 0.6\GeV$ to define
the basis, which illustrates how the value of $\lambda$ affects the convergence
of the expansion.

To quantify the uncertainties, we would like to have a systematic way to
estimate the error due to the neglected terms in the truncated sum in
\eq{truncate}. The truncation error is affected by the choice of $N$, $Y(x)$,
and $\lambda$, providing several handles on this uncertainty. A virtue of adding
one more term from our orthonormal basis, compared to adding one more parameter
to a generic model, is that due to the orthogonality of the basis functions the
additional parameter provides independent information and should avoid large
parameter correlations.  By varying $N$, one can change the number of
coefficients, and study how sensitive the results are to the truncation.  As a
consistency check, one can use a different basis, or change $Y(x)$ or $\lambda$
and check that the difference is within the previous truncation error estimate.

A feature of our construction is that the truncation error
can be estimated using $\sum_{n=0}^\infty c_n^2 = 1$. We define
\begin{align} \label{eq:fr_def}
\sqrt{\lambda\, \hF(\lambda\, x)} \equiv f(x) &= f^{(N)}(x) + c_r\, f_r(x)\,,
\end{align}
where
\begin{equation} \label{eq:crdef}
c_r = \biggl(1 - \sum_{n = 0}^N c_n^2 \biggr)^{1/2}
\,,\end{equation}
and the remainder function, $f_r(x)$, is orthogonal to $f_{0,\ldots,N}(x)$,
\begin{equation} \label{eq:frexp}
f_r(x) = \sum_{n = 1}^\infty r_n\, f_{N+n}(x)
\,,\end{equation}
and normalized
\begin{equation} \label{eq:frnorm}
\int_0^\infty \df x\, [f_r(x)]^2 = 1
\,,\qquad
\sum_{n = 1}^\infty r_n^2 = 1
\,.\end{equation}
In terms of $f_r(x)$, the truncation error from approximating $\hF(k)$ by
$\hF^{(N)}(k)$ is
\begin{align} \label{eq:Ftrunc}
\hF^{(N)}_\mathrm{trunc}[f_r](k)
&\equiv \bigl\lvert\hF(k) - \hF^{(N)}(k)\bigr\rvert \\
& =\frac{1}{\lambda}\, \biggl\lvert\,
2\,c_rf_r\Bigl(\frac{k}{\lambda}\Bigr)\, f^{(N)}\Bigl(\frac{k}{\lambda}\Bigr)
+ c_r^2 f_r^2\Bigl(\frac{k}{\lambda}\Bigr) \biggr\rvert \nn
\,.\end{align}

\begin{figure*}[t]
\includegraphics[width=\columnwidth]{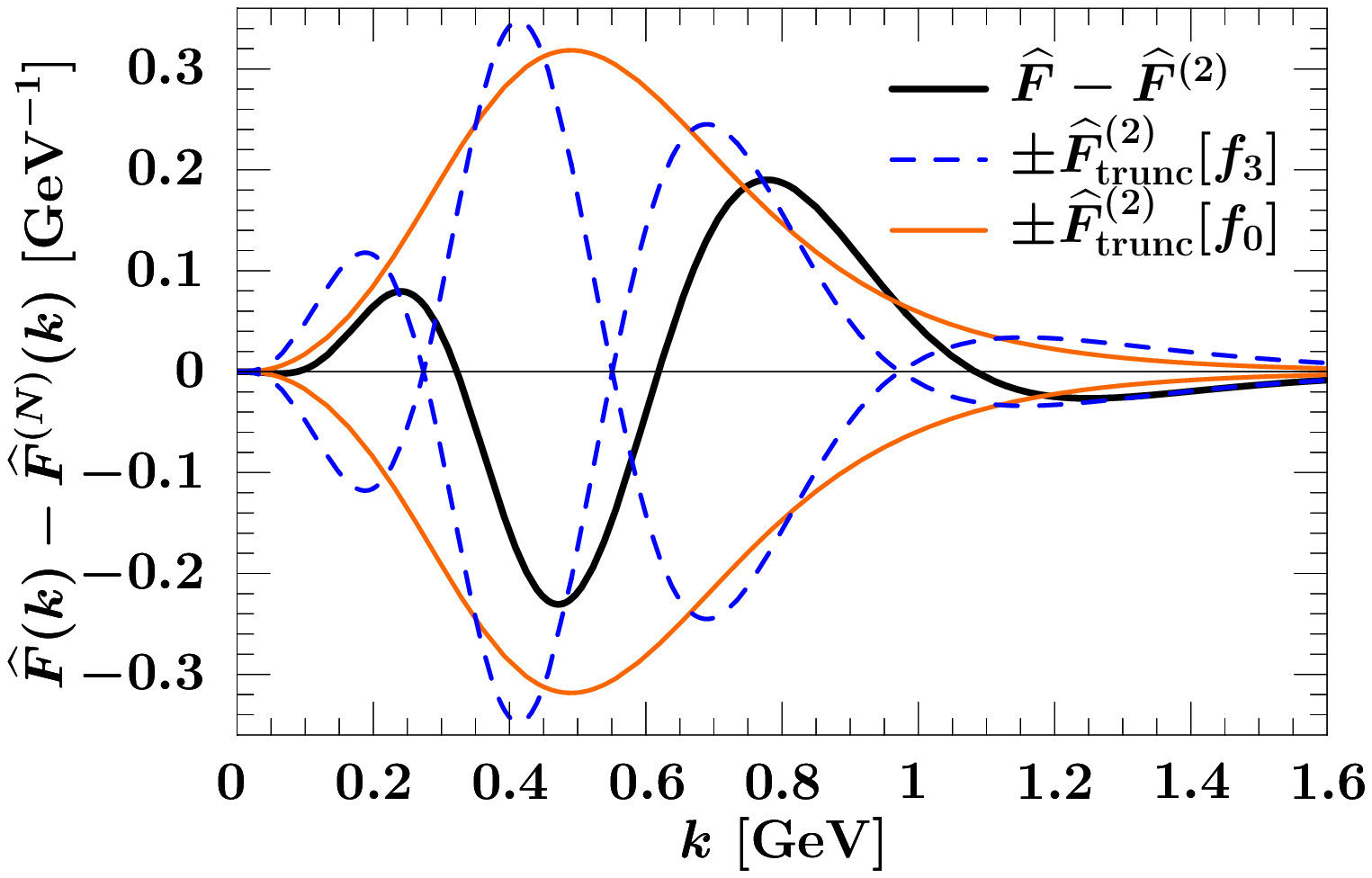}%
\hspace{\columnsep}%
\includegraphics[width=\columnwidth]{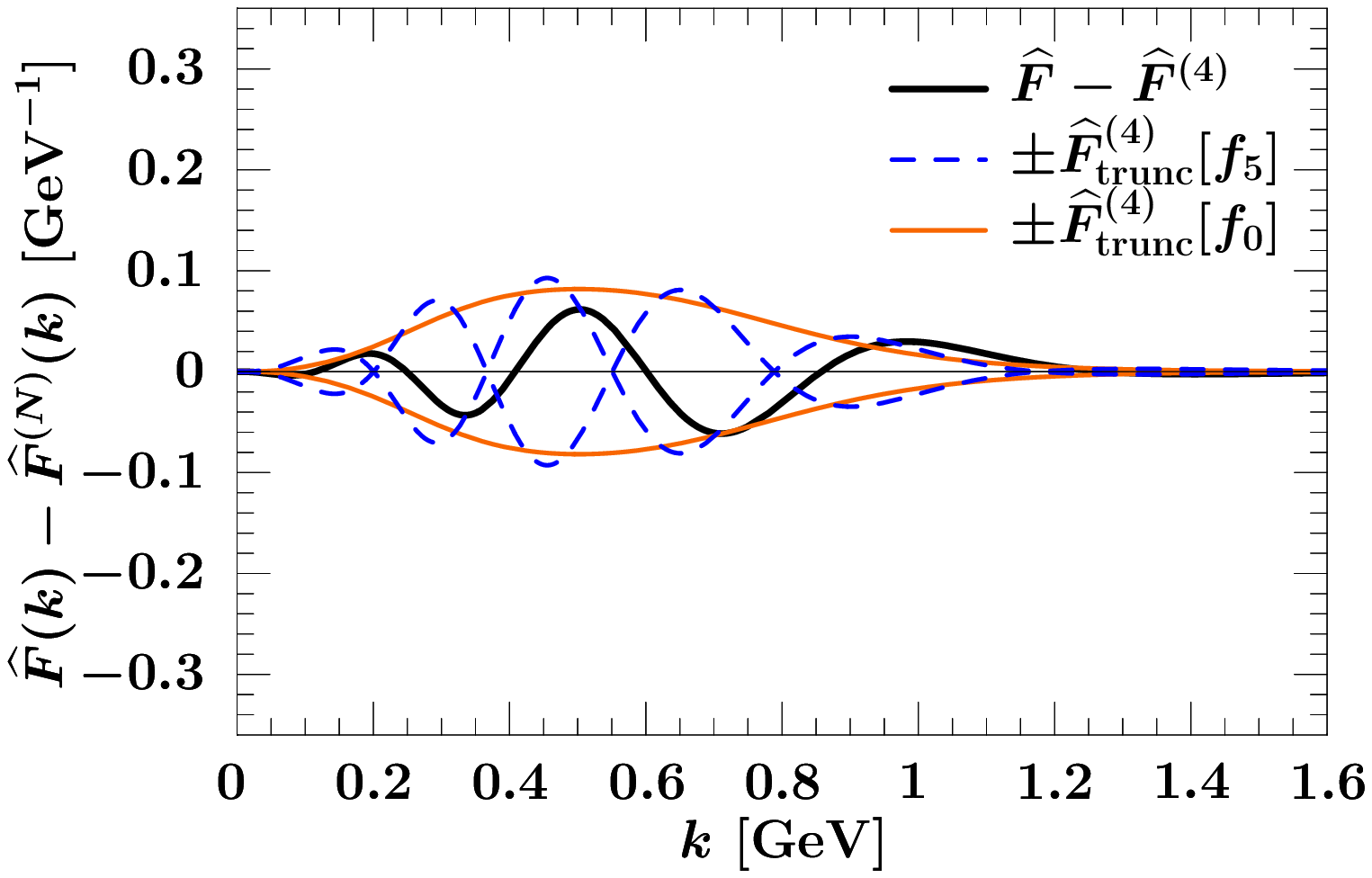}
\caption{Truncation error estimates.  The solid black curves show
$\hF^\mathrm{Gauss}(k) - \hF^{(N)}(k)$ corresponding to the right plot in
Fig.~\ref{fig:modeling}, for $N=2$ (left) and $N = 4$ (right). The dashed blue
curves show $\pm \max\big\{\hF^{(N)}_\mathrm{trunc}[\pm f_{N+1}]\big\}$, and the
solid orange curves show $\pm \max\big\{\hF^{(N)}_\mathrm{trunc}[\pm
f_{0}]\big\}$, i.e., the error estimates using Eqs.~(\ref{eq:fr1}) and
(\ref{eq:fr2}), respectively.}
\label{fig:truncate}
\end{figure*}

Although \eq{frnorm} implies $\abs{r_n} \leq 1$, it provides no bound on
$\sum_n\abs{r_n}$. Therefore, one cannot derive a rigorous bound on
$\abs{f_r(x)}$ from \eq{frexp} for all $x>0$, and it is not possible to obtain a
rigorous bound on the truncation error, $\hF^{(N)}_\mathrm{trunc}$, either.
However, for practical purposes, it is sufficient to estimate
$\hF^{(N)}_\mathrm{trunc}$. Its size is controlled by $c_r$ in \eq{crdef}, which
is determined by the first $N+1$ coefficients, and can be minimized by choosing
a good basis. Thus, as long as enough coefficients are included in the basis so
that the estimated truncation uncertainty is small compared to other
uncertainties, any reasonable estimate of $f_r(x)$ provides a useful estimate of
the truncation uncertainty.

For suitable choices of the basis, we expect the series to converge fairly
quickly, since the higher basis functions oscillate more and more rapidly, and
we do not expect significant structure in the momentum distribution of the $b$
quark in the $B$ meson over momentum scales $\ll \lqcd$. Hence, one way to
estimate $f_r(x)$ is to assume that one term in \eq{frexp} saturates the sum.
It is natural to take that to be the first (possibly the second) term, which
corresponds to
\begin{equation} \label{eq:fr1}
f_r(x) = \pm f_{N+1}(x)
\,.\end{equation}
Another possibility is to assume that $\abs{f_r(x)} \lesssim f_0(x)$, and take
\begin{equation} \label{eq:fr2}
f_r(x) = \pm f_0(x)
\,.\end{equation}
This is motivated by the fact that, except for small and large values of $x$,
$f_0(x)$ gives roughly an envelope for the higher basis functions, as can be
seen from Fig.~\ref{fig:basis}. In each case, the sign of $f_r(x)$ is
undetermined.  For small $c_r$, the second term in \eq{Ftrunc} can be neglected
and the sign of $f_r(x)$ is irrelevant.

To illustrate these methods for estimating the truncation error we consider
again the toy model used in Fig.~\ref{fig:modeling}, with
$\lambda=0.6$ in our basis as in the right panel. In Fig.~\ref{fig:truncate},
the solid black curve shows $\hF(k) - \hF^{(N)}(k)$ for $N=2$ (left
panel) and $N = 4$ (right panel). The dashed blue curves show the estimate using
\eq{fr1}, taking the point-by-point maximum of $\hF^{(N)}_\mathrm{trunc}[\pm
f_{N+1}](k)$, while the solid orange curves use \eq{fr2} instead. As expected,
the former correlates more with the absolute value of the point-by-point
deviations, while the latter roughly envelopes the difference.  As can be seen,
we obtain reasonable estimates of the deviation of $\hF^{(N)}(k)$ from
$\hF(k)$.  The overall sizes of the truncation errors are well estimated in both
plots, because they are proportional to $c_r$.  For the left plot $c_r = 0.088$
and for the right plot $c_r = 0.025$.  If the expansion of $\hF(k)$ converges
reasonably fast, then $f_r(x)$ will be an oscillatory function, as shown in
Fig.~\ref{fig:truncate}.  Ultimately we are interested in how the shape function
impacts the uncertainty in the extraction of $\abs{V_{ub}}$ or that in
determining the $B\to X_s\gamma$ event fraction above a certain photon energy
cut.  These depend on weighted integrals of $\hF(k)$, so their uncertainties
will be much smaller than the point-by-point errors in approximating $\hF(x)$ by
$\hF^{(N)}(x)$ shown in Fig.~\ref{fig:truncate}.

\subsection{Subleading shape functions}

Any precision analysis of differential spectra in $B\to X_s\gamma$ or $B\to
X_u\ell\bar\nu$ must incorporate power corrections that go beyond
Eq.~(\ref{eq:factor}).  At order $\lqcd/m_b$, six subleading shape functions
enter the description of the most general inclusive spectra in $B\to X_s\gamma$
and $B\to X_u\ell\bar\nu$~\cite{Bauer:2001mh, Leibovich:2002ys, Bauer:2002yu,
Lee:2004ja}. We refer to these as the primary subleading shape functions.  In
addition there are terms in $B\to X_s\gamma$ that enter from operators other
than $O_7$ and are related to the photon's hadronic structure, whose
contribution to the total rate is not calculable in the
OPE~\cite{Kapustin:1995fk,Lee:2006wn}. Considering only the primary subleading
shape functions and using suitably weighted integrals of the $B\to
X_u\ell\bar\nu$ differential rate together with $B\to X_s\gamma$, the ${\cal
O}(\lqcd/m_b)$ shape functions can be canceled in the determination of
$|V_{ub}|$~\cite{Lee:2008vs}. However the same drawbacks apply to the weighting
method at subleading order as those mentioned already in the introduction, so it
is interesting to consider how our analysis can be extended to include these
subleading shape functions.

The appropriate factorization theorem for the primary subleading shape
functions, analogous to Eq.~(\ref{eq:factor}), is known from
Ref.~\cite{Lee:2004ja}. Thus, when one-loop partonic calculations of these
functions and their corresponding jet functions are available, a construction
analogous to Eq.~(\ref{eq:F_def}) can be carried out to build in the proper
large-$\w$ tail, $\mu_\Lambda$ dependence, etc. However, it should be cautioned
that a large number of additional shape functions enter in the matching at
$\ord{\alpha_s\lqcd/m_b}$~\cite{Lee:2004ja, Beneke:2004in, Trott:2005vw},
so the utility of extending our full analysis to
this level is unclear.

Here, we simply discuss how a complete basis can be constructed for the primary
subleading shape functions.  We know less about these shape functions than about
the leading order one. Their moments are still related to HQET matrix elements.  In
particular, the zeroth moment of the $\ord{\lqcd/m_b}$ shape functions vanishes,
which means they must be negative for some values of $k$,
and their first moments either vanish or are given by linear combinations of
$\lambda_1$ and $\lambda_2$.  The zeroth moments no longer vanish for the
$\ord{\lqcd^2/m_b^2}$ shape functions and
beyond~\citetext{\citealp{Leibovich:2002ys}, \citealp{Mannel:2004as}, \citealp{Tackmann:2005ub}}.

For simplicity, we define the sign of each $\ord{\lqcd/m_b}$ shape function such that its first nonzero moment is positive. A convenient expansion for the functions with vanishing first moment is given
by
\begin{equation}
H(\lambda\,x) = -\frac\df{\df x}\, \biggl[ \sum_n  d_n\, f_n(x) \biggr]^2
\,,
\end{equation}
while the four-quark operator shape functions, whose zeroth
and first moments vanish, can be expanded as
\begin{equation}
H^{4q}(\lambda\,x) = \frac{\df^2}{\df x^2}\, \biggl[ \sum_n  d_n\, f_n(x) \biggr]^2
\,.\end{equation}
With this form, $\int_0^\infty \df k\, H^{(4q)}(k) = \int_0^\infty \df k\,k\,
H^{4q}(k) = 0$. The first nonzero moments provide constraints on the
coefficients, similar to \eq{cn_constraint}, for example
\begin{equation}
\intlim{0}{\infty}{k} k\, H(k) = \lambda^2 \sum_n  d_n^2
\,,\end{equation}
which can be set to the appropriate linear combinations of $\lambda_1$ and
$\lambda_2$. Note that a different set of coefficients, $d_n$, occurs for each
subleading shape function.

\section{\boldmath Extracting the \Fname\ Function and predicting decay rates}
\label{sec:fit}

For a given basis for the leading and subleading shape functions we want to
extract the basis coefficients, $c_i$ and $d_i$, including their uncertainties
and correlations. This extraction will use data on the $B\to X_s\gamma  $
and $B\to X_u\ell\bar\nu$ spectra and data that determines moments of the
shape functions from $B\to X_c\ell\bar\nu$. To simplify our discussion, we adopt
a notation that is suitable for the coefficients $\{c_0,\ldots,c_N\}$ of the
function $\hF(k)$ appearing in the leading shape
function. The generalization to incorporate subleading shape functions is
straightforward. All results in this section are exact for $N\to\infty$ and for
finite $N$ one has to take into account the truncation error discussed
in \subsec{truncation}.

Since the function $\hF(k)$ enters the decay spectra linearly, we can
independently compute the contributions of the product of any two basis
functions, $f_m(x) f_n(x)$, in the expansion of $\hF(k)$, which we denote by
$\df\Gamma_{mn}$.  The differential spectra $\df\Gamma_s = \df\Gamma_s/\df
E_\gamma$ or $\df\Gamma_u = \df\Gamma_u/\df E_\ell \df p_X^+ \df p_X^-$ are then
given by
\begin{equation} \label{eq:spec_expansions}
\df\Gamma = \sum_{m,n=0}^N c_m\, c_n\, \df\Gamma_{mn}
\,,\end{equation}
where from combining \eq{dG} and (\ref{eq:expdef}), the $\df\Gamma_{mn}$ are
\begin{align} \label{eq:Gammamn_def}
\df\Gamma_{mn} &= \Gamma_0\, H(p^-)\! \int_0^{p_X^+}\! \df k\
\frac{\hP(p^-,k)}{\lambda}
\nn\\ & \quad \times
f_m\Bigl(\frac{\pxp\!-\!k}{\lambda}\Bigr) \,f_n\Bigl(\frac{\pxp\!-\!k}{\lambda}\Bigr)
.\end{align}
For simplicity, we suppress the scale $\mu$ in the arguments of $H$ and $\hP$, and
the $u$ or $s$ subscripts on $H$. The $\df\Gamma_{mn}$ in \eq{Gammamn_def} act
as a basis for the physically measurable distributions, and the result in
\eq{spec_expansions} is a quadratic polynomial in each of the fit parameters, $c_i$.

In practice, experimental spectra are binned. Integrating \eq{spec_expansions}
the rate in the $i$-th bin is
\begin{equation} \label{eq:GammaFit}
\Gamma^i  = \sum_{m,n=0}^N c_m\, c_n\,\Gamma_{mn}^i
\,,\end{equation}
where $\Gamma_{mn}^i$ is the integral of $\df\Gamma_{mn}$ in \eq{Gammamn_def}
over the phase space region of the $i$-th bin.  In addition, we can impose
constraints on the moments of $\hF(k)$,
\begin{equation} \label{eq:MomentsFit}
M^j  = \int_0^\infty \! \df k\, k^j\, \hF(k) = \sum_{m,n=0}^N c_m\, c_n
\,M_{mn}^j
\,.\end{equation}
Here, $M^j$ are given in terms of experimental measurements of $\widehat\delta$
and $\widehat\lambda_1$ from $B\to X_c\ell\bar\nu$, while $M_{mn}^j$ is
\begin{equation}
M^j_{mn}  = \frac{1}{\lambda} \intlim{0}{\infty}{k} k^j\, f_m\Bigl(\frac{k}
{\lambda}\Bigr)\, f_n\Bigl(\frac{k}{\lambda}\Bigr)
\,.\end{equation}

One can determine the coefficients $\{c_0,\ldots,c_N\}$ by fitting
\eqs{GammaFit}{MomentsFit} to the experimental data.  It is straightforward to
combine several different spectra and measurements by different experiments.
Doing a simultaneous fit to \eqs{GammaFit}{MomentsFit} allows one to combine the
$B\to X_s\gamma$ shape information~\cite{Chen:2001fja, Abe:2008sx,
Aubert:2007my} and information on the $m_X$ spectrum in $B\to
X_u\ell\bar\nu$~\cite{Aubert:2004bq} with the information on $m_b$ and matrix
elements of local operators known from $B\to X_c\ell\nu$~\cite{Bauer:2004ve,
Buchmuller:2005zv, Barberio:2007cr}.  Given the theoretical input,
$\Gamma^i_{mn}$ and $M_{mn}^j$, computed in this paper, one needs to simply fit
quadratic polynomials in the fit parameters, $c_n$, to the data.  The
experimental uncertainties and correlations in $\Gamma^i$ and $M^j$ and the
theoretical uncertainties and correlations in $\Gamma_{mn}^i$ and $M_{mn}^j$
will translate to uncertainties of the coefficients $c_n$.  In this approach,
the uncertainty in the functional form of the $F$ function is automatically and
straightforwardly determined by the uncertainties of the coefficients $\{c_0,\ldots,c_N\}$,
and the truncation error from approximating $\hF(k)$ by $\hF^{(N)}(k)$
discussed in \subsec{truncation}.

\begin{figure}[t]
\includegraphics[width=\columnwidth]{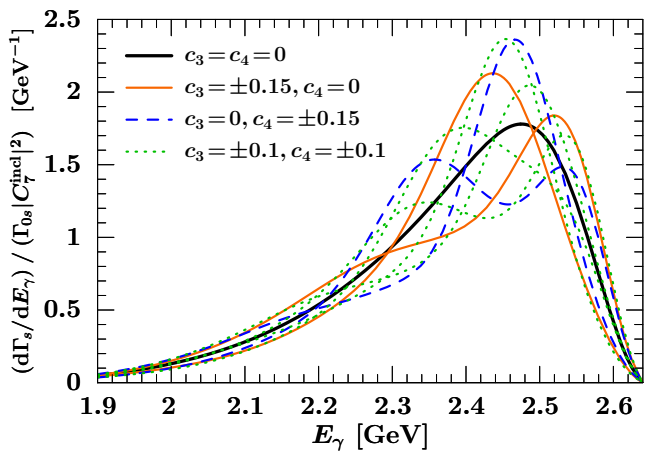}
\caption{The $B\to X_s\gamma$ photon spectrum in the $B$ rest frame using the nine shape
function models shown in Fig.~\ref{fig:expansionmom}.}
\vspace{-1ex}
\label{fig:spectrum}
\end{figure}

Finally, in Fig.~\ref{fig:spectrum} we combine our results to show the $B\to
X_s\gamma$ spectrum, $(\df\Gamma_s/\df E_\gamma) / \bigl[\Gamma_{0s}\, |C_7^{\rm
incl}(0)|^2\bigr]$, in the $B$ meson restframe.  The results are shown using the
short distance parameters $m_b^{1S}$ and $\lambda_1^\inv$, and the nine shape
function models plotted in Fig.~\ref{fig:expansionmom}, which have fixed zeroth,
first, and second moments.  The solid black curve is our default model,
$\hFmod(k)$ in Eq.~(\ref{eq:Fmodel}).
In the current and future experimental analyses $m_B - 2E_\gamma \lesssim 2 \GeV$.
Our formalism has the virtue that we do not need to distinguish regions 1) and 2) in Eq.~\eqref{Expansions},
i.e., it is simultaneously valid for both $m_B-2E_\gamma^\mathrm{cut} \sim\lqcd$ and $\gg\lqcd$.
The variation of the curves near maximal
$E_\gamma$ indicates that in the peak region the first few moments of the shape
function are not sufficient to predict \pagebreak the spectrum.  However, for $E_\gamma \lesssim
2.1\GeV$, the uncertainty in the prediction becomes significantly smaller, and the
leading order shape function model uncertainty diminishes. The spectrum in
this region is also affected by subleading shape functions and additional
perturbative corrections not studied here. (Note that the measured
spectrum is somewhat broadened by experimental effects.)
We found the same conclusion using
many other shape function models, constructed from different bases,
and expect that this will also be substantiated
by an actual experimental analysis in which the space of shape functions is
explored by allowing for more components in our orthonormal basis and a
simultaneous fit to all relevant data.  This will have important implications
for the $B\to X_s\gamma$ rate in the presence of an experimental cut, the
extraction of $|V_{ub}|$ from inclusive semileptonic $B$ decays, and the analysis of
the inclusive $B\to X_s \ell^+\ell^-$ rate in the small $q^2$ region.

\section{Conclusions}
\label{sec:conc}

In this paper, we introduced new methods for calculations relevant for inclusive
$B$ decays which involve the parton distribution function of the $b$ quark in
a $B$ meson, called the shape function.  On the theoretical side, our results allow for an
improved description of the decay rates and a more reliable assessment of the
uncertainties than earlier studies.  On the experimental side, they are
straightforward to implement and provide a transparent way to combine
constraints on the shape and moments of the shape function, with controllable
uncertainties.

The shape function is constrained by the
measurements of the shape of the $B\to X_s\gamma$ photon energy
spectrum~\cite{Chen:2001fja, Abe:2008sx, Aubert:2007my} and the $m_X$ spectrum
in $B\to X_u\ell\bar\nu$~\cite{Aubert:2004bq}, and its moments are related to
the $b$ quark mass, $m_b$, and nonperturbative matrix elements of local
operators in the OPE, which are constrained by fits to $B\to X_c \ell\bar\nu$
decay distributions~\cite{Bauer:2004ve, Buchmuller:2005zv, Barberio:2007cr}.

The first key ingredient in our analysis is the new description of the shape
function, given in \eq{S_construction} and discussed in detail in
\sec{SF}, which is by construction consistent with the renormalization
group evolution and the perturbative result for the tail of the shape function.
It allows combining all experimental information from the shape
of $B\to X_s\gamma$ and $B\to X_u\ell\bar\nu$ spectra, and $m_b$ and $\lambda_1$ constrained by $B\to
X_c\ell\bar\nu$ distributions. Any short distance scheme for the $b$ quark mass and
the kinetic energy matrix element can be implemented. We presented a simple formula
for the differential rates which incorporates resummed perturbative corrections,
with details of the derivation given in \apps{pert}{plus}.

The second key ingredient in our analysis is the expansion of the \Fname\ function,
describing the nonperturbative part of the shape function,
in a complete set of orthonormal functions in
\sec{basis}.  This gives a new way to quantify uncertainties in the functional
form of the shape function, which was previously not explored fully
systematically.  Choosing $Y(x)$ in \eq{Ydef} to coincide with any of
the models used in the literature~\cite{DeFazio:1999sv, Lange:2005yw, Andersen:2005mj, Gambino:2007rp, Aglietti:2007ik} gives an orthonormal basis for the shape function in which the
first function is the model and corrections can be studied.

One should use renormalon-free short distance definitions for input parameters, such
as the $b$ quark mass, $m_b$, or the kinetic energy matrix element,
$\lambda_1$.  We found that in our framework the kinetic scheme
definition of $\lambda_1^{\rm kin}$ seems to oversubtract from the
HQET definition of $\lambda_1$ in dimensional regularization, numerically similar to
using the $\overline{\rm MS}$ mass for inclusive spectra. To solve this problem, in
App.~\ref{app:invla1} we introduced a new short distance ``invisible" scheme,
$\lambda_1^\inv$, which is renormalon-free and only differs from the usual definition
starting at order $\alpha_s^2$.  While it does not improve the behavior of the perturbation series
decisively, being almost invisible, at least it does not make it worse.

It should be emphasized that all developments presented in this paper are
consistent with and incorporate all predictions that follow from QCD, without
recourse to models or relying on any ad hoc assumptions. In fact, any method
consistent with the factorization theorem and the OPE discussed in \sec{bsg}
can be cast in our framework.

Our results lead to the following strategy to determine $\abs{V_{ub}}$ with optimal
and reliable uncertainties:

\begin{enumerate}\vspace*{-1ex}
\itemsep -2pt

\item Fix a basis for the expansion of the
\Fname\ function by choosing a suitable $Y(x)$ and a value for $\lambda$.

\item Do a combined fit to the binned $B\to X_s\gamma$ and/or $B\to X_u\ell\bar\nu$
spectra, and to the
information on $m_b$ and $\lambda_1$ from $B\to X_c\ell\bar\nu$ (and possibly
higher moments) to extract the basis coefficients $\{c_0,\ldots, c_N\}$.

\item Verify that the truncation error is small compared to other uncertainties.

\item Use the values and covariance matrix of the extracted basis coefficients
to make predictions with reliable uncertainties.

\end{enumerate}\vspace*{-1ex}

More details on the shape function fitting procedure, on extracting $|V_{ub}|$
from various $B\to X_u\ell\bar\nu$ differential decay distributions, and
predictions for the $B\to X_s\gamma$ rate will be presented elsewhere.

\begin{acknowledgments}
  We thank Kerstin Tackmann for discussions and comments on the
  manuscript.  Z.L. and I.S. thank the CERN Theory Group and the Aspen Center for
  Physics for hospitality while parts of this work were completed. F.T. thanks the
  particle physics group at Humboldt University Berlin for its hospitality
  during the final stages of this work. This work was supported in part by the
  Director, Office of Science, Offices of High Energy and Nuclear Physics of the
  U.S.\ Department of Energy under the Contracts DE-AC02-05CH11231 (Z.L. and F.T.) and
  DE-FG02-94ER40818 (I.S.). I.S. was also supported in part by the DOE OJI program and
  by the Sloan Foundation.
\end{acknowledgments}

\appendix

\section{Perturbative Results}
\label{app:pert}

In this appendix we collect the known results for the hard functions
$H_{u,s}(\pxpm, \mu_i)$ in \eq{factor}, and derive an analytic result for the
function $P(p^-, p_X^+, \mu_i)$ in \eq{P_def}, which includes perturbative
corrections from the jet and soft functions as well as the shape function RGE.
In Ref.~\cite{Fleming:2007xt} it was shown in the context of dijet production
that the convolution form for the soft function analogous to \eq{F_def} allows
all factors associated with perturbative corrections and RGE to be evaluated
analytically. Here we show that this is also true for inclusive $B$ decays. Our
calculation differs from Ref.~\cite{Fleming:2007xt} in that we develop a method
that avoids using an imaginary part at intermediate steps. (An alternative
analytic method that avoids the plus distributions by using moments was
developed in Ref.~\cite{Neubert:2005nt}.)  Furthermore, with our basis for
$\hF(k)$, we show that the integrals over the basis function can be written in
terms of hypergeometric functions.

We discuss fixed order results in \subsec{fixedorder}, the RGE in \subsec{RGE},
and the rate in short distance schemes in \subsec{scheme}. The definitions of
the required plus distributions and many useful relations are collected in
\app{plus}.

\subsection{Fixed order results}
\label{subsec:fixedorder}

In \eqs{factor}{dG} the leading order hard function $H_s(\pxp, \mu_i)$ for $B\to X_s\gamma$  is
\begin{align} \label{eq:Hs}
H_s(p_X^+, \mu_i) &= \frac{(m_B - p_X^+)^3}{m_b^3}\,
  \abs{C_7^\mathrm{incl}(0)}^2 \nn\\*
& \quad \times h_s(m_b,\mu_b)\, U_H(m_b,\mu_b,\mu_i)
\,,
\end{align}
where the evolution factor $U_H(m_b,\mu_b,\mu_i)$ is given below in
Eq.~(\ref{eq:UH}), and has the boundary condition $U_H(m_b,\mu_b,\mu_b)=1$.  The
Wilson coefficient $C_7^\mathrm{incl}$ in $H_s$ contains the weak scale matching
of the full theory onto the effective Hamiltonian from which the $W$ and $t$ are
integrated out, and resums perturbative corrections between the weak scale and
the scale $\mu \sim m_b$. It does not depend on $\mu_b$ and is defined by the
split matching procedure in Ref.~\cite{Lee:2005pk}, which separates the
perturbation series above and below the scale $m_b$. It can be written
as~\cite{Lee:2006gs}
\begin{equation}
C_7^\mathrm{incl}(0) = \mathcal{C}_7 + F_7(0) + G_7(0)
\,,\end{equation}
with the central values $C_7^{\rm incl}(0) = -0.341 - 0.015 i$ and ${\cal C}_7=
-0.261$. The Wilson coefficient $\mathcal{C}_7$ and the functions $F_7(q^2)$ and
$G_7(q^2)$ are defined to be separately $\mu$ independent. Here, $\mathcal{C}_7$
contains the dependence on the $\overline{\rm MS}$ $b$ quark mass and the Wilson
coefficient $C_7(\mu)$ from the operator $O_7$ in the weak Hamiltonian,
\begin{align} \label{eq:cC7}
\mathcal{C}_7 &= C_7(\mu)\,\frac{\overline{m}_b(\mu)}{m_b}
 + C_i(\mu) \kappa_{i7}\nn\\
 & - \frac{\alpha_s(\mu)}{4\pi} \ln\frac{\mu}{m_b}
 \biggl[ \frac83 C_7(\mu) \frac{\overline{m}_b(\mu)}{m_b} -\frac{32}{9}
 C_8(\mu) \nn\\
 & + C_i(\mu) \Big( \gamma_{ij}^{(0)} \kappa_{j7} + \gamma_{i7}^{(0)}\Big)\biggr]
  + {\cal O}(\alpha_s^2)
\,.\end{align}
The Wilson coefficients $C_i(\mu)$, $\kappa_{i7}$, and the anomalous
dimensions $\gamma_{ij}^{(0)}$ are with respect to the operator basis of
Ref.~\cite{Chetyrkin:1996vx} and are given explicitly in Ref.~\cite{Lee:2006gs}.
The function $F_7(q^2)$ contains perturbative corrections from other operators
in the weak Hamiltonian, while $G_7(q^2)$ contains nonperturbative corrections
from intermediate $c\bar{c}$ states. They are given in Eqs.~(A5) and (A12) in
Ref.~\cite{Lee:2006gs}.

The coefficient $h_s(m_b, \mu_b)$ in \eq{Hs} corresponds to the matching coefficient
for the tensor current in SCET. At $\ord{\alpha_s}$ it was computed in
Ref.~\cite{Bauer:2000yr}.  To determine the hard matching coefficient at
$\ord{\alpha_s^2}$ we take the two-loop computation of the $|C_7|^2$ terms in
the $b\to s\gamma$ rate and spectrum from Refs.~\cite{Blokland:2005uk} and
subtract the terms in the partonic two-loop spectrum in SCET coming from the jet
and shape functions (given by $P(m_b,k,\mu_b)$ in Eq.~(\ref{Pmub}) below).
Due to the split matching, the $\mu_b$ dependence in $h_s(m_b, \mu_b)$ cancels
entirely against the $\mu_b$ dependence in $U_H(m_b, \mu_b, \mu_i)$ in \eq{Hs},
and does not depend on the $\mu$
dependence of the coefficients in the electroweak Hamiltonian.
We obtain
\begin{align} \label{hs}
& h_s(m_b,\mu_b)
\nn\\*
&= 1 - \frac{\alpha_s(\mu_b)}{\pi}\,C_F \biggl(
  \ln^2\!\frac{\mu_b}{m_b} + \frac52\ln\frac{\mu_b}{m_b} + 3 + \frac{\pi^2}{24} \biggr)
\nn\\ &\quad
   + \frac{\alpha_s^2(\mu_b)}{\pi^2}\, C_F \biggl\{
  \frac{1}{2}\,C_F \ln^4\!\frac{\mu_b}{m_b}
   + \Bigl(\frac{5}{2}\,C_F-\frac{1}{6}\,\beta_0 \Bigr) \ln^3\!\frac{\mu_b}{m_b}
\nn\\ &\quad
   + \biggl[ \Bigl(\frac{49}{8} +  \frac{\pi^2}{24}\Bigr)C_F
   + \Bigl(\frac{\pi^2}{12} - \frac13\Bigr)C_A
   -\frac{25}{24}\, \beta_0\biggr] \ln^2\!\frac{\mu_b}{m_b}
\nn\\ &\quad
   + \biggl[ \Bigl(\frac{117}{16} + \frac{17\pi^2}{48} - 3\zeta_3 \Bigr) C_F
   + \Bigl( \frac{\pi^2}{12}-\frac{29}{72} + \frac{11\zeta_3}{4}\Bigr)C_A
\nn\\ &\quad
   - \Bigl(\frac{341}{144}+\frac{\pi^2}{12}\Bigr)\beta_0 \biggr] \ln\frac{\mu_b}{m_b}
   + 3.88611\,C_F + 5.89413\,C_A
\nn\\ & \quad
   - \Bigl(\frac{7859}{3456} + \frac{109\pi^2}{576} + \frac{13\zeta_3}{48}\Bigr)\beta_0
   + \frac{3563}{1296} - \frac{29\pi^2}{108} - \frac{\zeta_3}{6}
  \biggr\}
\,.\end{align}
Our expressions here agree with the corresponding numerical results in
Eq.~$(99)$ of Ref.~\cite{Ali:2007sj}. Note that the inverse powers of $m_b$ in
Eqs.~\eqref{eq:Hs}, \eqref{eq:cC7}, and \eqref{eq:Hu} below cancel the $m_b^5$
factors in $\Gamma_{0s,u}$ in Eq.~(\ref{Gamma0su}), and any mass scheme can be
used. The $m_b$ appearing in $h_s(m_b,\mu_b)$ and $U_H(m_b,\mu_b,\mu_i)$ is
defined in the pole scheme. Here, the pole mass renormalon ambiguity is less
relevant, because $m_b$ only appears as a reference scale for $\mu_b$, so the
effect of changing $m_b$ is included in the variation of $\mu_b$. For numerical
calculations, we use $m_b  = 4.7\GeV$.

The hard function $H_u(E_\ell,p_X^-,p_X^+, \mu_i)$
for $B\to X_u\ell\bar\nu$ in \eqs{factor}{dG} is given by
\begin{align} \label{eq:Hu}
& H_u(E_\ell,p_X^-,p_X^+, \mu_i)
  = \frac{24}{m_b^5} \big(m_B-p_X^+\big) \big(2E_\ell + p_X^- -m_B\big)\nn\\
& \times\biggl\{ \big(2 m_B-p_X^+ -p_X^--2E_\ell\big)\, h_{u1}(m_b,p^-,\mu_b)
  \nn\\
&\quad +\big(m_B - p_X^+ -2E_\ell\big) \biggl[ h_{u2}(m_b,p^-,\mu_b) \nn\\
&\quad + \frac{m_B-p_X^+}{p_X^- -p_X^+}\, h_{u3}(m_b,p^-,\mu_b)\biggr] \biggr\}\,
  U_H(p^-, \mu_b, \mu_i)
\,,\end{align}
where the $h_{ui}(m_b, p^-,\mu_b)$ correspond to the matching coefficients of the $V-A$ current in SCET.
To order $\alpha_s$~\cite{Bauer:2000yr, Bauer:2003pi, Bosch:2004th}
\begin{align}
& h_{u1}(m_b,p^-,\mu_b)
\nn\\ & \quad
= 1 - \frac{\alpha_s(\mu_b)}{\pi}\,C_F
 \biggl[
  \ln^2\!\frac{\mu_b}{p^-} + \frac52\ln\frac{\mu_b}{m_b}
  + {\rm Li}_2\Big(1-\frac{p^-}{m_b}\Big)
\nn\\ & \qquad
  + \frac{3p^- - 2m_b}{2m_b - 2p^-}\, \ln\frac{p^-}{m_b}
  + \frac{\pi^2}{24} + 3 \biggr]
\,, \nn\\
& h_{u2}(m_b,p^-,\mu_b)
\nn\\&\quad
= \frac{\alpha_s(\mu_b)}{2\pi}\,C_F\, \frac{m_b}{m_b - p^-}\biggl(
\frac{p^-}{m_b-p^-}  \ln\frac{p^-}{m_b} + 1 \biggr)\,,
\nn\\
& h_{u3}(m_b,p^-,\mu_b)
\\ & \quad
=\frac{\alpha_s(\mu_b)}{2\pi}\,C_F \frac{p^-}{m_b- p^-}\biggl(
  \frac{m_b- 2 p^-}{m_b- p^-}  \ln\frac{p^-}{m_b} - 1 \biggr)
\nn\,.\end{align}
The corresponding two-loop results for $H_u$ will be easy to implement once they become
available.  Finally, the evolution factor $U_H(p^-, \mu_b, \mu_i)$ in \eq{Hs}
is identical for $H_s$ and $H_u$, and is given below in \eq{UH}.

We now turn to the function $P(p^-,k,\mu_i)$ in \eq{P_def}.
Changing variables $\w\to \w'+k -\w$, we have
\begin{align} \label{eq:P_conv}
P(p^-, k, \mu_i) &= \!\inte{\w} U_S(k - \w, \mu_i, \mu_\Lambda) \\*
&\quad \times \!\inte{\w'} p^- J[p^-(\w - \w'), \mu_i]\, C_0(\w', \mu_\Lambda)
\nn\,.\end{align}
To incorporate the fixed order $\alpha_s$ corrections to the jet function,
$J(p^2, \mu_i)$, and the shape function kernel, $C_0(\w,\mu_\Lambda)$, we will first carry
out the $\w'$ convolution integral. The evolution factor $U_S(\pxp, \mu_i,
\mu_\Lambda)$, which sums logarithms between the scales $\mu_i$ and
$\mu_\Lambda$, and the integral over $\w$ are discussed later.

To all orders in perturbation theory the jet function and shape function kernel
can be written as
\begin{align}
J(p^2, \mu_i)
&= \frac{1}{\mu_i^2} \sum_{n = -1}^\infty J_n[\alpha_s(\mu_i)]\, \cL_{n}(p^2/\mu_i^2)
\,,\nn\\
C_0(\w, \mu_\Lambda)
&= \frac{1}{\mu_\Lambda} \sum_{n = -1}^\infty S_n[\alpha_s(\mu_\Lambda)]\, \cL_{n}(\w/\mu_\Lambda)
\,,\end{align}
where the $\cL_n(x)$ are plus distributions defined in \eq{cLn_def} for $n \geq
0$ and in Eq.~(\ref{Lm1}) for $n=-1$, and the coefficients $J_n(\alpha_s)$ and
$S_n(\alpha_s)$ have expansions in $\alpha_s$. The jet function coefficients are
known at one~\cite{Bauer:2003pi} and two loops~\cite{Becher:2006qw}, and are
given by
\begin{align} \label{eq:Jn}
J_{-1}(\alpha_s) &=
1 + \frac{\alpha_s}{\pi}\,\Bigl(\frac{7}{4} - \frac{\pi^2}{4}\Bigr)C_F
\nn\\* &\quad
 + \frac{\alpha_s^2}{\pi^2} \biggl[
 \Bigl(\frac{205}{128}-\frac{67\pi^2}{96} + \frac{7\pi^4}{120} - \frac{9\zeta_3}{8} \Bigr) C_F^2
\nn\\* &\quad\quad
 + \Bigl(\frac{1417}{1728}-\frac{7\pi^2}{144}-\frac{17\pi^4}{2880} -\frac{9\zeta_3}{8}\Bigr) C_F C_A
\nn\\* & \qquad
  + \Bigl(\frac{4057}{3456}-\frac{17\pi^4}{144}-\frac{\zeta_3}{12}\Bigr) C_F \beta_0 \biggr]
\,,\nn\\
J_0(\alpha_s) &=
  - \frac{\alpha_s}{\pi}\, \frac{3}{4}\, C_F
  - \frac{\alpha_s^2}{\pi^2} \biggl[
    \Bigl(\frac{45}{32}-\frac{7\pi^2}{16} + \frac{\zeta_3}{2} \Bigr) C_F^2
\nn\\* &\quad
  + \Bigl(\frac{73}{144} - \frac{5\zeta_3}{2}\Bigr) C_F C_A
  + \Bigl(\frac{247}{288} - \frac{\pi^2}{24}\Bigr) C_F \beta_0 \biggr]
\,,\nn\\
J_1(\alpha_s) &= \frac{\alpha_s}{\pi}\,C_F
  + \frac{\alpha_s^2}{\pi^2} \biggr[ \Bigl(\frac{37}{16}-\frac{5\pi^2}{12} \Bigr) C_F^2
\nn\\* &\quad
  + \Bigl(\frac{1}{3}-\frac{\pi^2}{12}\Bigr) C_F C_A  + \frac{29}{48}\, C_F \beta_0 \biggl]
\,,\nn\\
J_2(\alpha_s) &=
  - \frac{\alpha_s^2}{\pi^2}\, \Bigl(\frac{9}{8}\, C_F^2 + \frac{1}{8}\, C_F \beta_0 \Bigr)
\,,\nn\\
J_3(\alpha_s) &=
  \frac{\alpha_s^2}{\pi^2}\, \frac{1}{2}\, C_F^2
\,.\end{align}
The shape function coefficients to one~\cite{Bauer:2003pi} and two
loops~\cite{Becher:2005pd} are
\begin{align} \label{eq:Sn}
S_{-1}(\alpha_s) &=
1 - \frac{\alpha_s}{\pi}\,\frac{\pi^2}{24}\,C_F
  - \frac{\alpha_s^2}{\pi^2} \biggr[ \Bigl(\frac{\pi^2}{12} + \frac{3\pi^4}{640} - 2\zeta_3\Bigr) C_F^2
\nn\\* &\quad
  + \Bigl( \frac{29}{108} + \frac{31\pi^2}{144} - \frac{67\pi^4}{2880} + \frac{9\zeta_3}{8}\Bigr) C_F C_A
\nn\\* & \quad
  + \Bigl(-\frac{1}{216} + \frac{5\pi^2}{576} - \frac{5\zeta_3}{48}\Bigr) C_F \beta_0 \biggl]
\,,\nn\\*
S_0(\alpha_s) &=
  - \frac{\alpha_s}{\pi}\,C_F
  + \frac{\alpha_s^2}{\pi^2} \biggr[ \Bigl(-\frac{7\pi^2}{24} + 4\zeta_3\Bigr) C_F^2
\nn\\* &\quad
  + \Bigl( \frac{11}{18} + \frac{\pi^2}{12} -\frac{9\zeta_3}{4} \Bigr) C_F C_A - \frac{1}{36}\, C_F \beta_0 \biggl]
\,,\nn\\
S_1(\alpha_s) &=
  - \frac{\alpha_s}{\pi}\,2C_F
  + \frac{\alpha_s^2}{\pi^2} \biggr[ \Bigl(1 - \frac{7\pi^2}{12} \Bigr) C_F^2
\nn\\* &\quad
  + \Bigl(-\frac{2}{3}+\frac{\pi^2}{6}  \Bigr) C_F C_A -\frac{1}{3}\, C_F \beta_0 \biggl]
\,,\nn\\
S_2(\alpha_s) &=
  \frac{\alpha_s^2}{\pi^2}\, \Bigl( 3 C_F^2 + \frac{1}{2}\,C_F \beta_0 \Bigr)
\,,\nn\\
S_3(\alpha_s) &=
  \frac{\alpha_s^2}{\pi^2}\, 2C_F^2
\,.\end{align}
All other coefficients in $J$ and $C_0$ start at higher orders in $\alpha_s$.

To convolute $\cL_n[p^-(\w-\w')/\mu_i^2]$ with $\cL_m(\w/\mu_\Lambda)$ we
first rescale them to have the same dimensionless arguments. Using \eq{cLn_rescale},
$J$ and $C_0$ satisfy the rescaling identities
\begin{align}
\label{eq:JC_rescale}
J(p^-\w, \mu) &= \frac{1}{p^-\xi} \sum_{n=-1}^\infty
  J_n\Bigl[\alpha_s(\mu), \frac{p^-\xi}{\mu^2}\Bigr]\,
  \cL_n\Bigl(\frac{\w}{\xi}\Bigr)
\,,\nn\\
C_0(\w, \mu) &= \frac{1}{\xi} \sum_{n=-1}^\infty
  S_n\Bigl[\alpha_s(\mu),\frac{\xi}{\mu}\Bigr]\,
  \cL_n\Bigl(\frac{\w}{\xi}\Bigr)
 \,.
\end{align}
where $\xi$ is an arbitrary dimension-one parameter that we will choose at our
convenience later on, and the rescaled coefficients are
\begin{align}\label{eq:JS2var}
J_{-1}(\alpha_s, x) &=
 J_{-1}(\alpha_s) + \sum_{n = 0}^\infty J_n(\alpha_s)\,\frac{\ln^{n+1} x}{n+1}
\,,\nn\\
J_n(\alpha_s, x) &=
\sum_{k = 0}^\infty \frac{(n+k)!}{n!\,k!}\, J_{n+k}(\alpha_s)\,\ln^k x
 \,,\nn\\
S_{-1}(\alpha_s, x) &=
 S_{-1}(\alpha_s) + \sum_{n = 0}^\infty S_n(\alpha_s)\,\frac{\ln^{n+1} x}{n+1}
\,,\nn\\
S_n(\alpha_s, x) &=
\sum_{k = 0}^\infty \frac{(n+k)!}{n!\,k!}\, S_{n+k}(\alpha_s)\,\ln^k x
\,.\end{align}
Using \eq{JC_rescale}, the convolution of $J$ and $C_0$ in \eq{P_conv} becomes
\begin{align} \label{eq:JC0}
&(J \otimes C_0)(\w, \mu_i, \mu_\Lambda)
\nn\\
&\equiv \!\inte{\w'} p^- J[p^-(\w - \w'), \mu_i]\, C_0(\w', \mu_\Lambda)
\nn\\
&= \sum_{m,n= -1}^\infty \!
  J_m\Bigl[\alpha_s(\mu_i), \frac{p^-\xi}{\mu_i^2}\Bigr]\,
  S_n\Bigl[\alpha_s(\mu_\Lambda), \frac{\xi}{\mu_\Lambda}\Bigr]
\nn\\ & \quad\times
  \frac{1}{\xi} \inte{x} \cL_m\Bigl(\frac{\w}{\xi} - x\Bigr)\, \cL_n(x)
\nn\\
&=  \sum_{\ell = - 1}^\infty\!
\sum_{\substack{m,n \geq -1 \\ m + n + 1 \geq \ell}}^\infty \!
  \V^{mn}_\ell
  J_m\Bigl[\alpha_s(\mu_i), \frac{p^-\xi}{\mu_i^2}\Bigr]\,
  S_n\Bigl[\alpha_s(\mu_\Lambda), \frac{\xi}{\mu_\Lambda}\Bigr]
\nn\\* & \quad\times
  \frac{1}{\xi}\, \cL_\ell\Bigr(\frac{\w}{\xi}\Bigr)
\,.\end{align}
In the last step we used \eq{cL_conv} to perform the $x$ integral, yielding a
sum of plus distributions, whose coefficients $V_\ell^{mn}$ are given in
\eq{Vkna_def}.

\subsection{Renormalization group evolution}
\label{subsec:RGE}

Next, we summarize results for the renormalization group evolution, and then
carry out the $\omega$ integral in \eq{P_conv}.  The factor
$U_H(p^-,\mu_b,\mu_i)$ which describes the evolution of the hard functions in
\eqs{Hs}{Hu} between the hard scale, $\mu_b$, and the jet scale, $\mu_i$,
is~\cite{Bauer:2000yr}
\begin{equation} \label{eq:UH}
U_H(p^-,\mu_b, \mu_i)
= e^{K_H(\mu_i, \mu_b)}\, \bigg( \frac{p^-}{\mu_b} \bigg)^{\eta(\mu_i, \mu_b)}
\,,\end{equation}
where definitions of $\eta(\mu_i,\mu_b)$ and $K_H(\mu_i,\mu_b)$ are given in
\eq{etaK_def} below.  The soft evolution factor
$U_S(\w, \mu_i, \mu_\Lambda)$ in \eq{P_conv} sums logarithms between the jet
scale, $\mu_i$, and soft scale, $\mu_\Lambda$. To all orders in perturbation
theory it can be written as
\begin{align} \label{eq:cUS}
&U_S(\w, \mu_i, \mu_\Lambda) = e^{K_S(\mu_i, \mu_\Lambda)}
  \frac{e^{- \gamma_E \eta }}{\Gamma(\eta)}
  \frac{1}{\mu_\Lambda}
  \biggr[ \frac{\theta(\w/\mu_\Lambda)}{ (\w/\mu_\Lambda)^{1-\eta} }
  \biggl]_+^\text{\footnotesize Ref.\,\cite{Fleming:2007xt}}
\nn\\& \quad
= E_S(\xi, \mu_i, \mu_\Lambda)\, \frac{1}{\xi}
  \Bigl[\eta\, \cL^\eta\Big(\frac{\w}{\xi}\Big) + \,
  \cL_{-1}\Big(\frac{\w}{\xi}\Big) \Bigr]
\,,\end{align}
where $\eta \equiv \eta(\mu_i, \mu_\Lambda)$, and we defined
\begin{equation} \label{eq:US}
E_S(\xi, \mu_i, \mu_\Lambda) = e^{K_S(\mu_i, \mu_\Lambda)}\,
  \Bigl(\frac{\xi}{\mu_\Lambda}\Bigr)^\eta\
  \frac{e^{-\gamma_E\, \eta}}{ \Gamma(1+\eta)}
\,.\end{equation}
To go from the first to the second line in \eq{cUS} we used
\eqs{cL_Flemingetal}{cLa_rescale}.  An expression for
$U_S(\w,\mu_i,\mu_\Lambda)$ was first found in Ref.~\cite{Balzereit:1998yf}
and extended to all orders in Ref.~\cite{Neubert:2004dd}. The form with the
plus distribution derived in Ref.~\cite{Fleming:2007xt} makes the formula valid
without having to require $\mu_i>\mu_\Lambda$.

In Eqs.~\eqref{eq:UH} to \eqref{eq:US}, $\eta(\mu,\mu_0)$ and  $K_x(\mu, \mu_0)$ for $x=H$ or $S$ are
given to all orders by
\begin{align} \label{eq:etaK_def}
\eta(\mu, \mu_0) &= 2\intlim{\alpha_s(\mu_0)}{\alpha_s(\mu)}{\alpha_s}
  \frac{\Gamma_\mathrm{cusp}(\alpha_s)}{\beta(\alpha_s)}
\,,\nn\\
K_x(\mu, \mu_0)&= -2\intlim{\alpha_s(\mu_0)}{\alpha_s(\mu)}{\alpha_s} \biggl[
\frac{\Gamma_\mathrm{cusp}(\alpha_s)}{\beta(\alpha_s)}
\nn\\*
&\qquad
\times \int_{\alpha_s(\mu_0)}^{\alpha_s} \frac{\df\alpha_s'}{\beta(\alpha_s')}
+ \frac{\gamma_x(\alpha_s)}{\beta(\alpha_s)} \biggl]
\,.\end{align}
Here, $\Gamma_\mathrm{cusp}$ is the universal cusp anomalous dimension.
The only difference between $K_H$ and $K_S$ are the hard and soft anomalous
dimensions, $\gamma_x = \gamma_S$ or $\gamma_H$. Expanding the $\beta$ function
and anomalous dimensions as usual,
\begin{align}
\beta(\alpha_s) &=
- 2 \alpha_s \sum_{n=0}^\infty \beta_n\Bigl(\frac{\alpha_s}{4\pi}\Bigr)^{n+1}
\,,\nn\\
\Gamma_{\rm cusp}(\alpha_s) &=
\sum_{n=0}^\infty \Gamma_n \Bigl(\frac{\alpha_s}{4\pi}\Bigr)^{n+1}
\,,\nn\\
\gamma_x(\alpha_s) &=
\sum_{n=0}^\infty \gamma^x_n \Bigl(\frac{\alpha_s}{4\pi}\Bigr)^{n+1}
\,,\end{align}
the integrals in \eq{etaK_def} to NNLL are
\begin{widetext}
\begin{align}
\eta(\mu,\mu_0) &=
 - \frac{ \Gamma_0  }{\beta_0}\, \biggl[ \ln r
 + \frac{\alpha_s(\mu_0)}{4\pi}\, \biggl(\frac{\Gamma_1 }{\Gamma_0 }
 - \frac{\beta_1}{\beta_0}\biggr)(r-1)
 + \frac{\alpha_s^2(\mu_0)}{16\pi^2} \biggl(
    \frac{\Gamma_2 }{\Gamma_0 } - \frac{\beta_1\Gamma_1 }{\beta_0 \Gamma_0 }
      + \frac{\beta_1^2}{\beta_0^2} -\frac{\beta_2}{\beta_0} \biggr) \frac{r^2-1}{2}
    \biggr]
\,, \nn\\
K_x(\mu,\mu_0) &= \frac{\Gamma_0}{2\beta_0^2}\,
\biggl\{ \frac{4\pi}{\alpha_s(\mu_0)}\, \Bigl(1 - \frac{1}{r} - \ln r\Bigr)
   + \biggl(\frac{\Gamma_1 }{\Gamma_0 } - \frac{\beta_1}{\beta_0}\biggr) (1-r+\ln r)
   + \frac{\beta_1}{2\beta_0} \ln^2 r + \frac{2\beta_0\gamma_0^x}{\Gamma_0}\,\ln r
\nn\\ & \quad
+ \frac{\alpha_s(\mu_0)}{4\pi}\, \biggl[
  \biggl(\frac{\beta_1^2}{\beta_0^2} - \frac{\beta_2}{\beta_0} \biggr) \Bigl(\frac{1 - r^2}{2} + \ln r\Bigr)
  + \biggl(\frac{\beta_1\Gamma_1 }{\beta_0 \Gamma_0 } - \frac{\beta_1^2}{\beta_0^2} \biggr) (1- r+ r\ln r)
  - \biggl(\frac{\Gamma_2 }{\Gamma_0} - \frac{\beta_1\Gamma_1}{\beta_0\Gamma_0} \biggr) \frac{(1- r)^2}{2}
\nn\\ & \qquad
+ \frac{2\beta_0\gamma_0^x}{\Gamma_0 }\,
  \biggl(\frac{\gamma_1^x}{\gamma_0^x} - \frac{\beta_1}{\beta_0}\biggr) (r- 1)
     \biggr] \biggr\}
\,,\end{align}
with $r = \alpha_s(\mu)/\alpha_s(\mu_0)$.
When using $\eta(\mu,\mu_0)$ and $K_x(\mu,\mu_0)$ at LL, NLL, and NNLL according to our conventions in \eq{RGEconventions}, we do not re-expand the results in \eqs{UH}{US}, but keep their full expressions everywhere. For the running coupling we always use the three-loop expression
\begin{equation}
\frac{1}{\alpha_s(\mu)} = \frac{X}{\alpha_s(\mu_0)}
  +\frac{\beta_1}{4\pi\beta_0}  \ln X
  + \frac{\alpha_s(\mu_0)}{16\pi^2} \biggr[
  \frac{\beta_2}{\beta_0} \Bigl(1-\frac{1}{X}\Bigr)
  + \frac{\beta_1^2}{\beta_0^2} \Bigl( \frac{\ln X}{X} +\frac{1}{X} -1\Bigr) \biggl]
\,,\end{equation}
\end{widetext}
where $X\equiv 1+\alpha_s(\mu_0)\beta_0 \ln(\mu/\mu_0)/(2\pi)$, and we evolve to
lower scales using the reference value $\alpha_s(\mu_0=4.7\,{\rm GeV})=0.2155$
as in Table~\ref{tab:num} with $n_f=4$.  Up to three-loop order, the
coefficients of the $\beta$ function in the $\overline {\rm MS}$ scheme are
\begin{align}
\beta_0 &= \frac{11}{3}\,C_A -\frac{2}{3}\,n_f
\,,\nn\\
\beta_1 &= \frac{34}{3}\,C_A^2  - \frac{10}{3}\,C_A\, n_f  - 2 C_F\, n_f
\,, \nn\\
\beta_2 &=
\frac{2857}{54}\,C_A^3 + \Bigl(C_F^2 - \frac{205}{18}\,C_F C_A
 - \frac{1415}{54}\,C_A^2 \Bigr)\, n_f
\nn\\ &\quad
 + \Bigl(\frac{11}{9}\, C_F + \frac{79}{54}\, C_A \Bigr)\, n_f^2
\,.\end{align}
The cusp~\cite{Korchemsky:1987wg, Moch:2004pa},
soft~\cite{Korchemsky:1992xv, Gardi:2005yi}, and hard anomalous dimension
coefficients are
\begin{align}
\Gamma_0 &= 4 C_F
\,,\nn\\
\Gamma_1 &= \Bigl( \frac{268}{9} -\frac{4\pi^2}{3} \Bigr)\,C_F C_A  -
   \frac{40}{9} C_F\, n_f
\,,\nn\\
\Gamma_2 &=
\Bigl(\frac{490}{3} -\frac{536 \pi^2}{27} + \frac{44 \pi ^4}{45}
  + \frac{88 \zeta_3}{3}\Bigr) C_F C_A^2
\nn\\* & \quad
  + \Bigl(\frac{80 \pi^2}{27} - \frac{836}{27} - \frac{112 \zeta_3}{3} \Bigr)C_FC_A\, n_f
\nn\\* & \quad
  + \Bigl(32 \zeta_3 - \frac{110}{3}\Bigr) C_F^2\, n_f
  - \frac{16}{27}\,C_F\, n_f^2
\,,\nn\\[1ex]
\gamma_0^S &= -2 C_F
\,,\nn\\
\gamma_1^S &=
\Bigl(\frac{110}{27} + \frac{\pi^2}{18} - 18\zeta_3 \Bigr) C_F C_A
 + \Bigl(\frac{4}{27} + \frac{\pi^2}{9}\Bigr)C_F\, n_f
\,,\nn\\[1ex]
\gamma_0^H &= 5 C_F
\,,\nn\\
\gamma_1^H &=
\Bigl(\frac{3}{2} - 2\pi^2 + 24\zeta_3\Bigr) C_F^2
- \Bigl(\frac{125}{27} + \frac{\pi^2}{3}\Bigr) C_F\, n_f
\nn\\* & \quad
+ \Bigl(\frac{1549}{54} + \frac{7\pi^2}{6} - 22\zeta_3 \Bigr) C_F C_A
\,.\end{align}
To determine $\gamma_1^H$ we used $\gamma_1^H= - \gamma_1^J-\gamma_1^S$, which
follows from the $\mu$ independence of $\df\Gamma_s/\df E_\gamma$. We use the
two-loop computation of $\gamma_1^J$ in Ref.~\cite{Becher:2006qw}.

The evolution of the shape function kernel $C_0(\w,\mu)$ from $\mu_\Lambda$ to
$\mu_i$ can be written as the sum of a finite number of terms,
\begin{align} \label{eq:C0_evolve}
C_0(\w, \mu_i)
&= \inte{\w'} U_S(\w-\w',\mu_i,\mu_\Lambda)\, C_0(\w',\mu_\Lambda)
\nn\\
&= E_S(\xi, \mu_i, \mu_\Lambda) \sum_{n = -1}^\infty
  S_n\Bigl[\alpha_s(\mu_\Lambda), \frac{\xi}{\mu_\Lambda}\Bigr]
\nn\\ & \quad\times
  \frac{1}{\xi}\, \biggl[\eta \inte{x}  \cL^\eta\Bigl(\frac{\w}{\xi} - x\Bigr)\,
  \cL_{n}(x) + \, \cL_n \Bigl(\frac{\w}{\xi}\Bigr) \biggr]
\nn\\
&= E_S(\xi, \mu_i, \mu_\Lambda)
  \sum_{n=-1}^\infty\, \sum_{\ell = -1}^{n+1} \V^n_\ell(\eta)
\nn\\ & \quad\times
  S_n\Bigl[\alpha_s(\mu_\Lambda), \frac{\xi}{\mu_\Lambda}\Bigr]\,
  \frac{1}{\xi}\,\cL_\ell^\eta\Bigl(\frac{\w}{\xi}\Bigr)
\,.\end{align}
In the last step, we used \eq{cL_conv} to perform the $x$ integral. The
coefficients of the resulting plus distributions, $V_\ell^n(\eta)$, are defined in
\eq{Vkna_def}. If $C_0(\w,\mu)$ is known to $\ord{\alpha_s^k}$ accuracy, then
the terms contributing in \eq{C0_evolve} are bounded by $n\le 2k-1$. To
obtain the shape function evolved up to $\mu_i$, $S(\w,\mu_i)$, we combine
Eq.~(\ref{eq:F_def}) with (\ref{eq:C0_evolve}) taking for convenience $\xi=\w$
and changing the integration variable to $z$ via $k=\w(1-z)$. This gives the
result for $S(\w,\mu_i)$ that is quoted in Eq.~\eqref{eq:Swrun} in the text.

Returning to $P(p^-, k, \mu_i)$ in \eq{P_conv}, to perform the $\w$ integral we can now apply
the same steps as in \eq{C0_evolve} to compute the convolution of $U_S(\w,
\mu_i, \mu_\Lambda)$ with the result of \eq{JC0}. This yields
\begin{align} \label{eq:P}
P(p^-, k, \mu_i)
&= \sum_{j= -1}^\infty P_j(p^-,\xi,\mu_i,\mu_\Lambda)\,
\frac{1}{\xi}\, \cL_j^\eta \Bigl(\frac{k}{\xi} \Bigr)
\,,\end{align}
where $\eta=\eta(\mu_i, \mu_\Lambda)$, and the coefficients are
\begin{align} \label{eq:Pj}
&P_j(p^-,\xi,\mu_i,\mu_\Lambda)
\nn\\* &\quad
= E_S(\xi, \mu_i, \mu_\Lambda)
\sum_{\substack{m,n\geq -1\\m + n + 2\geq j}}^\infty \biggl\{
\sum_{\substack{\ell \geq -1\\\ell \geq j-1}}^{m+n+1} \V^{mn}_\ell\, \V^\ell_j(\eta)
\nn\\* & \qquad\times
  J_m\Bigl[\alpha_s(\mu_i), \frac{p^-\xi}{\mu_i^2}\Bigr]
  S_n\Bigl[\alpha_s(\mu_\Lambda), \frac{\xi}{\mu_\Lambda}\Bigr]
 \biggr\}
\,,\end{align}
with $\V_j^\ell(\eta)$ and $\V_\ell^{mn}$ given in \eqs{Vkna_def}{Vkmn_def}.
The jet and shape function
coefficients, $J_m$ and $S_n$, have perturbative expansions in $\alpha_s(\mu_i)$
and $\alpha_s(\mu_\Lambda)$, respectively, given by \eqs{Jn}{Sn} together with
\eq{JC_rescale}. To $\ord{\alpha_s}$ we have $m, n\leq 1$, so dropping cross
terms of $\ord{\alpha_s^2}$ and higher, we need $-1\leq \ell\leq 1$ and $-1\leq
j\leq 2$. To $\ord{\alpha_s^2}$, we have $m, n\leq 3$, and we need cross terms
between $J_m$ and $S_n$ up to $-1\leq \ell\leq 3$ and $-1\leq j\leq 4$.

The final step now is to compute the convolution of $P(p^-, k, \mu_i)$ with some
function $F(k)$,
\begin{align} \label{PF}
&\inte{k} P(p^-, k, \mu_i)\, F(\pxp - k)
\\
 & \quad
= \sum_{j = - 1}^\infty P_j(p^-, \pxp, \mu_i,\mu_\Lambda)
  \int_0^1\! \df z\, \cL_j^\eta(z)\, F[\pxp(1-z)]
 \,.\nn
\end{align}
In both Eqs.~(\ref{eq:Swrun}) and (\ref{PF}) the remaining integral that needs
to be performed is $\cL_j^\eta(z)$ with $F[\w(1-z)]$.  For $j\ge 0$ it is
convenient to simplify this integral to
\begin{align} \label{cool}
 & \int_0^1\! \df z\, \cL_j^\eta(z)\, F[\w (1-z)]
 \\&\quad
  =\int_0^1\! \df z\, \frac{\ln^j z}{z^{1-\eta}}\,
  \bigl\{ F[\w(1-z)] - F(\w) \bigr\}
\nn\\ &\quad
 = \frac{1}{\eta^{1+j}} \int_0^1\! \df t\, \ln^jt\,
  \bigl\{ F\big[\w\big(1-t^{1/\eta}\big)\big] - F(\w) \bigr\}
\,,\nn
\end{align}
where the second line is valid for $\eta > -1$, but the last line only for $\eta>0$.
For $j=-1$ the integral is trivial since ${\cal L}_{-1}^\eta(z)=\delta(z)$.
Our basis for $F(k)$ involves a series of terms of the form
\begin{equation}
F(k) = \sum_{r,s} C_{rs} \Bigl(\frac{k}{\lambda}\Bigr)^r
\exp\Bigl(-s\, \frac{k}{\lambda}\Bigr)
\,,\end{equation}
where $r$ and $s$ are integers. For each term in the series the integral in
Eq.~(\ref{cool}) can be performed analytically using
\begin{align} \label{yuck}
& \intlim{0}{1}{z} \bigl[(1-z)^\ell e^{b\, z} - 1\bigr]\, z^{a+\eta-1}
  = -\frac{1}{a+\eta} \\*
&\qquad + \frac{\Gamma(a+\eta)\,\Gamma(1+\ell)}{\Gamma(1+a+\eta+\ell)}\,
  {}_1{}F_1(a + \eta, 1 + a + \eta + \ell, b) \nn
\,.\end{align}
By taking derivatives of this result with respect to $a$ one can obtain the
integral with $(\ln z)^j$ on the left-hand side. In practice, we find that the numerical
integration of Eq.~\eqref{cool} is sufficiently fast that using Eq.~\eqref{yuck}
is unnecessary.

Finally, we comment on a special case of the above results that was used to carry
out the $B\to X_s\gamma$ matching computation to derive Eq.~(\ref{hs}). Here we
used the SCET computation of the decay rate with $\mu_b=\mu_i=\mu_\Lambda=m_b$
and $\eta=0$, which depends on $P(m_b,p_X^+,m_b)$. For this special case, the
choice $\xi=m_b$ is convenient and Eqs.~(\ref{eq:P}) and (\ref{eq:Pj}) then
reduce to
\begin{align} \label{Pmub}
& P(m_b,p_X^+,m_b) = \sum_{j=-1}^\infty P_j(m_b,m_b,m_b,m_b) \frac{1}{m_b}
  {\cal L}_j\Big(\frac{p_X^+}{m_b}\Big)\,,
  \nn\\*
& P_j(m_b,m_b,m_b,m_b) =
  \!\!\!\!\!\!
  \sum_{\substack{m,n\geq -1\\m + n + 2\geq j}}^\infty
 \!\!\!\!\! \V^{mn}_j
  J_m\big[\alpha_s(m_b)\big] S_n\big[\alpha_s(m_b)\big]
 .
\end{align}
Thus $P_j(m_b,m_b,m_b,m_b)$ is the coefficient of the plus distribution or
$\delta$ distribution $\cL_j$ in the fixed-order computation evaluated at
$\mu=m_b$.

\subsection{Changing to short distance schemes}
\label{subsec:scheme}

To change the scheme from the pole scheme to a short distance scheme, we define
a perturbative function $\hP$ as in \eq{P_def}, but with $C_0$ replaced by
$\hC_0$, and in the differential spectrum $P$ and $F$ are replaced by $\hP$ and
$\hF$.  Displaying only the integration variables,
\begin{align} \label{eq:GSDsimple}
\df\Gamma_q = \Gamma_{0q}\, H_q \inte{k} \hP(k)\, \hF(\pxp - k)
\,,\end{align}
with
\begin{align}
\hP(k)
&= \inte{\w} U_S(k-\w)\inte{\w'} J(\w-\w')\,\hC_0(\w') \\
&= \biggl\{1 + \delta m_b\,\frac{\df}{\df k} + \biggl[\frac{(\delta m_b)^2}{2}
  - \frac{\delta\lambda_1}{6} \biggr]\, \frac{\df^2}{\df k^2} \biggr\}\, P(k)\nn
\,,\end{align}
and $\delta m_b$ and $\delta\lambda_1$ defined in Eq.~(\ref{delmbl1}).  This
result has exactly the same form as Eq.~(\ref{hatC0}), which relates $\widehat
C_0(\omega)$ and $C_0(\omega)$, so the analysis we carry out below also holds
for determining $\widehat P(k)$ and $\widehat C_0(\omega)$. To be definite,
we use $\hP(k)$ below. The results for $\hC_0(\w)$ are obtained by
substituting $P(k)\to C_0(\w)$.

To ensure the proper cancellation of renormalon ambiguities, the perturbative
series in $\delta m_b$, $\delta\lambda_1$, and $P(k)$ have to be re-expanded to
the desired order.  Denoting
\begin{align}
P(k) &= \sum_{n=0}^\infty \e^n\, P^{(n)}(k)
\,,\nn\\*
\delta m_b &= \sum_{n=1}^\infty \e^n\, \delta m_b^{(n)}(\mu_\Lambda)
\,,\nn\\*
\delta \lambda_1 &= \sum_{n=1}^\infty \e^n\, \delta \lambda_1^{(n)}(\mu_\Lambda)
\,,\end{align}
where the dummy variable $\e=1$ counts the order in the perturbative expansion.
As indicated, the scale for $\alpha_s$ in $\delta m_b$ and $\delta\lambda_1$ must
be set to $\mu_\Lambda$ to ensure that renormalons cancel. To illustrate this,
the $\mu_\Lambda$ variation in Fig.~\ref{fig:1Sschemepeak} yields
the numbers $0.74,\, 0.76,\, 0.72\GeV^{-1}$ for $S(\w = 0.5\GeV,\mu)$ at NNLL. However,
holding $\mu_\Lambda$ in $\delta m_b$ fixed at the central value
the result becomes $0.81,\,0.76,\,0.68\GeV^{-1}$, with a much larger scale dependence.

For $P(k)$, we count both $\alpha_s(\mu_i)\sim\e$ and $\alpha_s(\mu_\Lambda) \sim \e$ in $S_n$ and
$J_m$, i.e., we expand the cross terms between $J_m$ and $S_n$. Note that we do not
expand the cross terms in the product of $H_q$ and $\hP$ in \eq{GSDsimple}. Then, to $\ord{\e^2}$
\begin{align}
&\hP(k) = \e^0\, P^{(0)}(k) + \e^1\, P^{(1)}(k) + \e^2\, P^{(2)}(k)\\*
& \quad + \ \biggl[ \delta m_b^{(1)} \frac{\df}{\df k}
  - \frac{\delta\lambda_1^{(1)}}{6}\, \frac{\df^2}{\df k^2} \biggr]
  \Bigl[ \e\, P^{(0)}(k) + \e^2\, P^{(1)}(k) \Bigr] \nn\\*
& \quad + \e^2\, \biggl\{ \delta m_b^{(2)} \frac{\df}{\df k}
  + \biggl[\frac{(\delta m_b^{(1)})^2}{2} - \frac{\delta\lambda_1^{(2)}}{6}
  \biggr] \frac{\df^2}{\df k^2}\biggr\}\, P^{(0)}(k)
\nn\,,\end{align}
and integrating by parts we can move the derivatives to act on $\widehat F(k)$,
\begin{align}
&\inte{k} \hP(k)\, \hF(\pxp-k) \nn\\
&= \inte{k} \Big[ P^{(0)}(k) +\e\, P^{(1)}(k) +\e^2\, P^{(2)}(k)  \Big] \,\hF(\pxp-k)
\nn\\* & \quad
 +  \biggl[ \Bigl(\e\,\delta m_b^{(1)} + \e^2\, \delta m_b^{(2)} \Bigr)\, P^{(0)}(k)
\nn\\* & \qquad
 + \e^2\, \delta m_b^{(1)} P^{(1)}(k) \biggr]\, \hF^{\,\prime}(\pxp-k)
\nn\\ & \quad
+  \biggl\{ \biggl[ -\e\, \frac{\delta\lambda_1^{(1)}}{6}
  + \e^2\, \frac{ 3\big(\delta m_b^{(1)}\big)^2-\delta\lambda_1^{(2)} } {6}\biggr]
  P^{(0)}(k)
\nn\\ & \qquad
- \e^2\, \frac{\delta\lambda_1^{(1)}}{6}\, P^{(1)}(k) \biggr\}\, \hF^{\,\prime\prime}(\pxp-k)
\,.\end{align}

For $m_b$, we mostly use the $1S$ scheme~\cite{Hoang:1998ng}, in which
\begin{align} \label{md1}
\delta m_b^{(1S,1)} &= R^{1S}\, \frac{\alpha_s(\mu)\, C_F}8\,,\\
\delta m_b^{(1S,2)} &= R^{1S}\, \frac{\alpha_s^2(\mu)\, C_F}{8\, \pi}
  \biggl[ \Bigl( \ln\frac{\mu}{R^{1S}} + \frac{11}6 \Bigr) \beta_0 - \frac{4}{3} C_A \biggr]
\,,\nn
\end{align}
where $R^{1S} = m_b^{1S}\, \alpha_s(\mu)\, C_F$.
For $\lambda_1$, we use our  ``invisible scheme", in which
\begin{align} \label{lam1inv}
\delta\lambda_1^{({\rm i},1)} &= 0 \,,\\
\delta\lambda_1^{({\rm i},2)} &= R^2\, \frac{\alpha_s^2(\mu)}{\pi^2}\,
  \frac{C_FC_A}{4}\,\Big(\frac{\pi^2}{3}-1 \Big) \,.
\nn
\end{align}
By default we take $R=1\,{\rm GeV}$.

We also list the corresponding expressions in the kinetic
scheme~\cite{Czarnecki:1997sz}, defining $\lambda_1^{\rm kin} = -\mu_\pi^2$ and
using $R$ for the momentum cutoff,
\begin{align} \label{eq:delkin}
\delta\lambda_1^{({\rm kin},1)} &= R^2\, \frac{\alpha_s(\mu)}{\pi}\, C_F
\,,\nn\\*
\delta\lambda_1^{({\rm kin},2)} &= R^2\, \frac{\alpha_s^2(\mu)}{\pi^2}\,\frac{C_F}{2}
  \biggl[ \Bigl(\ln\frac{\mu}{2R} + \frac{13}{6}\Bigr) \beta_0
\nn\\ & \quad
  + \Bigl(\frac{13}{6} - \frac{\pi^2}{3} \Bigr) C_A \biggr]
\,,\nn\\*
\delta m_b^{({\rm kin},1)} &= R\, \frac{\alpha_s(\mu)}{\pi}\,\frac{4}{3}\,C_F
  + \frac{\delta\lambda_1^{({\rm kin},1)}}{2m_b}
\,,\nn\\
\delta m_b^{({\rm kin},2)} &= R\, \frac{\alpha_s^2(\mu)}{\pi^2}\, \frac{2}{3}\,C_F
  \biggl[ \Bigl(\ln\frac{\mu}{2R} + \frac{8}{3}\Bigr) \beta_0
\nn\\* &\quad
  + \Bigl(\frac{13}{6} - \frac{\pi^2}{3}\Bigr) C_A \biggr]
  + \frac{\delta\lambda_1^{({\rm kin},2)}}{2m_b}
\,.\end{align}
By default in the kinetic scheme $R=1\,{\rm GeV}$.

\section{Plus Distributions and Convolutions}
\label{app:plus}

We define a general plus distribution for some function $q(x)$, which is
less singular than $1/x^2$ as $x\to 0$, as
\begin{align} \label{eq:genplus}
\bigl[q(x)\bigr]_+^{[x_0]} &\equiv \bigl[\theta(x) q(x)\bigr]_+^{[x_0]} \\
& = \lim_{\e\to 0}\, \frac{\df}{\df x} \bigl[ \theta(x - \e)\, Q(x,x_0) \bigr]
\nn\\[3pt]
& = \lim_{\e\to 0}\,  \bigl[ \theta(x - \e)\, q(x) + \delta(x-\e)\, Q(x,x_0)
\bigr] \nn
\,,\end{align}
with
\begin{equation}
Q(x,x_0) = \intlim{x_0}{x}{x'} q(x')
\,.\end{equation}
Since $Q(x_0,x_0)=0$, the point $x_0$ should be thought of as a boundary
condition for the plus distribution. Integrating against a test function $f(x)$,
we have
\begin{align} \label{plusonf}
& \intlim{-\infty}{x_\mathrm{max}}{x} [\theta(x)\, q(x)]_+^{[x_0]}\, f(x)
  \\* &\quad
= \intlim{0}{x_\mathrm{max}}{x} q(x)\,[f(x) - f(0)] + f(0)\, Q(x_\mathrm{max},x_0)
\,. \nn
\end{align}
Taking $f(x)=1$ in Eq.~(\ref{plusonf}) one sees that the integral of the
plus distribution vanishes only when integrated over a range with $x_{\rm
max}=x_0$.  Plus distributions with different boundary conditions are related to
each other by
\begin{equation}
\label{eq:x0shift}
\bigl[\theta(x)\, q(x)\bigr]^{[x_0]}_+ = \bigl[\theta(x)\, q(x)\bigr]^{[x_1]}_+
   + \delta(x)\, Q(x_1,x_0)
\,.\end{equation}
We will almost exclusively use the boundary condition $x_0 = 1$, and
will drop the superscript $[x_0]$ on the plus distributions when
this default choice is used.

Taking the special case $q(x) = 1/x^{1-a}$ with $a>-1$, we define
\begin{equation}
\label{eq:cLa_def}
\cL^a(x) = \biggl[\frac{\theta(x)}{x^{1-a}} \biggr]_+
= \lim_{\e\to 0}\, \frac{\df}{\df x}
  \biggl[ \theta(x - \e)\, \frac{x^a - 1}{a} \biggr]
\,.\end{equation}
With our boundary condition, $\cL^a(x)$ for $a = 0$ reduces to the standard
definition of $[\theta(x)/x]_+$. For $q(x) = \ln^n x/x$ with integer $n \geq 0$
we define
\begin{equation}
\label{eq:cLn_def}
\cL_n(x) = \biggl[\frac{\theta(x)\ln^n x}{x}\biggr]_+
= \lim_{\e\to 0}\, \frac{\df}{\df x} \biggl[
  \theta(x - \e)\, \frac{\ln^{n+1} x}{n+1} \biggr]
\,.\end{equation}
Since both $\cL^a(x)$ and $\cL_n(x)$ are defined with the same boundary
condition $x_0 = 1$, they are related by
\begin{equation}
\cL_n(x) = \frac{\df^n}{\df a^n}\, \cL^a(x)\bigg\vert_{a = 0}
\,.\end{equation}
This makes it easy to derive identities involving $\cL_n(x)$ from identities
involving $\cL^a(x)$ by taking derivatives with respect to $a$.

The definitions in \eqs{cLa_def}{cLn_def} can be contrasted with those in
Ref.~\cite{Fleming:2007xt}, where the same boundary condition $x_0=1$ is used
for $[\theta(x) \ln^n x/x]_+ = {\cal L}_n(x)$, but $x_0=\infty$ and $x_0 = 0$
are used for $[\theta(x)/x^{1-a}]_+$ with $a<0$ and $a>0$, respectively.  This
is the form appearing in the soft evolution factor, \eq{US}, where it comes
multiplied by a factor of $a$, so the limit for $a\to 0$ exists.  Using
\eq{x0shift} to convert to our definitions, we have
\begin{align} \label{eq:cL_Flemingetal}
\biggl[\frac{\theta(x)}{x^{1-a}} \biggr]_+^\text{\footnotesize Ref.\,\cite{Fleming:2007xt}}
 &= \cL^a(x) + \frac{1}{a}\, \delta(x)
\,.
\end{align}

Finally, we define the ``mixed'' distribution
\begin{equation} \label{eq:cLna_def}
\cL_n^a(x) = \biggl[\frac{\theta(x)\ln^n x}{x^{1-a}}\biggr]_+
  = \frac{\df^n}{\df b^n}\, \cL^{a+b}(x)\bigg\vert_{b = 0}
\,,\end{equation}
which satisfies $\cL_n^0(x) \equiv \cL_n(x)$ and $\cL_0^a(x) \equiv \cL^a(x)$.
For convenience we also define
\begin{equation} \label{Lm1}
\cL_{-1}(x) \equiv \cL_{-1}^a(x) \equiv \delta(x)
\,.\end{equation}
The following identities are useful
\begin{align}
\cL_{m+n}(x) &= \frac{\df^m}{\df a^m} \, \cL^a_n(x) \bigg\vert_{a = 0}
\,,\\[2pt]
\cL_{m+n+1}(x) &= (m+1) \frac{\df^m}{\df a^m}\,
  \frac{\cL_n^a(x) - \cL_n(x)}{a} \bigg\vert_{a = 0}
\,,\nn\\[2pt]
\cL^a_{m+1}(x) &= (m+1) \frac{\df^m}{\df b^m}\,
  \frac{\cL^{a+b}(x) - \cL^a(x)}{b} \bigg\vert_{b = 0}
\,. \nn
\end{align}

\begin{widetext}
The $\cL^a(x)$ satisfies the rescaling identity (for $\lambda > 0$)
\begin{equation} \label{eq:cLa_rescale}
\lambda\,\cL^a(\lambda x)
= \lim_{\e\to 0}\, \frac{\df}{\df x} \biggl[ \theta(x - \e)\,
  \frac{(\lambda x)^a - 1}{a} \biggr]
= \lambda^a \cL^a(x) + \frac{\lambda^a - 1}{a}\,\delta(x)
\,,\end{equation}
from which we can obtain the rescaling identity for $\cL_n(x)$,
\begin{equation} \label{eq:cLn_rescale}
\lambda\, \cL_n(\lambda x)
= \frac{\df^n}{\df a^n} \lambda^a \cL^a(x)\bigg\vert_{a = 0}
  + \frac{\ln^{n+1}\! \lambda}{n+1}\, \delta(x)
= \sum_{k = 0}^n \binom{n}{k} \ln^k\!\lambda\, \cL_{n-k}(x)
  + \frac{\ln^{n+1}\!\lambda}{n+1}\, \delta(x)
\,.\end{equation}
This agrees with the result in Eq.~(C3) of Ref.~\cite{Fleming:2007xt}.  We will
also need convolutions of two plus distributions,
\begin{align}
\inte{y} \cL^a(x - y)\, \cL^b(y)
&= \lim_{\e\to 0}\frac{\df}{\df x}\biggl\{ \theta(x - \e) \biggl[
\frac{x^{a+b}}{a+b} \V(a,b) + \frac{x^a - 1}{a}\,\frac{x^b - 1}{b}
  \biggr]\biggr\}
\nn\\[1ex]
&= \Bigl(\cL^{a+b}(x) + \frac{\delta(x)}{a+b} \Bigr) \V(a,b)
  + \Bigl(\frac{1}{a}+\frac{1}{b}\Bigr) \cL^{a+b}(x)
  - \frac{1}{b}\, \cL^a(x) - \frac{1}{a}\, \cL^b(x)
\,.\end{align}
In the second step we used the definition in \eq{cLa_def}. Here $V(a,b)$ is
defined by
\begin{equation}
\V(a,b) = \frac{\Gamma(a)\,\Gamma(b)}{\Gamma(a+b)} - \frac{1}{a} - \frac{1}{b}
\,,\end{equation}
which satisfies $\V(0,0) = 0$.  Taking derivatives with respect to $a$ and $b$ we
can get the corresponding formulas for convolutions involving $\cL_n$,
\begin{align} \label{eq:cL_conv}
\inte{y} \cL^a(x - y)\, \cL_n(y)
&= \frac{\df^n}{\df b^n}\biggl(\cL^{a+b}(x) + \frac{\delta(x)}{a+b} \biggr)
  \V(a,b) \bigg\vert_{b = 0}
  + \frac{\cL_{n+1}^a(x)}{n+1} + \frac{\cL_n^a(x) - \cL_n(x)}{a}
\nn\\*
&\equiv
 \frac{1}{a}
  \sum_{k = -1}^{n+1} \V^n_k(a)\, \cL_k^a(x)
\ - \ \frac{1}{a}\,\cL_n(x)
\,,\nn\\
\inte{y} \cL_m(x - y)\, \cL_n(y)
&= \frac{\df^m}{\df a^m}\, \frac{\df^n}{\df b^n}
  \biggl(\cL^{a+b}(x) + \frac{\delta(x)}{a+b} \biggr)
 \V(a,b) \bigg\vert_{a = b = 0} + \Bigl(\frac{1}{m+1} + \frac{1}{n+1}\Bigr)\,
  \cL_{m+n+1}(x)
\nn\\*
&\equiv
\sum_{k = -1}^{m+n+1} \V^{mn}_k\, \cL_k(x)
\,.\end{align}
The result in the second line of \eq{cL_conv} reproduces a result given in
Eq.~(B6) of Ref.~\cite{Jain:2008gb}.  The coefficients $\V_k^n(a)$ and $\V_k^{mn}$
are related to the Taylor series expansion of $V(a,b)$ around $a = 0$ and $a = b
= 0$.  The nonzero terms for $n\ge 0$ are
\begin{align} \label{eq:Vkna_def}
\V_k^n(a) &= \begin{cases}
   \displaystyle a\, \frac{\df^n}{\df b^n}\,\frac{\V(a,b)}{a+b}\bigg\vert_{b = 0}\,,
  &   k=-1\,, \\[10pt]
   \displaystyle  a\, \binom{n}{k}   \frac{\df^{n-k}}{\df b^{n-k}}\, \V(a,b)
   \bigg\vert_{b = 0} + \delta_{kn} \,, \qquad
  & 0\le k\le n \,,  \\[10pt]
  \displaystyle  \frac{a}{n+1} \,,
  & k=n+1  \,.
\end{cases}
\end{align}
The term $\delta_{kn}$ in $\V_k^n(a)$ and the last coefficient $\V_{n+1}^n(a)$
arise from the boundary terms in the convolution integral. The $\V_k^{mn}$ are symmetric in
$m$ and $n$, and the nonzero terms for $m,n\ge 0$ are
\begin{align} \label{eq:Vkmn_def}
\V_k^{mn} &= \begin{cases}
 \displaystyle \frac{\df^m}{\df a^m}\, \frac{\df^n}{\df b^n}\,\frac{\V(a,b)}{a+b}\bigg\vert_{a = b = 0} \,,
   & k=-1\,, \\[10pt]
 \displaystyle  \sum_{p = 0}^m \sum_{q = 0}^n \delta_{p+q,k}\,\binom{m}{p} \binom{n}{q}
\frac{\df^{m-p}}{\df a^{m-p}}\, \frac{\df^{n-q}}{\df b^{n-q}} \ \V(a,b)
  \bigg\vert_{a = b = 0}\,, \qquad
   & 0\le k \le m+n \,,\\[15pt]
 \displaystyle  \frac{1}{m+1} + \frac{1}{n+1}\,,
   & k=m+n+1 \,.
\end{cases}
\end{align}
The last coefficient $\V_{m+n+1}^{mn}$ again contains the boundary term.  Using
Eq.~\eqref{Lm1} we can extend the results in \eq{cL_conv} to include the cases
$n = -1$ or $m = -1$.  The relevant coefficients are
\begin{equation}
\V_{-1}^{-1}(a) = 1
\,,\qquad
\V_0^{-1}(a) = a
\,,\qquad
\V_{k \geq 1}^{-1}(a) = 0
\,,\qquad
\V^{-1,n}_k = \V^{n,-1}_k = \delta_{nk}
\,.\end{equation}
\end{widetext}

\section{\boldmath The invisible scheme for $\lambda_1$}
\label{app:invla1}

In this appendix we define a new scheme for $\lambda_1$, which we call the
``invisible scheme''.  It is a short distance scheme, free of the $u=1$
renormalon ambiguity, and it only deviates from the standard HQET definition of
$\lambda_1$ at ${\cal O}(\alpha_s^2)$.  Since the renormalon in $\lambda_1$
depends on the regularization scheme and in particular its
symmetries~\cite{Martinelli:1995zw, Neubert:1996zy}, it is desirable to define
$\lambda_1$ using a scheme which has the same symmetries as the multiloop
dimensional regularization calculations of its coefficient function.

For a general ultraviolet (UV) regulator $\Lambda_{\rm UV}$, the bare kinetic
energy operator is
\begin{equation} \label{Okinbare}
\bigl[ \bar b_v (i D_\perp)^2\, b_v \bigr]_{\rm bare}
  = Z_1\, \bar b_v b_v
  + Z_{\rm kin}\,  \bar b_v (i D_\perp)^2\, b_v + \ldots  \,,
\end{equation}
where $Z_{\rm I}\propto \Lambda_{\rm UV}^2$ and the ellipses denote higher
dimension operators whose coefficients vanish as $\Lambda_{\rm UV}\to \infty$.
Usually the kinetic energy matrix element in HQET is defined by
\begin{equation} \label{lam1}
  \lambda_1 = \mae{B}{\bar b_v (i D_\perp)^2\, b_v}{B} \,,
\end{equation}
where UV divergences are regulated in dimensional regularization (we follow our
convention of using full $B$ states even if this is not always the practice in
HQET). In that case, power divergences do not appear, so in Eq.~(\ref{Okinbare})
$Z_1=0$.  Furthermore, since this scheme respects reparametrization
invariance~\cite{Luke:1992cs}, $Z_{\rm kin}=1$ and $\lambda_1$ is
$\mu$ independent.  Nonperturbatively $\lambda_1$ can still be sensitive to the
quadratic UV divergence of $Z_1$ through a $u=1$ renormalon.  The presence of
this renormalon implies that there is an ${\cal O}(\Lambda_{\rm QCD}^2)$
ambiguity in the definition in Eq.~(\ref{lam1}).  In observables like a decay
rate this $\lambda_1$ ambiguity cancels against a corresponding infrared
renormalon ambiguity in the large-order behavior of the perturbation series in
the leading order Wilson coefficients~\cite{Beneke:1994bc}.

\begin{figure}[t]
\includegraphics[width=\columnwidth]{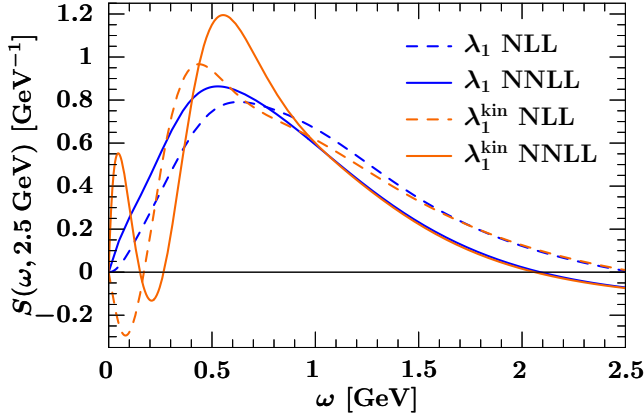}
\caption{Effect of the short distance subtractions
$\delta\lambda_1^\mathrm{kin}$ for the kinetic scheme at NLL (dashed) and NNLL
(solid) on $S(\w,2.5\GeV)$ for $\mu_\Lambda = 1.0\GeV$. The blue (dark) lines correspond to using
$m_b^\mathrm{kin}$ together with $\lambda_1^\mathrm{pole}$, while orange (light)
lines use both $m_b^\mathrm{kin}$ and $\lambda_1^\mathrm{kin}$.}
\label{fig:kinlambda1}
\end{figure}

We use the notation $\widehat \lambda_1$ for a generic short distance
definition, which does not suffer from the renormalon ambiguity. Any $\widehat
\lambda_1$ can be related to $\lambda_1$ by a perturbative series
$\delta\lambda_1\sim \alpha_s+\alpha_s^2 + \ldots$, where $\widehat\lambda_1 =
\lambda_1 - \delta \lambda_1$.  For the kinetic scheme the $u=1$ renormalon
ambiguity is avoided by defining $\lambda_1^{\rm kin}$ using the second moment
of a time-ordered product of currents with an explicit hard cutoff regulator
$\mu_f$~\cite{Bigi:1994ga}. Here $Z_1$ and $\delta\lambda_1\propto \mu_f^2
\alpha_s$.  Since the kinetic energy operator mixes into $\bar b_v b_v$, it
modifies the perturbation series multiplying $\bar b_v b_v$, and it is believed
that this removes the corresponding $u=1$ infrared renormalon in the Wilson
coefficients. In Ref.~\cite{Bosch:2004th} a ``shape function'' scheme for
$\lambda_1$ was introduced based on the second moment of $S(\omega,\mu)$ with a
cutoff, which also has $\delta \lambda_1\propto \mu_f^2 \alpha_s$. A potential
problem with these schemes is that they resolve an issue with the large-order
behavior of perturbation theory by introducing a series that starts with a term
$\delta\lambda_1\sim \alpha_s$.  While the inclusion of an $\ord{\alpha_s}$
term is known to provide good numerical stability when removing the $u=1/2$
renormalon from the pole mass, far less numerical analysis has been done on the
low-order impact of the $u=1$ renormalon in $\lambda_1$.  If the low-order
series is not yet influenced by the $u=1$ renormalon, then schemes with an
$\ord{\alpha_s}$ correction may oversubtract, and not improve the
perturbation series.  In Fig.~\ref{fig:kinlambda1} we show that in our
shape function analysis there is evidence that this is the case.\footnote{This
observation relies on our use of $F_1(\w)=\widehat F(\w)$ in
Eq.~(\ref{hatC0}) to determine $\delta C_0(\w)$. It would be interesting
to explore if a different choice would change the conclusions drawn from
Fig.~\ref{fig:kinlambda1}.} The dashed and solid
curves are the NLL and NNLL results, respectively. The curves show
$S(\w,\mu_i=2.5\GeV)$ with the kinetic mass scheme, but use either
$\lambda_1^{\rm kin}$ (blue curves) or $\lambda_1$ from Eq.~(\ref{lam1}) (orange
curves). In the $\lambda_1^{\rm kin}$ scheme the oscillatory behavior near the
origin at both NLL and NNLL order is indicative of an oversubtraction.

Some understanding of the weakness of the $u=1$ renormalon can be obtained from
analytic computations. For Lorentz invariant regulators the $u=1$ renormalon for
$\lambda_1$ is ``invisible''~\cite{Martinelli:1995zw, Neubert:1996zy}, namely
the ambiguity for $\lambda_1$ is smaller than dimensional analysis indicates. In
perturbation theory renormalon ambiguities appear as a divergent series with
terms $\sim n!\, \alpha_s^{n+1}$.  However, the leading renormalon series $\sim
n!\,\beta_0^n \alpha_s^{n+1}$ is absent, hence the renormalon ambiguity at
lowest order in $\lambda_1$ is ``invisible"~\cite{Martinelli:1995zw}.
Correspondingly, at one-loop order $Z_{\rm I}=0$ for Lorentz invariant
regulators (even for a hard cutoff), and generically $Z_1\propto
C_A\alpha_s^2$~\cite{Neubert:1996zy}.  In the kinetic scheme the regulator is
not Lorentz invariant, and hence not suppressed by ``invisibility''. The same
holds for definitions of $\lambda_1$ involving a lattice spacing
regulator~\cite{Martinelli:1995zw}.

To avoid oversubtractions from $\lambda_1$ at low orders in perturbation theory,
we would like to define a short distance scheme with $\delta\lambda_1\sim
\alpha_s^2$. This can be achieved by finding a scheme that is consistent with
the suppression indicated by the invisible renormalon. To construct an
``invisible scheme'' for $\lambda_1$ we define
\begin{align} \label{lam1i}
 \lambda_1^\inv(R) = \lambda_1 - \delta \lambda_1^\inv(R) \,,
\end{align}
where the series in $\alpha_s$ is obtained by evaluating the matrix element
\begin{align} \label{dellam1i}
  \delta\lambda_1^\inv(R) = \mae{b_v}{ \bar b_v (i D_\perp)^2\, b_v }{b_v} \Big|_{R}
\,.\end{align}
Here, $R$ is a Lorentz invariant hard cutoff UV regulator, ensuring that
$\delta\lambda_1^\inv(R)\sim R^2\alpha_s^2$. The definition in Eq.~(\ref{lam1i})
states that the invisible scheme $\lambda_1^\inv(R)$ is the kinetic energy of
the $b$ quark in the $B$ meson minus the free kinetic energy of the $b$ quark.
Since the ${\cal O}(\Lambda_{\rm QCD}^2)$ ambiguity in $\lambda_1$ is a UV
effect caused by the mixing of the kinetic operator into $\bar b_v b_v$, it is
the same for the $B$ meson and $b$ quark states, and cancels out in the
difference. This can be seen explicitly by using Eq.~(\ref{Okinbare}) and noting
that our states are normalized so that $\langle B_v |\bar b_v b_v |B_v\rangle =
\langle b_v | \bar b_v b_v | b_v \rangle = 1$. The ${\cal O}(\Lambda_{\rm
QCD}^2)$ ambiguity is universal to the definition of $\lambda_1$ and independent
of $R$. Although a precise definition of $R$ is needed to define the scheme, the
$u=1$ renormalon ambiguity cancels out in $\lambda_1-\delta\lambda_1^\inv$ for
any such regulator. We adopt a definition that allows us to use the computation in
Ref.~\cite{Neubert:1996zy}, where it was shown that $Z_1\ne 0$ at ${\cal
O}(\alpha_s^2)$ in a Lorentz invariant cutoff scheme. We define
\begin{align} \label{dlam1Rdef}
\delta \lambda_1^\inv(R)
&= \lim_{v'\to v} A_{vv'}(0, 0) \Big|_R
\\
&= \lim_{v'\to v} \int_0^R\! \int_0^R\!
  \frac{\df w\, \df w'}{(2\pi i)^2\, w\, w'}\,
  \mbox{\raisebox{-7pt}{$\stackrel{\displaystyle \rm Disc}{w}$}}\,
  \mbox{\raisebox{-9.5pt}{$\stackrel{\displaystyle \rm Disc}{w'}$}}\, A_{vv'}(w,w')
\nn\,,\end{align}
where
\begin{equation}
A_{vv'}(v\cdot k,v'\cdot k') = \frac{3\, \mae{b_{v'}(k')}{\bar b_{v'}
  v_\mu v'_\nu\, ig\, G^{\mu\nu} b_v }{b_v(k)}} {(v\cdot v')^2 - 1}
\,,\end{equation}
and as usual the discontinuity of a function is given by ${\rm Disc}_w\, f(w)
= \lim_{\e\to 0} \big[ f(w+i\e)-f(w-i\e)\big]$. Note that it is sufficient to
consider the $\bar b_{v'} ig\, G^{\mu\nu} b_v$ operator due to the virial
theorem in HQET, which relates $\lim_{v'\to v} A_{vv'}(0,0)$ to the
corresponding matrix element of the kinetic energy operator in Eq.~\eqref{dellam1i}.  The result for the
cutoff matrix element from Ref.~\cite{Neubert:1996zy} implies
\begin{align} \label{dlamiR}
\delta\lambda_1^\inv(R)
&= R^2\, C_F C_A \biggl(\frac{\pi^2}{3}-1\biggr)\,\frac{\alpha_s^2(\mu)}{4\pi^2}
\nn\\*
&= 0.232\, R^2\, \alpha_s^2(\mu) \,.
\end{align}
Equation~\eqref{dlamiR} gives the relation of the invisible $\lambda_1^\inv$ to the
HQET $\lambda_1$ in Eq.~\eqref{lam1}, which was used in the text. We use
$R=1\GeV$ as our default value. Equations~\eqref{lam1i} and \eqref{dlam1Rdef}
provide a $\mu$ independent definition for $\lambda_1^\inv(R)$, so the
$\mu$ dependence in $\alpha_s^2(\mu)$ will cancel against a higher order $R^2
\alpha_s^3(\mu)\ln(\mu/R)$ term in $\delta\lambda_1^\inv(R)$.


\end{document}